\documentclass[12pt]{amsart}
\usepackage{feynmp}
\usepackage{latexsym}
\usepackage{amsfonts}
\usepackage{amssymb}
\usepackage{amscd}
\usepackage{amsbsy}
\usepackage{amsmath}
\usepackage[dvips]{graphicx}
\usepackage{epsfig}    
\usepackage[cmtip,arrow]{xy}
\usepackage{pb-diagram,pb-xy}
    \usepackage{Bdiagx}

\usepackage[dvips]{graphicx}
\usepackage[dvips]{color}

\def\etc{{\it etc.\ }}
\def\ie{{\it i.e.\ }}
\def\eg{{\it e.g.\ }}
\def\cf{{\it cf.\ }}
\def\rhs{{\it r.h.s.\ }}

\def\End{\mathop{{\rm End}}\nolimits}
\def\Der{\mathop{{\rm Der}}\nolimits}
\def\Hom{\mathop{{\rm Hom}}\nolimits}

\def\im{\mathop{{\rm im}}\nolimits}
\def\Tr{\mathop{{\rm Tr}}\nolimits}

\def\xspan{\mathop{{\rm span}}\nolimits}

\def\p{^{\prime}}

\def\absv#1{ |#1| }
\def\numv#1{ \absv{#1} }

\def\del{ \partial }

\def\End{\mathop{{\rm End}}\nolimits}

\def\pr#1#2{ \noindent{\em Proof of #1~\ref{#2}.} }
\def\proof{ \noindent{\em Proof.} }
\def\skproof{ \noindent{\em Sketch of the Proof.} }

\def\qed{ \hfill $\Box$ }

\def\lrbc#1{ \left( #1 \right) }
\def\lrbs#1{ \left[ #1 \right] }

\def\atv#1#2{ \left. #1\right|_{#2} }

\def\mtr#1{ ||#1|| }

\def\prb#1{ \{ #1 \} }

\def\invv#1#2{ \lrbc{#1}^{#2} }
\def\invG#1{ \invv{#1}{\mfG} }
\def\invT#1{ \invv{#1}{\mfT} }

\def\stdf#1#2{ \{ #1\;|\; #2 \} }
\def\stdfl#1#2{ \left\{\left. #1\;\right|\; #2 \right\} }
\def\xspanv#1#2{ \xspan\left(\left.#1\;\right|\;#2\right) }
\def\mltiv#1#2{ (#1)_{#2} }

\def\corr#1{ \langle #1 \rangle }
\def\corri#1#2{ \corr{#1}_{#2} }

\def\Srs{ {\rm S} }
\def\Urs{ {\rm U} }

\def\Wrs{ {\textstyle\bigwedge} }
\def\Wrss{ {\scriptstyle\wedge} }

\def\Sg{ \Srs\mfg }
\def\Sgs{ \Srs\mfgs }
\def\Ug{ \Urs\mfg }
\def\Wg{ \Wrs\mfg }
\def\Wgs{ \Wrs\mfgs }

\def\xmap#1#2#3{ #1\colon #2\longrightarrow #3 }
\def\ximap#1#2#3{ #1\colon #2\hookrightarrow #3 }
\def\xmapt#1{ \mathop{{\longmapsto}}^{#1} }

\def\xumap#1#2#3{ \begin{CD} #1 @>{#2}>> #3 \end{CD} }

\def\smon{ \sum_{m=1}^n }
\def\smono{ \sum_{m=1}^{n+1} }
\def\smoi{ \sum_{m=1}^\infty }

\parskip=\medskipamount                 
\thicklines                     
\newcommand{\bimn}[7]{\bibitem{#1}#2,
{\em #3}, { #4}\hspace{0.25em}{\bf
#5}\hspace{0.25em}(#6)\hspace{0.25em}{#7}.}

\def\inbar{\vrule height1.5ex width.4pt depth0pt}
\def\IC{\relax\,\hbox{$\inbar\kern-.3em{\rm C}$}}
\def\IN{\relax{\rm I\kern-.18em N}}
\def\IQ{\relax\,\hbox{$\inbar\kern-.3em{\rm Q}$}}
\def\IR{\relax{\rm I\kern-.18em R}}
\def\ZZ{\relax{\sf Z\kern-.4em Z}}

\def\cA{{\cal A}}

\def\etc{{\it etc\/}}

\newtheorem{theorem}{Theorem}[section]
\newtheorem{proposition}[theorem]{Proposition}
\newtheorem{corollary}[theorem]{Corollary}
\newtheorem{conjecture}[theorem]{Conjecture}
\newtheorem{lemma}[theorem]{Lemma}

\newtheorem{remark}[theorem]{Remark}

\catcode`\@=11

\newif\if@fewtab\@fewtabtrue

\catcode`\@=11

\newif\if@fewtab\@fewtabtrue

{\count255=\time\divide\count255 by 60
\xdef\hourmin{\number\count255} \multiply\count255
by-60\advance\count255 by\time
\xdef\hourmin{\hourmin:\ifnum\count255<10 0\fi\the\count255}}
\def\ps@draft{\let\@mkboth\@gobbletwo
    \def\@oddhead{}
    \def\@oddfoot
      {\hbox to 7 cm{\footnotesize {\em Draft of \jobname:} \draftdate
       \hfil}\hskip -7cm\hfil\rm\thepage \hfil}
    \def\@evenhead{}\let\@evenfoot\@oddfoot}

\def\ceqno{\global\@fewtabfalse
    \ifcase\@eqcnt \def\@tempa{& & &}\or \def\@tempa{& &}
      \or \def\@tempa{&}
      \or\def\@tempa{}\fi\@tempa
{\rm(\theequation)}}

\def\aeqno#1{\global\@fewtabfalse
    \ifcase\@eqcnt \def\@tempa{& & &}\or \def\@tempa{& &}
      \or \def\@tempa{&}
      \or\def\@tempa{}\fi\@tempa
{\rm(\theequation,#1)}}

\def\label#1{\ifnum\draftcontrol=1
 \global\def\draftnote{$\scriptstyle #1$}\fi
 \@bsphack\if@filesw {\let\thepage\relax
   \def\protect{\noexpand\noexpand\noexpand}%
\xdef\@gtempa{\write\@auxout{\string
      \newlabel{#1}{{\@currentlabel}{\thepage}}}}}\@gtempa
   \if@nobreak \ifvmode\nobreak\fi\fi\fi
  \@esphack}

\def\alabel#1#2{\label{#1}\global\@fewtabfalse
    \ifcase\@eqcnt \def\@tempa{& & &}\or \def\@tempa{& &}
      \or \def\@tempa{&}
      \or\def\@tempa{}\fi\@tempa
{\hbox to 3cm{\phantom{\rm(\theequation,#2)} \draftnote
\hfil}\hskip -3cm {\rm(\theequation,#2)}}}

\def\clabel#1{\label{#1}\global\@fewtabfalse
    \ifcase\@eqcnt \def\@tempa{& & &}\or \def\@tempa{& &}
      \or \def\@tempa{&}
      \or\def\@tempa{}\fi\@tempa
{\hbox to 3cm{\phantom{\rm(\theequation)} \draftnote \hfil}\hskip
-3cm{\rm(\theequation)}}}

\def\eqnarray{\def\draftnote{{}}\global\@fewtabtrue
\stepcounter{equation}\let\@currentlabel=\theequation
\global\@eqnswtrue
\global\@eqcnt\z@\tabskip\@centering\let\\=\@eqncr
$$\halign to \displaywidth\bgroup\@eqnsel\hskip\@centering\@eqcnt\z@
  $\displaystyle\tabskip\z@{##}$&\global\@eqcnt\@ne
  \hskip 1\arraycolsep \hfil$\displaystyle{##}$\hfil
  &\global\@eqcnt\tw@ \hskip 1\arraycolsep
$\displaystyle\tabskip\z@{##}$ \hfil
\tabskip\@centering&\global\@eqcnt\thr@@\llap{##}\tabskip\z@ \cr}

\def\endeqnarray{\@@eqncr\egroup
      \global\advance\c@equation\m@ne$$\global\@ignoretrue}

\def\@eqnnum{\hbox to 3cm{\phantom{\rm(\theequation)} \draftnote
                         \hfil}\hskip -3cm {\rm(\theequation)}}

\def\@@eqncr{\let\@tempa\relax
    \ifcase\@eqcnt \def\@tempa{& & &}\or \def\@tempa{& &}
      \or \def\@tempa{&}
      \or\def\@tempa{}
\fi\@tempa \if@eqnsw \if@fewtab\@eqnnum\fi
\stepcounter{equation}\fi\global
\@eqnswtrue\global\@eqcnt\z@\global\@fewtabtrue\cr}

\def\draftcite#1{\ifnum\draftcontrol=1#1\else{}\fi}

\def\@lbibitem[#1]#2{\item{}\hskip -3cm \hbox to 2cm
{\hfil$\scriptstyle\draftcite{#2}$}\hskip
1cm[\@biblabel{#1}]\if@filesw
     {\def\protect##1{\string ##1\space}\immediate
      \write\@auxout{\string\bibcite{#2}{#1}}}\fi\ignorespaces}

\def\@bibitem#1{\item\hskip -3cm \hbox to 2cm
{\hfil $\scriptstyle\draftcite{#1}$}\hskip 1cm \if@filesw
\immediate\write\@auxout
       {\string\bibcite{#1}{\the\value{\@listctr}}}\fi\ignorespaces}

\def\nsection#1{\section{#1}\setcounter{equation}{0}}

\def\draftdate{\number\month/\number\day/\number\year\ \ \ \hourmin }
 \global\def\draftcontrol{0}
\catcode`\@=12

\def\theequation{{\thesection.\arabic{equation}}}

\def\qq{\begin{eqnarray}}
\def\qqq{\end{eqnarray}}
\def\rx#1{~(\ref{#1})}

\def\ex#1{eq.\hspace*{-3pt}\rx{#1}}
\def\eex#1{eqs.\hspace*{-3pt}\rx{#1}}
\def\cx#1{~\cite{#1}}
\def\rw#1{~\ref{#1}}

\def\fg#1{Fig.~\ref{#1}}
\def\fgs#1{Figs.~\ref{#1}}

\newlength{\shiftwidth}
\def\shift#1{&&\hbox to \shiftwidth{\hfill $\displaystyle#1$}}
\newlength{\sshiftwidth}
\def\sshift#1{\lefteqn{\hbox to
\sshiftwidth{\hfill$\displaystyle#1$}}}

\def\qbezier{\bezier{120}}
\def\DottedCircle{
\bezier{4}(0.966,-0.259)(1.04,0)(0.966,0.259)
\bezier{4}(0.966,0.259)(0.897,0.518)(0.707,0.707)
\bezier{4}(0.707,0.707)(0.518,0.897)(0.259,0.966)
\bezier{4}(0.259,0.966)(0,1.04)(-0.259,0.966)
\bezier{4}(-0.259,0.966)(-0.518,0.897)(-0.707,0.707)
\bezier{4}(-0.707,0.707)(-0.897,0.518)(-0.966,0.259)
\bezier{4}(-0.966,0.259)(-1.04,0)(-0.966,-0.259)
\bezier{4}(-0.966,-0.259)(-0.897,-0.518)(-0.707,-0.707)
\bezier{4}(-0.707,-0.707)(-0.518,-0.897)(-0.259,-0.966)
\bezier{4}(-0.259,-0.966)(0,-1.04)(0.259,-0.966)
\bezier{4}(0.259,-0.966)(0.518,-0.897)(0.707,-0.707)
\bezier{4}(0.707,-0.707)(0.897,-0.518)(0.966,-0.259) }
\def\Endpoint[#1]{
\ifcase#1 \put(1,0){\circle*{0.15}}
\or\put(0.866,0.5){\circle*{0.15}}
\or\put(0.5,0.866){\circle*{0.15}} \or\put(0,1){\circle*{0.15}}
\or\put(-0.5,0.866){\circle*{0.15}}
\or\put(-0.866,0.5){\circle*{0.15}} \or\put(-1,0){\circle*{0.15}}
\or\put(-0.866,-0.5){\circle*{0.15}}
\or\put(-0.5,-0.866){\circle*{0.15}} \or\put(0,-1){\circle*{0.15}}
\or\put(0.5,-0.866){\circle*{0.15}}
\or\put(0.866,-0.5){\circle*{0.15}} \fi}
\def\Arc[#1]{
\thicklines         
\ifcase#1 \bezier{25}(0.966,-0.259)(1.04,0)(0.966,0.259) \or
\bezier{25}(0.966,0.259)(0.897,0.518)(0.707,0.707) \or
\bezier{25}(0.707,0.707)(0.518,0.897)(0.259,0.966) \or
\bezier{25}(0.259,0.966)(0,1.04)(-0.259,0.966) \or
\bezier{25}(-0.259,0.966)(-0.518,0.897)(-0.707,0.707) \or
\bezier{25}(-0.707,0.707)(-0.897,0.518)(-0.966,0.259) \or
\bezier{25}(-0.966,0.259)(-1.04,0)(-0.966,-0.259) \or
\bezier{25}(-0.966,-0.259)(-0.897,-0.518)(-0.707,-0.707) \or
\bezier{25}(-0.707,-0.707)(-0.518,-0.897)(-0.259,-0.966) \or
\bezier{25}(-0.259,-0.966)(0,-1.04)(0.259,-0.966) \or
\bezier{25}(0.259,-0.966)(0.518,-0.897)(0.707,-0.707) \or
\bezier{25}(0.707,-0.707)(0.897,-0.518)(0.966,-0.259) \fi}
\def\DottedArc[#1]{
\ifcase#1 \bezier{4}(0.966,-0.259)(1.04,0)(0.966,0.259) \or
\bezier{4}(0.966,0.259)(0.897,0.518)(0.707,0.707) \or
\bezier{4}(0.707,0.707)(0.518,0.897)(0.259,0.966) \or
\bezier{4}(0.259,0.966)(0,1.04)(-0.259,0.966) \or
\bezier{4}(-0.259,0.966)(-0.518,0.897)(-0.707,0.707) \or
\bezier{4}(-0.707,0.707)(-0.897,0.518)(-0.966,0.259) \or
\bezier{4}(-0.966,0.259)(-1.04,0)(-0.966,-0.259) \or
\bezier{4}(-0.966,-0.259)(-0.897,-0.518)(-0.707,-0.707) \or
\bezier{4}(-0.707,-0.707)(-0.518,-0.897)(-0.259,-0.966) \or
\bezier{4}(-0.259,-0.966)(0,-1.04)(0.259,-0.966) \or
\bezier{4}(0.259,-0.966)(0.518,-0.897)(0.707,-0.707) \or
\bezier{4}(0.707,-0.707)(0.897,-0.518)(0.966,-0.259) \fi}
\def\Chord[#1,#2]{
\thinlines \ifnum#1>#2\Chord[#2,#1] \else\ifnum#1<#2 \ifcase#1
\ifcase#2 \or\qbezier(1,0)(0.516,0.138)(0.866,0.5)
\or\qbezier(1,0)(0.45,0.26)(0.5,0.866)
\or\qbezier(1,0)(0.327,0.327)(0,1)
\or\qbezier(1,0)(0.179,0.311)(-0.5,0.866)
\or\qbezier(1,0)(0.0536,0.2)(-0.866,0.5) \or\put(1, 0){\line(-2,
0){2}} \or\qbezier(1,0)(0.0536,-0.2)(-0.866,-0.5)
\or\qbezier(1,0)(0.179,-0.311)(-0.5,-0.866)
\or\qbezier(1,0)(0.327,-0.327)(0,-1)
\or\qbezier(1,0)(0.45,-0.26)(0.5,-0.866)
\or\qbezier(1,0)(0.516,-0.138)(0.866,-0.5) \fi \or\ifcase#2\or
\or\qbezier(0.866,0.5)(0.378,0.378)(0.5,0.866)
\or\qbezier(0.866,0.5)(0.26,0.45)(0,1)
\or\qbezier(0.866,0.5)(0.12,0.446)(-0.5,0.866)
\or\qbezier(0.866,0.5)(0,0.359)(-0.866,0.5)
\or\qbezier(0.866,0.5)(-0.0536,0.2)(-1,0) \or\put(0.866,
0.5){\line(-5, -3){1.73}}
\or\qbezier(0.866,0.5)(0.146,-0.146)(-0.5,-0.866)
\or\qbezier(0.866,0.5)(0.311,-0.179)(0,-1)
\or\qbezier(0.866,0.5)(0.446,-0.12)(0.5,-0.866)
\or\qbezier(0.866,0.5)(0.52,0)(0.866,-0.5) \fi \or\ifcase#2\or\or
\or\qbezier(0.5,0.866)(0.138,0.516)(0,1)
\or\qbezier(0.5,0.866)(0,0.52)(-0.5,0.866)
\or\qbezier(0.5,0.866)(-0.12,0.446)(-0.866,0.5)
\or\qbezier(0.5,0.866)(-0.179,0.311)(-1,0)
\or\qbezier(0.5,0.866)(-0.146,0.146)(-0.866,-0.5) \or\put(0.5,
0.866){\line(-3, -5){1}} \or\qbezier(0.5,0.866)(0.2,-0.0536)(0,-1)
\or\qbezier(0.5,0.866)(0.359,0)(0.5,-0.866)
\or\qbezier(0.5,0.866)(0.446,0.12)(0.866,-0.5) \fi
\or\ifcase#2\or\or\or \or\qbezier(0,1.)(-0.138,0.516)(-0.5,0.866)
\or\qbezier(0,1.)(-0.26,0.45)(-0.866,0.5)
\or\qbezier(0,1.)(-0.327,0.327)(-1,0)
\or\qbezier(0,1.)(-0.311,0.179)(-0.866,-0.5)
\or\qbezier(0,1.)(-0.2,0.0536)(-0.5,-0.866) \or\put(0, 1){\line(0,
-2){2}} \or\qbezier(0,1.)(0.2,0.0536)(0.5,-0.866)
\or\qbezier(0,1.)(0.311,0.179)(0.866,-0.5) \fi
\or\ifcase#2\or\or\or\or
\or\qbezier(-0.5,0.866)(-0.378,0.378)(-0.866,0.5)
\or\qbezier(-0.5,0.866)(-0.45,0.26)(-1,0)
\or\qbezier(-0.5,0.866)(-0.446,0.12)(-0.866,-0.5)
\or\qbezier(-0.5,0.866)(-0.359,0)(-0.5,-0.866)
\or\qbezier(-0.5,0.866)(-0.2,-0.0536)(0,-1) \or\put(-0.5,
0.866){\line(3, -5){1}}
\or\qbezier(-0.5,0.866)(0.146,0.146)(0.866,-0.5) \fi
\or\ifcase#2\or\or\or\or\or
\or\qbezier(-0.866,0.5)(-0.516,0.138)(-1,0)
\or\qbezier(-0.866,0.5)(-0.52,0)(-0.866,-0.5)
\or\qbezier(-0.866,0.5)(-0.446,-0.12)(-0.5,-0.866)
\or\qbezier(-0.866,0.5)(-0.311,-0.179)(0,-1)
\or\qbezier(-0.866,0.5)(-0.146,-0.146)(0.5,-0.866) \or\put(-0.866,
0.5){\line(5, -3){1.73}} \fi \or\ifcase#2\or\or\or\or\or\or
\or\qbezier(-1,0)(-0.516,-0.138)(-0.866,-0.5)
\or\qbezier(-1,0)(-0.45,-0.26)(-0.5,-0.866)
\or\qbezier(-1,0)(-0.327,-0.327)(0,-1)
\or\qbezier(-1,0)(-0.179,-0.311)(0.5,-0.866)
\or\qbezier(-1,0)(-0.0536,-0.2)(0.866,-0.5) \fi
\or\ifcase#2\or\or\or\or\or\or\or
\or\qbezier(-0.866,-0.5)(-0.378,-0.378)(-0.5,-0.866)
\or\qbezier(-0.866,-0.5)(-0.26,-0.45)(0,-1)
\or\qbezier(-0.866,-0.5)(-0.12,-0.446)(0.5,-0.866)
\or\qbezier(-0.866,-0.5)(0,-0.359)(0.866,-0.5) \fi
\or\ifcase#2\or\or\or\or\or\or\or\or
\or\qbezier(-0.5,-0.866)(-0.138,-0.516)(0,-1)
\or\qbezier(-0.5,-0.866)(0,-0.52)(0.5,-0.866)
\or\qbezier(-0.5,-0.866)(0.12,-0.446)(0.866,-0.5) \fi
\or\ifcase#2\or\or\or\or\or\or\or\or\or
\or\qbezier(0,-1.)(0.138,-0.516)(0.5,-0.866)
\or\qbezier(0,-1.)(0.26,-0.45)(0.866,-0.5) \fi
\or\ifcase#2\or\or\or\or\or\or\or\or\or\or
\or\qbezier(0.5,-0.866)(0.378,-0.378)(0.866,-0.5) \fi\fi\fi\fi}
\def\FullChord[#1,#2]{
\Endpoint[#1] \Endpoint[#2] \Arc[#1] \Arc[#2] \Chord[#1,#2] }
\def\EndChord[#1,#2]{
\Endpoint[#1] \Endpoint[#2] \Chord[#1,#2] }
\def\Picture#1{
\begin{picture}(2,1)(-1,-0.167)
#1
\end{picture}
}
\def\DottedChordDiagram[#1,#2]{
\Picture{\DottedCircle \FullChord[#1,#2]} }
\def\ZZ{ \mathbb{Z} }
\def\IQ{ \mathbb{Q} }
\def\IC{ \mathbb{C} }
\def\IR{ \mathbb{R} }
\def\IP{ \mathbb{P} }

\def\vl{ \vec{\lambda} }

\def\mfg{ \mathfrak{g} }
\def\mfh{ \mathfrak{h} }
\def\mfr{ \mathfrak{r} }
\def\mfv{ \mathfrak{v} }
\def\mfG{ \mathfrak{G} }
\def\mfT{ \mathfrak{T} }

\def\cA{ {\cal A} }

\def\tcA{ \tilde{\cA} }

\def\cA{ \mathcal{A} }

\def\smvi#1{ \sum_{#1}^\infty }

\def\snoi{ \smvi{n=1} }

\def\smoi{ \smvi{m=1} }

\def\hlfv{ {1\over 2} }

\def\xP{ P }

\def\WZW{WZW}
\def\WZWm{\WZW\ model}
\def\WZWms{\WZWm s}

\def\nbh{neighborhood}
\def\nbhs{\nbh s}

\def\ctgrf{categorification}

\def\Grsm{Grassmannian}
\def\Grsms{\Grsm s}

\def\Dbr{$D$-brane}

\def\Am{A-model}
\def\Ams{\Am s}
\def\tAm{topological \Am}
\def\tAms{\tAm s}

\def\ws{world-sheet}
\def\wss{\ws s}
\def\sgr{seam graph}
\def\Rs{Riemann surface}
\def\Rss{\Rs s}
\def\sRs{\smd\ \Rs}
\def\sRss{\sRs s}
\def\smd{\sm ed}
\def\sm{seam}

\def\lg{local graph}
\def\lgs{\lg s}

\def\cprd{cross-product}
\def\cprds{\cprd s}

\def\Kh{K\"{a}hler}
\def\Khm{\Kh\ manifold}
\def\Khms{\Khm s}

\def\Lg{Lagrangian}
\def\Lgsm{\Lg\ submanifold}
\def\Lgsms{\Lgsm s}

\def\CFT{CFT}
\def\CFTs{\CFT s}
\def\QFT{QFT}
\def\QFTs{\QFT s}

\def\nU{ U }
\def\nUv#1{ \nU_{#1} }
\def\nUP{ \nUv{\Pt} }
\def\nUPp{ \nUv{\Ptp} }

\def\mf{ f }
\def\mfv#1{ \mf_{#1} }
\def\mfP{ \mfv{\Pt} }
\def\mfot{ \mfv{12} }

\def\Grl{ {\rm Gr} }
\def\Grv#1{ \Grl_{#1} }
\def\Grvv#1#2{ \Grv{#1,#2} }
\def\Grvn#1{ \Grvv{#1}{n} }
\def\Grmn{ \Grvn{m} }
\def\Grnmn{ \Grvn{n-m} }
\def\bGrmn{ \overline{\Grl}_{m,n} }
\def\Grmin{ \Grvn{\msi} }

\def\ICv#1{ \IC^{#1} }
\def\ICmi{ \ICv{\msi} }
\def\ICn{ \ICv{n} }

\def\msi{ m_i }

\def\Gra{ \Gamma }
\def\Grap{ \Gra\p }
\def\Grgra{ \Gra_\gra }
\def\gra{ \gamma }
\def\grav#1{ \gra_{#1} }
\def\gravert{ \grav{\vert} }
\def\graP{ \grav{\Pt} }
\def\graPv#1{ \gra(\Pt_{#1}) }
\def\WS{ \Sigma }
\def\WSi{ \WS_i }
\def\WSgra{ \WS_\gra }
\def\dlWS{ \del\WS }
\def\WSp{ \Sigma^\prime }

\def\bsRv#1#2{ (#1,#2) }
\def\bsRSG{ \bsRv{\WS}{\Gra} }
\def\bsRSGp{ \bsRv{\WSp}{\Grap} }
\def\bsRSgra{ \bsRv{\WSgra}{\Grgra} }

\def\Th{ T }
\def\Thi{ \Th_i }
\def\Tho{ \Th_1 }
\def\Tht{ \Th_2 }
\def\Thb{ \bar{\Th} }
\def\Thu{ \Th_{\up} }
\def\Thd{ \Th_{\down} }
\def\Thdb{ \Thb_{\down} }

\def\Pt{ P }
\def\Ptp{ \Pt\p }
\def\Pto{ \Pt_1 }
\def\Ptt{ \Pt_2 }

\def\pV{ V }
\def\pVo{ \pV_1 }
\def\pVt{ \pV_2 }

\def\Op{ O }
\def\Opv#1{ \Op_{#1} }
\def\Opa{ \Opv{\ia} }
\def\Opb{ \Opv{\ib} }
\def\OpP{ \Op(\Pt) }

\def\Opvv#1#2{ \Op_{#1}(#2) }
\def\OpomP{ \Opvv{\com}{\Pt} }
\def\OpomPv#1{ \Opvv{\com_{#1}}{\Pt_{#1}} }

\def\sOp{ \mathfrak{O} }
\def\sOpv#1{ \sOp_{#1} }
\def\sOpgra{ \sOpv{\gra} }

\def\ia{ a }
\def\ib{ b }

\def\mOp{ g }
\def\mOpdv#1{ \mOp_{#1} }
\def\mOpuv#1{ \mOp^{#1} }
\def\mOpdab{ \mOpdv{\ia\ib} }
\def\mOpuab{ \mOpuv{\ia\ib} }

\def\com{ \omega }
\def\Kom{ \omega_{\rm K} }
\def\Komi{ \Kom^{(i)} }
\def\CohH{ {\rm H} }
\def\Cohvv#1#2{ \CohH^{#1}(#2) }

\def\Cohgv#1{ \Cohvv{\ast}{#1} }
\def\CohgXi{ \Cohgv{\KXi} }
\def\CohgXLe{ \Cohgv{\XLe} }
\def\CohgMgP{ \Cohgv{\MgP} }
\def\CohgMg{ \Cohgv{\Mgra} }

\def\Cohtvv#1#2#3{ \CohH^{#1}_{#2}(#3) }
\def\Cohtv#1#2{ \Cohtvv{\ast}{#1}{#2} }
\def\CohALe{ \Cohtv{\cnAed}{\XLe} }

\def\CohgAMgra{ \Cohtv{\cnAgra}{\Mgra} }

\def\fp{ p }
\def\fpe{ \fp_{\ed} }
\def\fpi{ \fp^{-1} }
\def\fpie{ \fpi_{\ed} }

\def\fP{ P }
\def\fPe{ \fP_{\ed} }
\def\fPei{ \fPe^{-1} }
\def\fPi{ \fP_i }
\def\fPvert{ \fP_\vert }

\def\Mv#1{ \mathcal{M}_{#1} }
\def\MgP{ \Mv{\graP} }
\def\MgPv#1{ \Mv{\graPv{#1}} }
\def\Mgra{ \Mv{\gra} }
\def\Mgvert{ \Mv{\gravert} }
\def\MbsRSG{ \Mv{0,\bsRSG} }

\def\up{ {\rm up} }
\def\down{ {\rm down} }

\def\siabsOp{ \sum_{\ia,\ib\in\sOp} }
\def\siabsOpgra{ \sum_{\ia,\ib\in\sOpgra} }

\def\KX{ X }
\def\KXi{ \KX_i }
\def\KXe{ \KX_{\ed} }
\def\KXgra{ \KX_{\gra} }
\def\KXgvert{ \KX_{\gravert} }
\def\KXbsRSG{ \KX_{\bsRSG} }

\def\XL{ L }
\def\tXL{ \tilde{\xL} }
\def\XLe{ \XL_{\ed} }
\def\tXLe{ \tXL_{\ed} }
\def\XLi{ \XL_i }

\def\ed{ e }
\def\Ned{ N_{\ed} }
\def\cnA{ A }
\def\cnAed{ \cnA_{\ed} }
\def\cnAgP{ \cnA_{\graP} }
\def\cnAgra{ \cnA_{\gra} }
\def\tcnA{ \tilde{\cnA} }
\def\tcnAed{ \tcnA_{\ed} }
\def\vert{ v }
\def\xne{ n_{\ed} }
\def\xngra{ n_{\gra} }
\def\xmgra{ m_{\gra} }
\def\xnS{ n_{\WS} }

\def\llgr#1#2#3{ #1_{#3_1},\ldots,#1_{#3_{#2}} }
\def\lpgr#1#2#3{ #1_{#3_1}\times\cdots\times#1_{#3_{#2}} }
\def\ligr#1#2#3{ #1_{#3_1}\cap\cdots\cap#1_{#3_{#2}} }
\def\lopgr#1#2#3{ #1_{#3_1}\oplus\cdots\oplus#1_{#3_{#2}} }

\def\rmU{ {\rm U} }
\def\rmu{ {\rm u} }
\def\Uv#1{ \rmU(#1) }
\def\uv#1{ \rmu(#1) }
\def\Un{ \Uv{n} }

\def\phf{ \phi }
\def\phfv#1{ \phf_{#1} }
\def\phfi{ \phfv{(i)} }

\def\corrci#1#2{ \corri{#1}{0,#2} }
\def\corrl{ \OpomPv{1}\cdots\OpomPv{n} }

\def\xprv#1{ F_{#1} }
\def\xprsv#1{ \xprv{#1,\ast} }
\def\xprpv#1{ \xprv{\Pt_{#1}} }
\def\xprpsv#1{ \xprsv{\Pt_{#1}} }
\def\xprP{ \xprv{\Pt} }

\def\rmT{ {\rm T} }
\def\rmTx{ \rmT_x }
\def\rmHom{ {\rm Hom} }

\def\xV{ V }
\def\xVort{ \xV^{\perp} }

\def\yA{ A }
\def\yB{ B }

\def\bm{ \mathbf{m} }
\def\XLbm{ \XL_{\bm} }
\def\ym{ m }
\def\ymij{ \ym_{ij} }
\def\yV{ V }
\def\yVij{ \yV_{ij} }

\def\oik{ 1\leq i\leq k }
\def\ojN{ 1\leq j\leq N }

\def\xS{ S }
\def\xSv#1{ \xS_{#1} }
\def\xSo{ \xSv{1} }
\def\xSt{ \xSv{2} }

\def\xL{ L }

\begin{document}

\setlength{\unitlength}{3947sp}
\setlength{\unitlength}{1mm}

\begin{titlepage}
\vfill
\begin{center}

{\large \bf
Topological A-models on seamed Riemann surfaces.}\\

\bigskip

\bigskip
\centerline{L. Rozansky\footnote{ This work was supported by NSF
Grants DMS-0196235 and DMS-0196131} }

\centerline{\em Department of Mathematics, University of North
Carolina} \centerline{\em CB \#3250, Phillips Hall}
\centerline{\em Chapel Hill, NC 27599} \centerline{{\em E-mail
address:} {\tt rozansky@math.unc.edu}}

\vfill
{\bf Abstract}

\end{center}
\begin{quotation}

We define a class of topological A-models on a collection of
Riemann surfaces, whose boundaries are sewn together along the seams.
The target spaces for the Riemann surfaces are the \Grsms\
$\Grmin$ with the common value of $n$, and the boundary
conditions at the seams demand that the spaces $\ICmi\subset\ICn$
present the orthogonal decomposition of $\ICn$. The whole
construction is a \QFT\ interpretation of a part of Khovanov's
\ctgrf\ of the $sl(3)$ HOMFLY polynomial.

\end{quotation} \vfill \end{titlepage}

\pagebreak


\hyphenation{Re-she-ti-khin}
\hyphenation{Tu-ra-ev}
\hyphenation{sub-stan-ti-al}
\hyphenation{in-va-ri-ant}

\section{A \QFT\ on a \sRs}

The idea of defining a 2-dimensional theory on a `seamed' \ws\
is not exactly new. String theories, in which strings formed
`networks', were considered by many authors (see, \eg\cx{AHK},\cx{GZ},\cx{Sen} and
references therein). A similar idea was floated recently
in\cx{BBDO} in relation to the study of the boundary conditions in
\CFTs\ (see also the review\cx{S}). The present paper was inspired by M.~Khovanov's
construction\cx{Kh1}, which includes both seamed surfaces (seamed along
disjoint circles and called `foam') and a supply of the boundary
conditions
sufficient for defining an interesting \Am.

\subsection{Seamed Riemann surfaces}

Here is the definition of a seamed Riemann surface. First, we define it as a topological space.
Let $\WS$ be an oriented compact 2-dimensional manifold
(perhaps consisting of several connected components) with a
boundary $\dlWS$, which is a disjoint union of circles $S^1$. Let $\Gamma$
be a graph, which we will call a \emph{\sgr}. $\Gra$ may contain disjoint circles.
A cycle on $\Gra$ is either a disjoint circle or
a finite sequence of edges, such that the beginning of the
next edge coincides with the end of the previous edge, and the end
of the final edge coincides with the beginning of the first one. A
\emph{\sRs} $\bsRSG$ is constructed as a topological space
by gluing the circles of
$\dlWS$ to some cycles of $\Gra$.
We assume that every edge of
$\Gra$ is glued to at least one circle of $\dlWS$, otherwise it
can be removed from $\Gra$ without affecting the construction.
A simple example of a \sRs\ is depicted in \fg{f1}.

\begin{figure}[hb]
\input fig1v2.pstex_t
\caption{An example of a \sRs. Every triangle in this picture represents a connected
component $\WSi$ of a \Rs\ $\WS$.}
\label{f1}
\end{figure}

Next, we endow $\bsRSG$ with a complex structure by choosing
a \nbh\ $\nUP\subset\bsRSG$ of every point
$\Pt\in\bsRSG$ and specifying, which complex-valued functions on those \nbhs\ are called
analytic. This can be done by selecting the maps
\qq
\xmap{\mfP}{\nUP}{\IC}
\label{1x}
\qqq
and then defining the analytic functions
on $\nUP$ as pull-backs of the analytic functions on $\IC$. Of
course, these definitions must be consistent on the intersections
$\nUP\cap\nUPp$.

\begin{figure}[hbt]
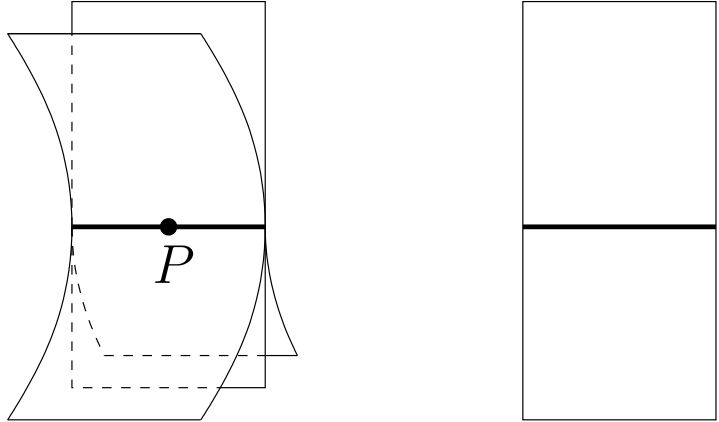

\input fig2wv1.pstex_t
\caption{A neighborhood of an edge and the real line in the complex
plane}
\label{f2}
\end{figure}

There are three different types of points of $\bsRSG$ depending on the structure
of their \nbhs:
the internal points of $\WS$, the internal points of the edges of
$\Gra$ and the vertices of $\Gra$.
A complex structure in the \nbhs\ of all internal points of
$\WS$ is defined simply by selecting a complex structure on
$\WS\setminus\dlWS$ compatible with its orientation.
If
$\Pt$ is an internal point of an edge $\ed$ of $\Gra$, then its \nbh\ is
depicted in \fg{f2}.
We draw the attached strips either above or below $\ed$ depending
on the orientation that they induce on it.
Hence
$\mfP$ maps the upper strips to the upper half-plane of
$\IC$, while mapping the lower strips to the lower half-plane of
$\IC$.

Note that for two strips $\xSo,\xSt$ attached to $\ed$, the map
$\mfP$ defines locally an analytic map $\xmap{\mfot}{\xSo}{\xSt}$
by the condition that $\mfP(\mfot(\Pto)) = \mfP(\Pto)$ for any point $\Pto\in\xSo$.
Obviously, $\mfot(\del\xSo)=\del\xSt$. Although different maps
$\mfP$ may lead to the same complex structure on $\nUP$, the map
$\mfot$ is determined by that complex structure uniquely, because
it is analytic and its value on the boundary $\del\xSo$ is
prescribed by the gluing.

\begin{figure}[hb]
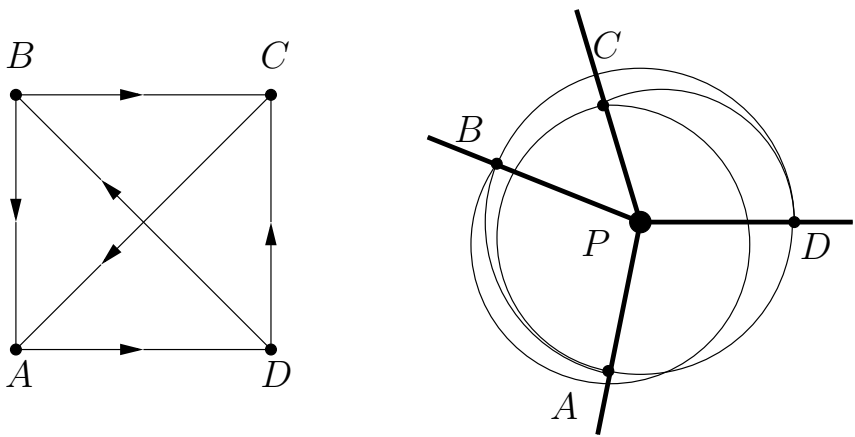

\input fig2xv3.pstex_t
\caption{A \lg\ and its image in the complex plane.}
\label{f2x}
\end{figure}

 The definition of the complex structure in the \nbh\ of a
vertex of $\Gra$ is slightly more complicated. We will use a
general construction, which is also consistent with the definition of
the complex structure for the first two types of points.
For a point $\Pt\in\bsRSG$ we define
its \emph{\lg} $\graP$
as the intersection between $\bsRSG$ and a small sphere centered
at $\Pt$.
If $\Pt$ is an internal point of $\WS$, then $\graP$ is
a circle. If $\Pt$ is an internal point of an edge $\ed$ of $\Gra$, then
$\graP$ has two vertices coming from the intersection of the small sphere with $\ed$,
and the edges of $\graP$ correspond to the strips of $\WS$ glued
to $\ed$. If $\Pt$ is a vertex of $\Gra$, then $\graP$ is a
graph, whose vertices correspond to the edges of $\Gra$ incident
to $\Pt$ and whose edges correspond to the strips of $\WS$ glued
to the edges of $\Gra$ incident to $\Pt$. The edges of $\gra$ are oriented according to the
orientation of the corresponding strips of $\WS$.

The \nbh\ $\nUP$ is
isomorphic to the cone of $\graP$, $\Pt$ being its vertex. The
map $\mfP$ maps this cone to $\IC$ as depicted in \fg{f2x}: the vertex of the cone maps
to the origin of $\IC$, the
cones of the vertices of $\graP$ map to the rays emanating from
the origin of $\IC$, and the cones of the edges of $\graP$ map to
the sectors of $\IC$ bounded by the corresponding rays, so that the
orientation of all the edges is counterclockwise (note that a
sector may, in principle, wind many times around the origin).

\subsection{Boundary conditions}

Most types of 2-dimensional \QFTs, such as \CFTs, N=2 sigma-models
and topological sigma-models, have two important properties. First, the theories
of the same type can be `crossed-multiplied', that
is, if we put two theories $\Tho$ and $\Tht$ of the same type on the same \ws, then
the resulting theory is again of the same type, and we denote it as $\Tho\times\Tht$.
Second, a theory can
be complex-conjugated into a theory of the same type, that is,
the original theory $\Th$ can be equivalently described as a (possibly different) theory
$\Thb$ of
the same type defined relative to the conjugated complex structure
on the same \ws.
In case of the \tAms,
the crossing of theories results in
the cross-product of the target spaces and the complex conjugation
of a theory is equivalent to the conjugation of the complex
structure of the target manifold.

Suppose that the \Rs\ $\WS$ splits into a union of connected \Rss\
$\WSi$. To every connected component $\WSi$ we assign a
2-dimensional \QFT\ $\Thi$ of the same type.
%
%
%
%
Now we have to formulate the
boundary conditions, which serve as a `glue' holding the parts of
the disjoint circles of $\dlWS$ together at the edges of the \sgr\ $\Gra$.
Since formulating an admissible boundary condition is a local problem, we
consider a neighborhood of an internal point of an edge of $\Gra$,
which is depicted in \fg{f2}.
The analytic map $\mfP$ identifies each strip of $\WS$ attached to $\ed$
with either the upper or the lower half-plane of $\IC$, while
preserving the gluing at $\ed$.
Thus
instead of looking for the boundary conditions of the \QFTs\ on
separate surface strips bounding the same edge,
we may equivalently consider a sewing condition at the real axis of $\IC$ for two
\QFTs\ $\Thu$ and $\Thd$, which are the \cprds\ of the theories
assigned to the strips that map to the corresponding half-planes.
Now we can `fold' the lower half-plane
(see, \eg\cx{BBDO}) by conjugating its complex structure and then
identifying it with the upper half-plane by the map $z\mapsto \bar{z}$, thus
equating the sewing condition for $\Thu$ and $\Thd$ to the
boundary condition for the single theory $\Thu\times\Thdb$ defined on the upper half-plane.

\subsection{Observables}

There are three ways, in which a point-like operator-observable can
be placed on a \sRs: it can be placed either at an internal point
of $\WS$, or at an internal point of an edge of $\Gra$, or at a
vertex of $\Gra$. In the first two cases, the list of admissible
operators is well-known from the study of \QFTs\ on Riemann
surfaces with boundaries (in case of an operator on an edge, one
has to consider the boundary operators of the theory
$\Thu\times\Thdb$). The operators at the vertices may require a
separate study (although a particular case of a vertex is well-known: it is a point
on the boundary of $\WS$, which separates two different
$D$-branes). Since the list of admissible
operators depends on the local properties of the theory, then in all three cases it
should be determined by the \lg\ of the point.
%

\subsection{Factorization}

The topological 2-dimensional \QFTs\ have a simple and important
factorization property. Namely, suppose that a Riemann surface
$\WS$ has two marked points $\Pto,\Ptt$. Let $\WSp$ be another Riemann surface
constructed from $\WS$ by cutting two small disks centered at
$\Pto$ and $\Ptt$ and gluing together the cutting boundaries. Then
a correlator on $\WSp$ splits into a sum of correlators on $\WS$
with pairs of operators inserted at $\Pto$ and $\Ptt$. More
precisely, if $\stdf{\Opa}{\ia\in\sOp}$ is
a basis in the space of all admissible operators at a point of $\WS$
($\sOp$ being the set, whose elements index these operators), then the
sphere correlator defines a scalar product
\qq
\mOpdab = \corri{\Opa\Opb}{S^2},
\label{1}
\qqq
and the factorization property reads
\qq
\corri{\cdots}{\WSp} = \siabsOp \mOpuab
\corri{\cdots\,\Opa(\Pto)\,\Opb(\Ptt)}{\WS},
\label{2}
\qqq
where $\Op(\Pt)$ denotes an operator $\Op$ placed at a point $\Pt$.

A similar factorization property should hold for \sRss.
If we cut out a small
neighborhood of a point of $\bsRSG$, then the boundary of the cut
is its \lg.
Suppose
that for two points $\Pto,\Ptt\in\bsRSG$ their \lgs\ are
isomorphic, and we denote them as $\gra$.  Then we can cut out
small neighborhoods of $\Pto$ and $\Ptt$ and glue the matching cut
boundaries together, thus forming a new \sRs\ $\bsRSGp$.
Let $\stdf{\Opa}{\ia\in\sOpgra}$ be a basis in the space of
admissible operators for the \lg\ $\gra$. In order to define a scalar product
on this space, we construct a special \sRs\ $\bsRSgra$ by taking
the cross-product $\gra\times[0,1]$ and contracting its boundaries
$\gra\times\prb{0}$ and $\gra\times\prb{1}$ to the points $\pVo$
and $\pVt$, which become the two vertices of $\Grgra$. Thus the \sgr\ $\Grgra$ consists of two vertices
$\pVo,\pVt$ connected by the edges, one edge per vertex of $\gra$, and
the \Rs\
$\WSgra$ consists of disjoint disks, one disk per edge of $\gra$.
The \lgs\ of $\pVo$ and $\pVt$ are isomorphic to $\gra$, so we can
insert the operators $\Opa$ there and define
\qq
\mOpdab = \corri{\Opa(\pVo)\;\Opb(\pVt)}{\bsRSgra}.
\label{3}
\qqq
Now the factorization property reads
\qq
\corri{\cdots}{\bsRSGp} =
\siabsOpgra \mOpuab
\corri{\cdots\,\Opa(\Pto)\,\Opb(\Ptt)}{\bsRSG}.
\label{4}
\qqq

\section{A-models on \sRss}
\label{s2}
\subsection{Boundary conditions}
Let us apply a general setup of a \QFT\ on a \sRs\ to
topological A-models.
A model of this type is specified by the choice of a compact \Kh\
manifold $\KX$ as a target space, so we assign  compact \Kh\
manifolds $\KXi$ to the connected components $\WSi$ of a \Rs\ $\WS$,
which is a part of a \sRs\ $\bsRSG$. A boundary condition for an
A-model was established by Witten in\cx{Wi1}: the boundary of a
\ws\ must be mapped to a Lagrangian submanifold $\XL\in\KX$. Thus,
if $\xne$ \ws\ strips joining the edge $\ed$ of the \sgr\
$\Gra$ carry the A-models with target spaces
$\KX_{i_1},\ldots,\KX_{i_{\xne}}$, then the boundary condition at
that edge is the selection of the Lagrangian submanifold
$\XLe\subset\KXe=\KX_{i_1}\times\cdots\times\KX_{i_{\xne}}$ (actually,
the \Kh\ manifolds assigned to the strips approaching the edge
`from below' have to be complex-conjugated, which means that their
\Kh\ forms change sign). Of course, $\XLe$ may factorize:
$\XLe=\XL_{i_1}\times\cdots\times\XL_{i_{\xne}}$, where
$\XLi\subset\KXi$ are Lagrangian subspaces, but then the gluing of
the \ws\ strips at $\ed$ is purely formal, and this case is not
interesting.

\subsection{Observables}

If $\Pt$ is an internal point of $\WSi$, then it was established in\cx{Wi2}
that for any cohomology class $\com\in\CohgXi$ there is an
admissible (BRST-closed) point-like operator-observable $\OpomP$.
Let us ignore the instanton corrections to the BRST operator. Then
the analysis of\cx{Wi1} indicates that if $\Pt$ is an internal
point of an edge $\ed\in\Gra$, then the admissible operators
$\OpomP$ are determined by the cohomology classes of the
corresponding Lagrangian submanifold: $\com\in\CohgXLe$.
Similar considerations indicate that if $\Pt$ is a vertex of $\Gra$,
then the operators $\OpomP$ are again determined by the cohomology
classes $\com\in\CohgMgP$, where $\graP$ is the \lg\ of $\Pt$ and $\Mgra$ is a special manifold
determined by the \lg\ $\gra$ in the following way. Recall that
the edges of $\gra$ correspond to the \ws\ strips of $\WS$ and
hence they are associated the \Khms\ $\KXi$. Let
$\KXgra=\lpgr{\KX}{\xngra}{i}$ be the \Khms\ associated to $\xngra$ edges of
$\gra$. The vertices of $\gra$ correspond to the edges of $\Gra$,
so let $\llgr{\XL}{\xmgra}{j}$ be the \Lgsms\ corresponding to
$\xmgra$ vertices of $\gra$. Suppose that a vertex $\vert\in\gra$
corresponds to an edge $\ed\in\Gra$, then there is an obvious
projection
\qq
\xmap{\fpe}{\KXgra}{\KXe}
\label{5}
\qqq
`forgetting' about the factors of $\KXgra$, which
do not participate in $\KXe$. Let $\tXLe = \fpie(\XLe)$
denote the pre-image of $\XLe\subset\KXe$. Then
$\Mgra\subset\KXgra$ is defined as the intersection
\qq
\Mgra=\ligr{\tXL}{\xmgra}{j}.
\label{6}
\qqq
Note that according to this definition, if $\Pt$ is an internal
point of $\WSi$, then $\MgP=\KXi$, and if $\Pt$ is an internal
point of an edge $\ed\in\Gra$, then $\MgP=\XLe$, so the
identification between the space of admissible operators $\OpP$
and the cohomology space $\CohgMgP$ works for all three types of
points of a \sRs.

Following\cx{Wi1}, we can also include the Chan-Paton factors associated with the
edges of $\Gra$. Suppose that we assign $\Ned$ Chan-Paton labels to an edge $\ed$.
This means that we introduce a flat
connection $\cnAed$
in the associated $\Uv{\Ned}$ bundle over the \Lgsm\ $\XLe$, whose
fiber is $\uv{\Ned}$ (the switching of the orientation of $\ed$ changes the sign of
this connection). 
If $\Pt$ is an internal point of $\ed$, then
the space of admissible operators $\OpomP$ is parametrized by the
elements of the twisted cohomology $\com\in\CohALe$, which is
defined on the sections of the bundle relative to the twisted
differential $d + \cnAed$.


The spaces of observables at the vertices of $\Gra$ admit a
similar description.
Let us orient all edges of $\Gra$.
Let $\gra$ be the \lg\ of a vertex of $\Gra$, and
consider the sequence of
maps
\qq
\Mgra\;\hookrightarrow \tXLe \xrightarrow{\fpe} \XLe,
\label{7}
\qqq
where the first map is a natural embedding in view of \ex{6} and
the second map is the restriction of\rx{5} to $\tXLe$. The
composition of the maps\rx{7} allows us to pull back the
connection $\cnAed$ on $\XLe$ to the connection $\tcnAed$ on
$\Mgra$. Now we introduce the connection $\cnAgP =
\lopgr{\tcnA}{\xmgra}{j}$ in the associated $\Uv{N_{e_{j_1}}}\times\cdots\times \Uv{N_{e_{j_{\xmgra}}}}$
bundle, whose fiber is the tensor product of the fundamental
representations of these groups (in fact, the fundamental representation must be
conjugated, if the oriented edge is directed into the vertex). The admissible operators at the
vertex
are parametrized by the corresponding twisted cohomology
$\CohgAMgra$.

\subsection{Correlators}

According to\cx{Wi2}, a correlator of a \tAm\ is determined by the
contributions of the sets of holomorphic maps
\qq
\xmap{\phfi}{\WSi}{\KXi},
\label{8}
\qqq
which satisfy the boundary conditions at the \sm\ edges.
We will neglect the Chan-Paton factors
and provide the geometric interpretation for the contribution of
the constant maps $\phfi$. We denote this contribution as
$\corrci{\corrl}{\bsRSG}$.
Recall that
\qq
\corrci{\corrl}{S^2} & = &
\int_{\KX} \com_1\wedge\cdots\wedge\com_n.
\label{9}
\qqq
We assume for simplicity that the
fermionic fields $\psi_{z}^{\bar{i}},\psi_{\bar{z}}^i$ have no
zero modes on the \sRs\ $\bsRSG$. Hence the calculation of the contribution of the constant
maps to the correlator is reduced to the integration over the
constant
modes of $\chi^{i},\chi^{\bar{i}}$ and over the moduli space $\MbsRSG$ of the
constant maps\rx{8}. This means that the general correlator is
again an intersection number:
\qq
\corrci{\corrl}{\bsRSG} & = &
\int_{\MbsRSG}\xprpsv{1}\,\com_1\wedge\cdots\wedge\xprpsv{n}\,\com_n,
\label{10}
\qqq
where $\xprpsv{i}\,\com_i$ is the pull-back of $\com_i$ by a map
\qq
\xmap{\xprpv{i}}{\MbsRSG}{\MgPv{i}},
\label{11}
\qqq
%
which can be easily constructed
for all three types of points
$\Pt\in\bsRSG$ in the following way. Let $\KXbsRSG =
\KX_1\times\cdots\times\KX_{\xnS}$ be the product of all target
spaces corresponding to the connected components
$\WS_1,\ldots,\WS_{\xnS}$ of $\WS$. Then for every component
$\WSi$ there is a natural projection
\qq
\xmap{\fPi}{\KXbsRSG}{\KXi}.
\label{12}
\qqq
Let $\ed$ be an edge of $\Gra$. We assume for simplicity
that every component $\WSi$ bounds $\ed$ at most once, so there is
another natural projection
\qq
\xmap{\fPe}{\KXbsRSG}{\KXe},
\label{13}
\qqq
which forgets about the factors of $\KXbsRSG$, whose \wss\ $\WSi$
do not bound $\ed$. Then $\MbsRSG$ is the intersection of the
pre-images of the \Lgsms\ $\XLe\subset\KXe$:
\qq
\MbsRSG = \bigcap_{\ed\in E(\Gra)} \fPei(\XLe)\subset\KXbsRSG.
\label{14}
\qqq
Thus if $\Pt$ is an internal point of $\WSi$, then we define
$\xprP$ as the composition of maps
\qq
\xprP:\;\MbsRSG\hookrightarrow\KXbsRSG\xrightarrow{\fPi}\KXi,
\label{15}
\qqq
and if $\Pt$ is an internal point of an edge, then $\xprP$ is the
composition of maps
\qq
\xprP:\;\MbsRSG\hookrightarrow\fPei(\XLe)\xrightarrow{\fPe}\XLe.
\label{16}
\qqq

Now let $\Pt$ be a vertex $\vert$ of $\Gra$. Assume for simplicity that
every component $\WSi$ bounds $\vert$ at most once. Then there is
a natural projection $\xmap{\fPvert}{\KXbsRSG}{\KXgvert}$ and
$\fPvert(\MbsRSG)\subset \Mgvert$, so in this case we define
$\xprP$ as the restriction $\fPvert|_{\MbsRSG}$.

Note that the \Lgsms\ $\XLe\subset\KXe$ must be selected together
with their orientation. The \Khms\ $\KXi$ also have natural
orientation coming from their complex structure. Hence the
moduli space of constant maps $\MbsRSG$ receives an orientation
from the formula\rx{14}, so the integral in the \rhs of \ex{10} is
well-defined, provided that we choose the order, in which we
intersect the \Lgsms.

\section{\Grsms}

The general construction of 2-dimensional theories on \sRss\ looks
rather abstract, unless we provide interesting boundary
conditions, which mix multiple theories of the same class. Luckily,
a wide class of \Lgsms\ in the products of \Khms\ is implied by the construction in
M.~Khovanov's paper\cx{Kh1}, which deals with the categorification
of the $sl(3)$ HOMFLY polynomial.

\subsection{\Lgsms}
\label{ss3.1}
A complex \Grsm\ is the `moduli space' of $m$-dimensional
subspaces of $\ICn$. A \Grsm\ can be endowed with a \Kh\ structure.
Namely, suppose that $\ICn$ has the standard hermitian scalar
product. Now if $\xV\subset\ICn$ ($\dim V=m$) represents a point
$x\in\Grmn$, then the tangent space $\rmTx\Grmn$ is canonically isomorphic to the space of
linear maps $\rmHom(\xV,\xVort)$ and the \Kh\ form $\Kom$ evaluated
on two tangent vectors $\yA,\yB\in\rmHom(\xV,\xVort)$ is
\qq
\Kom(\yA,\yB) = {1 \over 2i}\,\Tr_{\xV} (\yB^{\ast}\yA -
\yA^{\ast}\yB).
\label{17}
\qqq

Consider a set
of \Grsms\ $\Grvn{m_1},\ldots,\Grvn{m_k}$, such that
\qq
m_1+\cdots+m_k = n.
\label{16x2}
\qqq
Think of a point of their cross-product
\qq
\KX = \Grvn{m_1}\times\cdots\times\Grvn{m_k}
\label{16x3}
\qqq
as a set of subspaces $V_1,\ldots,V_{k}\subset\ICn$ ($\dim
V_i=m_i$) of the \emph{same} complex space $\ICn$ endowed with an hermitian
scalar product. Then the condition that the subspaces $V_i$ form an
orthogonal decomposition of $\ICn$, specifies a \Lgsm\
$\XL\subset\KX$.

The \Lg\ nature of $\XL$ can be verified by a straightforward
calculation.
If the subspaces $\xV_i$ ($\dim V_i = m_i$) form an orthogonal
decomposition of $\ICn$, then $\xV_i^\perp=\bigoplus_{j\neq i}
\xV_j$, so that
$\rmHom(\xV_i,\xV_i^\perp) = \bigoplus_{j\neq
i}\rmHom(\xV_i,\xV_j)$
and the \Kh\ form on the tangent space of $\Grmin$ is
\qq
\Komi(\yA,\yB) = {1\over 2i}\,\sum_{j\neq i}\Tr_{\xV}
(\yB^{\ast}_{ij}\yA_{ij} - \yA^{\ast}_{ij}\yB_{ij}),\qquad
\yA_{ij},\yB_{ij}\in\rmHom(\xV_i,\xV_j).
\label{18}
\qqq
The tangent space to the surface $\XL\subset\KX$ is specified by the
conditions
\qq
\yA_{ij} = \yA^{\ast}_{ji}\qquad\mbox{for all $1\leq i\neq j\leq
k$}.
\label{19}
\qqq
It is easy to see that these conditions halve the real dimension of the original
manifold and make the total \Kh\ form
$\Kom = \sum_{i=1}^k \Komi$ zero, so $\XL$ is indeed
a \Lgsm.

Note that the condition\rx{19} implies a simple model for the
complex-conjugated \Grsm\ $\bGrmn$. Namely, a map
$\Grmn\rightarrow\Grnmn$, which maps every $m$-dimensional
subspace $V\subset\ICn$ into its orthogonal complement, is an
anti-holomorphic isomorphism, hence
\qq
\bGrmn\cong\Grnmn.
\label{19x}
\qqq

A generalized version of the \Lgsm\ $\XL\subset\KX$ exists, if
instead of \ex{16x2} we have
\qq
m_1 + \cdots + m_k = Nn,
\label{20}
\qqq
where $N$ is a positive integer. In this case a \Lgsm\ $\XLbm$ is
specified by a list of non-negative numbers
\qq
\bm = (\ymij,\;\oik,\,\ojN  \,|\,
\ymij\geq 0,\;\sum_{j=1}^{N} \ymij = m_i).
\label{21}
\qqq
A point of
$\Grvn{m_1}\times\cdots\times\Grvn{m_k}$, specified by the set of
$k$ subspaces $V_i\subset\ICn$ ($1\leq i\leq k$), belongs to $\XL$,
if there exist the subspaces $\yVij\subset\ICn$ ($\oik$, $\ojN$), such that
$\dim\yVij=\ymij$ and the spaces $\yVij$ ($\oik$) form an orthogonal
decomposition of $\ICn$ for every fixed $j$, while the spaces $\yVij$
($\ojN$) form an orthogonal decomposition of $V_i$ for every fixed
$i$.

\subsection{\Ams}

The \Lgsms\ $\XLbm$ allow us to construct topological \Ams\ on \sRss\ along
the lines of Section\rw{s2}. First, we pick a value of $n$. Then
the \Khms\ $\KXi$ are the \Grsms\ $\Grmin$ for positive integers
$m_i<n$, and the boundary conditions at the seam edges of a \sRs\ are
specified with the help of \Lgsms\ $\XLbm$.

\begin{figure}[hb]
\input fig3w.pstex_t
\caption{The cube graph produces a singular space $\Mgra$.}
\label{f3}
\end{figure}

A particular feature of this model is that all moduli spaces
related to it have a $\Un$ symmetry, which acts on the `master-space' $\ICn$.
This $\Un$ symmetry acts
transitively on each \Grsm\ $\Grmin$, hence the maps\rx{5}
and\rx{12} are fiber-bundle projections. At the same time, the
\Grsm\ model indicates that some considerations of
Subsection\rw{2} are too naive: Khovanov and Kuperberg observed that the
spaces $\Mgra$ may be singular. His simplest example is the moduli
space associated with the `cube' graph of \fg{f3} for the case of
$n=3$ when the projective spaces $\IC\IP^2$ are associated to
every edge. As a result of this singularity, the
Poincare duality essential for the
factorization property\rx{4} is broken. This means that if $\Mgra$
is singular, then the identification of the space of observables with
the cohomology space $\CohgMg$ has to be reconsidered.


\section{Conclusion}

One of the major selling points of the string theory is that its
interactions are not arbitrary, but rather come from
natural geometric principles, such as the `pants' \ws, describing the
triple interaction between three closed strings. From this point
of view, the idea of a \sRs\ does not seem to be very attractive,
since it reminds us of the intersecting world-lines of old
\QFTs. Those intersections of world-lines were responsible for the
wild arbitrariness of the interaction coupling constants.
However, the theories on \sRss, in which the seams
are due to non-factorizable \Dbr\ type boundary conditions, seem
geometric enough in order to be considered seriously. Moreover,
the \Lgsms\ of subsection\rw{ss3.1} provide enough boundary
conditions in order to put the \Grsm-based \Ams\ on complicated
\sRss. Thus, one might conclude that \sRss\ are as good as
familiar \Rss\ for the purpose of building 2-dimensional \QFTs.

A mathematical implication of the \Grsm-based \Ams\ on \sRss\ is that they
lead to a Fukaya category on a family of manifolds rather than on a single
\Khm. One might expect that, due to the mirror symmetry, similar
construction could exist for the categories of coherent sheaves.
Also, it is worth noting that the \Grsm-based \Ams\ have an
equivalent description as Landau-Ginzburg models and as $G/G$
\WZWms\ (see, \eg\cx{Wi3}). It would be interesting to find the
boundary conditions of those models, which correspond to the \Lgsms\
$\XLbm$.

\subsection*{Acknowledgements}
I am indebted to Mikhail Khovanov for sharing the results of
his research prior to the publication of\cx{Kh1}. I am also very
thankful to P.~Aspinwall, D.~Bar-Natan, P.~Belkale, A.~Kapustin and A.~Vaintrob
for their patient explanations of the properties of \Ams\ and
\Grsms\ and the intricacies of Khovanov's theory.





\end{document}

\nsection{Graph algebras}

\subsection{Jacobi graph algebra $\cBL$}

\subsubsection{Pre-Jacobi graph algebra $\tcBL$}

Let $\xL$ be a possibly infinite set, its elements will be called
\emph{labels}. The set $\xL$ splits into a union $\xL = \xLU\cup\xLS\cup\xLW$
of respectively $\Urs$-type, $\Srs$-type and $\Wrs$-type labels. The the sets
$\xLS$ and $\xLW$ in turn split into subsets $\xLSu,\xLWu$ of \emph{upper}
(\ie contravariant)
and $\xLSd,\xLWd$ of \emph{lower} (\ie covariant) labels.
%
A
graph $\xD$ is called \emph{\xLld} if
\begin{itemize}
\item
1-vertices of $\xD$ are labeled by elements of $\xL$;
\item
1-vertices carrying the same $\Urs$ label are linearly ordered;
\end{itemize}
An \xLld\ graph is called \emph{\xrd} if
\begin{itemize}
\item
a cyclic order is chosen for incidents edges at every 3-vertex;
\item
all vertices carrying $\Wrs$ labels are linearly ordered.
\end{itemize}

Let $\sGRL$ be the set of all
(1,3)-valent \xLld\ and \xrd\ graphs as well as their disjoint unions with a finite
number of copies of the `special graph' $\tcrl$,
which is not a graph but a circle. The space $\tcBL$ is a formal span
of all elements of $\sGRL$ modulo the following relations:
\begin{itemize}
\item
if two graphs differ only by a cyclic order at a 1-vertex, then they
represent the opposite elements of $\tcBL$ (AS relation);
\item
if two graphs differ only by the linear order of their $\Wrs$-labeled
1-vertices then they represent either the same or the opposite elements of
$\tcBL$ depending on whether the permutation of 1-vertices which relates
both linear orders, is even or odd.
\end{itemize}
%

\begin{remark}
\rm
If $\xL$ contains only $\Urs$ labels, then $\tcBL$ is usually called $\cA$ (or, in
our notations, $\tcA$). The $\Wrs$ labels are necessary for the introduction
of differential forms.
\end{remark}

The space $\tcBL$ has an algebra structure. The product of two graphs
$\xDt\xDo$ is defined as their disjoint union, whereas the linear orders of
1-vertices are arranged in the following way: for the same $\Urs$ label the
1-vertices of $\xDo$ precede the 1-vertices of $\xDt$ and all $\Wrs$
1-vertices of $\xDo$ precede the $\Wrs$ 1-vertices of $\xDt$. If
$\xLU=\emptyset$, then $\tcBL$ is (super-)commutative.

$\tcBL$ is a graded algebra with multiple gradings. In particular, for a
graph $\xD\in\sGRL$ and for a label $\lA\in\xLS\cup\xLW$ we define
$\dgrA(\xD)$ as the number of $\lA$-colored 1-vertices of $\xD$. We also
define the Euler grading
\qq
\dgrE(\xD) & = & \absv{\xED} - \absv{V_{\Srs}(\xD)} - \absv{V_{\Wrss}(\xD)}
\nonumber\\
& = & - \chi(\xD) + \absv{V_{\Urs}(\xD)},
\label{2.1}
\qqq
where $\chi(\xD)$ is the Euler characteristic of $\xD$ considered as a $CW$-complex.
%


Besides the algebra multiplication, $\tcBL$ has an important operation of leg
gluing. For two labels $\lA,\lB\in\xLS$ (or $\xLW$) of opposite levels
we define an operation of gluing $m$ legs,
which is a linear map $\xmap{\glABm}{\tcBL}{\tcBL}$. This map is defined by
its action on a graph $\xD\in\sGRL$:
\qq
\glABm(\xD) = \siochAB \ysi \xDi,
\label{2.1x}
\qqq
where $\xDi$ ($1\leq i\leq{\dgrA\choose m}{\dgrB\choose m}$) are the graphs
constructed in all possible ways by the pairwise gluing of $m$
1-vertices labeled $\lA$ with $m$
1-vertices labeled $\lB$ and then dissolving the resulting 2-vertices, while
$\ysi$ are the sign factors: if $\lA,\lB\in\xLS$, then $\ysi=1$, if
$\lA,\lB\in\xLW$, then $\ysi$ is the sign of the permutation of $\xLW$
1-vertices that
would place the participating $\lB$ 1-vertices immediately after the $\lA$ 1-vertices to which
they are being glued, and if $\lA\in\xLS$, $\lB\in\xLW$, then
$\ysi=0$ if $m>1$
$\ysi$ is the sign of the permutation of $\xLW$ 1-vertices that
would place the participating $\lB$ 1-vertex at the beginning of
the ordered list of $\xLW$ vertices.




Note that $\glABm = (\glABo)^m/m!$.
In addition to $\glABm$ we define a gluing $\glLRAB$ which glues
all $\lA$ 1-vertices with all $\lB$ 1-vertices (if their numbers are not
equal, then the result is zero).
\begin{remark}
\rm
Since there are only finitely many ways in cutting a particular graph at $m$
points on its edges, then gluing $\glABm$ is always well-defined.
However, one has to check that an application of $\glLRAB$ to an
infinite linear combinations of graphs is well-defined.
\end{remark}

Sometimes it is convenient to think of
$\glABm$ (and $\glLRAB$) as a binary operation defined by the formula
\qq
\xDo\,\glABm\,\xDt = \gl{\lA[\xDo],\lB[\xDt]}{m}(\xDo\xDt),
\label{2.x1}
\qqq
where $\xDt\xDo$ is the $\tcBL$ algebra
product of $\xDt$ and $\xDo$ and
the notation $\gl{\lA[\xDo],\lB[\xDt]}{m}$ means that only
$\lA$ 1-vertices of $\xDo$ and $\lB$ 1-vertices of $\xDt$ are
glued.

The gluing operations $\glABm$ and $\glLRAB$ satisfy an important combinatorial
property, whose formulation would require us to introduce the
\emph{\hbrd} graphs.

A \hbrd\ graph is a graph, whose vertices are
called either ordinary or \blbd. The ordinary vertices are either
1-valent or 3-valent. Ordinary 1-valent vertices are \xLld\ and
\xrd, and the edges incident to an ordinary 3-valent vertex are
cyclicly ordered. Each \blbd\ vertex is labeled by a (1,3)-valent
graph with cyclic order of incident edges at every 3-valent
vertex. The incident edges at a \hbrd\ vertex are in one-to-one
correspondence with the 1-vertices of the (1,3)-valent graph which
labels it. The graphs of $\sGRL$ are the particular examples of hybrid
graphs, all of whose vertices are \xrd.

Let $\xD$ be a \hbrd\ graph. Its \unblbng\ $\funblb(\xD)$ is the
graph constructed by replacing the \hbrd\ vertices with the graphs
which label them. Let $\xD\in\sGRL$ have $m$ 1-valent vertices,
then its \blbng\ $\fblb(\xD)$ is the \hbrd\ graph which consists of a single
$m$-valent \blbd\ vertex labeled by $\xD$ and connected to $m$
ordinary 1-vertices which are labeled in the same way as those of
$\xD$. Obviously, $\funblb(\fblb(\xD))=\xD$.

Finally, if $x$ is a linear combination of \hbrd\ graphs $x =
\sum_{\xD} c_{\xD}\xD$, then its \emph{\cnctpt} is the sum of
connected graphs
$\cnctv{x}= \sum_{\text{$\xD$-connected}} c_{\xD}\xD$.
\begin{theorem}
\label{t.expgl}
Let $x\in\cBCL$ be a linear combination
of non-empty connected graphs, and let $\glABarb$ be
either $\glABm$ or $\glLRAB$.
Then
\qq
\glABarb(e^x) = \exp\Bigg(\funblb\Big(\cnctv{
\glABarb\Big(\exp(\fblb(x))\Big)
}\Big)\Bigg).
\label{2.1x1}
\qqq
\end{theorem}

In an important case when $x$ is a linear combination of non-empty connected
graphs, \ex{2.1x1} takes a simpler form
\qq
\glABarb(e^x) = \exp\Big(\cnctv{\glABarb(e^x)}  \Big).
\label{2.1x2}
\qqq
%








\subsubsection{Weight system}
\label{ss2.1.2}

Let $\mfg$ be a simple Lie algebra. For every label $\lA\in\xL$ we associate
an algebra $\aTA$ according to the following rule: $\aTA=\Ug$ if $\lA\in\xLU$
and
\qq
\aTA =
\left\{
\begin{array}{ll}
\Sg,&\mbox{if $\lA\in\xLSu$}
\\
\Sgs,&\mbox{if $\lA\in\xLSd$}
\\
\Wg,&\mbox{if $\lA\in\xLWu$}
\\
\Wgs,&\mbox{if $\lA\in\xLWd$}.
\end{array}
\right.
\label{2.1*}
\qqq

The \emph{weight system} associated with $\mfg$ is an algebra homomorphism
%
\qq
\xmap{\Tg}{\tcBL}{\TxLg},
\qquad\mbox{where $\TxLg = \invG{\bigotimes_{\lA\in\xL} \aTA}$,}
\label{2.2}
\qqq
%
while $\mfG$ is the Lie group associated with
$\mfg$. This homomorphism is constructed in
the following way.
Let $\hmfg\in \Srs^2\mfgs$ be the Killing form of $\mfg$ normalized
by the condition that the length of the short roots is $\sqrt{2}$.
The tensor $\himfg\in\Srtg$
denotes its inverse. Also let
$\fmfg\in\wdgtg$ be the structure constant tensor of $\mfg$. Now for a graph
$\xD\in\sGRL$ we consider the tensor product of spaces
\qq
\TgD=\big(\wdgtg\big)^{\otimes\numv{\grvth(\xD)}}\otimes\lrbc{\Srtg}^
{\otimes\numv{\gre(\xD)}}
\otimes (\Srtgs)^{\otimes\numv{\grvd(\xD)}},
\label{2.4}
\qqq
in which every space $\wdgtg$ is associated with a 3-vertex of $\xD$, every
space $\Srtg$ is associated with an edge of this graph and every space
$\Srtgs$ is associated with a lower 1-vertex.
We apply the natural `contraction of indices' maps
\qq
\mfg\otimes\mfgs\longrightarrow\IR
\label{2.4*}
\qqq
to
the pairs of spaces $\mfgs$ of $\wdgtg$, $\Srtgs$
and $\mfg$ of $\Srtg$ according to
the incidence of edges and vertices, thus obtaining a `contraction' map
\qq
\xmap{\txCD}{\TgD}{\invG{\mfg^{\otimes(
\numv{\grvU(\xD)}+
\numv{\grvSu(\xD)}+\numv{\grvWu(\xD)}}
\otimes
(\mfgs)^{\otimes(\numv{\grvSd(\xD)}+\numv{\grvWd(\xD)}}
}},
\label{2.5}
\qqq
in which every factor $\mfg$ and $\mfgs$ of
the target space is naturally associated with a
1-vertex of $\xD$. Composing
the map $\txCD$ with the symmetrization and anti-symmetrization among the factors
$\mfg$ and $\mfgs$ associated
with $\Srs$ and $\Wrs$
1-vertices carrying the same label, we arrive at the map
\qq
\lefteqn{
\xmap{\xCD}{\TgD}{\Big(\bigotimes_{\lX\in\xLSu}\Srs^{\dgrXD}\mfg\;
\otimes\;
\bigotimes_{\lX\in\xLWu}\Wrs^{\dgrXD}\mfg
}
}
\nonumber\\
&&\hspace{2in}\otimes\;
\bigotimes_{\lX\in\xLSd}\Srs^{\dgrXD}\mfgs\;
\otimes\;
\bigotimes_{\lX\in\xLWd}\Wrs^{\dgrXD}\mfgs
\Big)^{\mfG}\subset\TxLg.
\label{2.6}
\qqq
Now we define a special tensor $\tgD\in\TgD$ associated with the graph $\xD$
\qq
\tgD = \fmfg^{\otimes\numv{\grvth(\xD)}}\otimes(\himfg)^
{\otimes\numv{\gre(\xD)}}
\otimes \hmfg^{\otimes\numv{\grvd(\xD}}
\label{2.6*}
\qqq
and then define the weight system map by the formula
\qq
\Tg(\xD) =
\xCD\lrbc{\tgD
}
\in\TxLg.
\label{2.7}
\qqq
We also define $\Tg(\tcrl) = \dim \mfg$.


\subsubsection{IHX and STU quotient}

\begin{figure}[hbt]
\begin{fmffile}{f1_1}
\begin{displaymath}
\begin{array}{cccc}
\begin{fmfgraph*}(20,20)
\fmfleft{i1,i2} \fmfright{o1,o2}
\fmf{plain,tension=0.3}{i1,v,o2}
\fmf{plain,tension=0.3}{i2,v,o1}
\fmfdot{v}
\end{fmfgraph*}
, &
\begin{fmfgraph*}(20,20)
\fmfleft{i1,i2} \fmfright{o1,o2}
\fmf{plain,tension=0.3}{i1,v1,o1}
\fmf{plain,tension=0.3}{i2,v2,o2}
\fmf{plain,tension=0.3}{v1,v2}
\fmfdot{v1,v2}
\end{fmfgraph*}
, &
\begin{fmfgraph*}(20,20)
\fmfleft{i1,i2} \fmfright{o1,o2}
\fmf{plain,tension=0.3}{i1,v1,i2}
\fmf{plain,tension=0.3}{o1,v2,o2}
\fmf{plain,tension=0.3}{v1,v2}
\fmfdot{v1,v2}
\end{fmfgraph*}
, &
\begin{fmfgraph*}(20,20)
\fmfleft{i1,i2} \fmfright{o1,o2}
\fmf{plain,tension=1}{i1,v1}
\fmf{plain,tension=0.3}{v1,o2}
\fmf{plain,tension=1}{o1,v2}
\fmf{plain,tension=0.3}{v2,i2}
\fmf{plain,tension=0.3}{v1,v2}
\fmfdot{v1,v2}
\end{fmfgraph*}
\\
\xD & \xD_1 &\xD_2 &\xD_3
\end{array}
\end{displaymath}
\end{fmffile}
\caption{$\IHX$ graphs}
\label{f1.1}
\end{figure}


\begin{figure}[hbt]
\begin{fmffile}{f1_2}
\begin{displaymath}
\begin{array}{cccc}
\begin{fmfgraph*}(35,15)
\fmfleft{i1,i2} \fmfright{o1,o2}
\fmfpen{thin}
\fmf{plain,tension=0.3}{i2,v,o2}
\fmf{plain,tension=300,width=thick}{o1,v}
\fmf{plain_arrow,tension=300,width=thick}{v,i1}
\fmfdot{v}
\end{fmfgraph*}
, &
\begin{fmfgraph*}(35,15)
\fmfleft{i1,i2} \fmfright{o1,o2}
\fmf{plain,tension=300,width=thick}{o1,v1}
\fmf{plain_arrow,tension=300,width=thick}{v1,i1}
\fmf{plain,tension=0.3}{i2,v2,o2}
\fmf{plain,tension=0.3}{v1,v2}
\fmfdot{v1,v2}
\end{fmfgraph*}
, &
\begin{fmfgraph*}(35,15)
\fmfleft{i1,i2} \fmfright{o1,o2}
\fmf{plain,tension=300,width=thick}{o1,v1}
\fmf{plain,tension=300,width=thick}{v1,v2}
\fmf{plain_arrow,tension=300,width=thick}{v2,i1}
\fmf{plain,tension=0.3}{i2,v2}
\fmf{plain,tension=0.3}{o2,v1}
\fmfdot{v1,v2}
\end{fmfgraph*}
, &
\begin{fmfgraph*}(35,15)
\fmfleft{i1,i2} \fmfright{o1,o2}
\fmf{plain,tension=300,width=thick}{o1,v1}
\fmf{plain,tension=300,width=thick}{v1,v2}
\fmf{plain_arrow,tension=300,width=thick}{v2,i1}
\fmf{plain,tension=0.3}{i2,v1}
\fmf{plain,tension=0.3}{o2,v2}
\fmfdot{v1,v2}
\end{fmfgraph*}
\\
\xD & \xD_1 &\xD_2 &\xD_3
\end{array}
\end{displaymath}
\end{fmffile}
\caption{$\STU$ graphs}
\label{f1.2}
\end{figure}

The homomorphism\rx{2.2} suggests, that the algebras $\tcBL$ could
serve as `universal' models of tensor algebras built upon $\mfg$.
However, $\Tg$ has a large kernel due to the Jacobi identity
satisfied by the structure constants $\fmfg$ and due to the
defining relation $XY-YX=[X,Y]$ of the universal enveloping
algebra. Therefore the structure of tensor algebras built upon
$\mfg$ is better captured by the quotient algebras
\qq
\cBL = \tcBL/\tcBLIHX,
\label{2.7*}
\qqq
where \emph{IS} stands for \emph{IHX and STU}, while the ideal $\tcBLIHX$ is defined in the
following way. Let $\xD$ be an \xLld\ (1,3,4)-valent graph with a single
4-valent vertex. We define
$$\fIHX(\xD) = \sIHXs{\xDv},$$
where
$\xD_i$ ($1\leq i\leq 3$) are the graphs of $\sGRL$ constructed from $\xD$ by
replacing its 4-valent vertex according to \fg{f1.1}. Now let $\xDp$ be an
\xLld\ \xrd\ (1,2,3)-valent graph with a single 2-valent vertex. We assume that
this 2-vertex is labeled by an element of $\Urs$ and is linearly ordered
together with other 1-vertices carrying the same label. We define
$$\fSTU(\xDp) = \sIHXs{\xDv},$$
where $\xD_i$ ($1\leq i\leq 3$)
are the graphs of $\sGRL$ constructed from $\xD$ by replacing its 2-valent
vertex according to \fg{f1.2}. The ideal $\tcBLIHX$ is the span of
$\fIHX(\xD)$ and $\fSTU(\xDp)$ for all possible graphs $\xD,\xDp$.

The structure properties of $\tcBL$ descend to $\cBL$.
\begin{theorem}
\label{t2.1}
$\cBL$ is an algebra with the gradings $\dgrA$
($\lA\in\xLS\cup\xLW$) and $\dgrE$, with the gluing
\qq
\xmap{\glABm}{\cBL}{\cBL}
\label{2.8}
\qqq
and with the weight systems
\qq
\xmap{\Tg}{\cBL}{\TxLg}.
\label{2.9}
\qqq
Moreover, if $\xL\subset\xLp$, then the natural injection
$\tcBL\hookrightarrow\tcBLp$ descends to the injection
$\cBL\hookrightarrow\cBLp$.
\end{theorem}
\proof
The gradings of $\tcBL$ survive the
quotienting\rx{2.7*}, because the (1,3,4)-valent graph of \fg{f1.1} and its
resolutions $\xDo,\xDt,\xDth$ have the same definite gradings, and the same
is true for the graphs of \fg{f1.2}. The gluing of legs descends to $\cBCL$, because
$\glABm(\tcBLIHX)\subset\tcBLIHX$ in view of the commutativity of the diagram
\qq
\begin{diagram}
\node{\xspan(\text{(1,3,4)-valent graphs})}
  \arrow{e,t}{\glABm}
  \arrow{s,r}{\fIHX,\fSTU}
\node{\xspan(\text{(1,3,4)-valent graphs})}
  \arrow{s,r}{\fIHX,\fSTU}
\\
\node{\tcBCL}
  \arrow{e,t}{\glABm}
\node{\tcBCL}
\end{diagram}
\qqq
The weight system homomorphism descends to $\cBCL$ because of the Jacobi
identity satisfied by the bracket of $\mfg$ and because of the defining
relation $XY-YX=[X,Y]$ in $\Urs\mfg$.
\qed

\subsection{Cartan decomposition and Jacobi graph algebra $\cBCL$}

\subsubsection{Pre-Jacobi graph algebra $\tcBCL$}

A formal graph description of a coadjoint orbit requires us to make a
distinction between Cartan subalgebra elements and root space elements
`moving' along the edges of \xLld\ graphs. Therefore we split
the subsets of $\Srs$ and $\Wrs$ labels into subsets of
total, Cartan and root labels
%
\qq
\xLS = \xLtS\cup\xLCS\cup\xLrS,\qquad \xLW = \xLtW\cup\xLCW\cup\xLrW.
\label{2.10}
\qqq
These smaller subsets split further into the subsets of upper and lower
labels.

A graph is called \xCLld\ if it is \xLld\, and every edge
is declared either root or Cartan.
A \xCLld\ graph is called a leg-vertex type mismatch (\lvtm) if
a Cartan edge is incident to a root 1-vertex or a root edge is incident to
a Cartan 1-vertex.
The set $\sCGRL$ consists of all
(1,3)-valent \xCLld\ \xrd\ graphs which are not \lvtm, and of their disjoint
unions with finite numbers of `special graphs',
which are a Cartan circle $\ccrl$ and a root circle $\rcrl$.
The algebra $\tcBCL$ is a
formal linear span of the elements of $\sCGRL$ modulo the relations.
\begin{remark}
\rm
We may think of \lvtm\ graphs as 0-vectors in $\tcBCL$.
\end{remark}
In order to establish a relation between $\tcBL$ and $\tcBCL$, it
is convenient to introduce an auxiliary type of edge which we call
\emph{total}. By definition, a graph with a total edge represents
an element of $\tcBCL$, which is the sum of two graphs constructed
from the original one by replacing the total edge with either the
Cartan edge or the root edge. Thus if a total leg is incident to a
Cartan or a root 1-vertex, then it is automatically replaced by
the Cartan or the root leg, since we eliminate the \lvtm\ graphs.

The space $\tcBCL$ has an algebra structure defined in exactly the
same way as that of $\tcBL$. The definition of the leg gluing
operation $\glABm$ is also the same as in $\tcBL$, except two extra
conditions:
\begin{itemize}
\item
gluing together a Cartan 1-vertex and a root 1-vertex
results in a 0-vector of $\tcBCL$;
\item
gluing together two total 1-vertices, one of which is incident to a root
leg and the other to a Cartan leg also results in a 0-vector.
\end{itemize}
The definition of a total edge implies that a gluing between two total 1-vertices
incident to total legs results in a total edge.

\subsubsection{Weight system}

In subsection\rw{ss2.1.2} for a label $\lA\in\xLU$ we defined the
algebra $\aTA=\Ug$. We replace the definition\rx{2.1*} with
\qq
\aTA = \left\{
\begin{array}{ll}
\Sg,&\mbox{if $\lA\in\xLtSu$}
\\
\Sgs,&\mbox{if $\lA\in\xLtSd$}
\\
\Wg,&\mbox{if $\lA\in\xLtWu$}
\\
\Wgs,&\mbox{if $\lA\in\xLtWd$}.
\end{array}
\right.
,
\aTA = \left\{
\begin{array}{ll}
\Sh,&\mbox{if $\lA\in\xLCSu$}
\\
\Shs,&\mbox{if $\lA\in\xLCSd$}
\\
\Wh,&\mbox{if $\lA\in\xLCWu$}
\\
\Whs,&\mbox{if $\lA\in\xLCWd$}.
\end{array}
\right.
,
\aTA = \left\{
\begin{array}{ll}
\Sro,&\mbox{if $\lA\in\xLrSu$}
\\
\Sros,&\mbox{if $\lA\in\xLrSd$}
\\
\Wro,&\mbox{if $\lA\in\xLrWu$}
\\
\Wros,&\mbox{if $\lA\in\xLrWd$}.
\end{array}
\right.
\label{2.11}
\qqq

Since $\mfg = \mfh \oplus \mfr$ is an orthogonal decomposition
with respect to the Killing metric, then $\hmfg\in\Srtgs$ has a block
structure:
\qq
\hmfg = \hmfh\oplus\hmfr,\qquad\mbox{where $\hmfh\in\Srths\subset\Srtgs$ and
$\hmfr\in\Srtrs\subset\Srtgs$}.
\label{2.12}
\qqq
We define $\himfh\in\Srth$, $\himfr\in\Srtr$ is a similar way so that
$\himfg = \himfh \oplus \himfr$.
Let $\xD\in\sCGRL$. For an edge $\edg\in\gre(\xD)$ and for a lower 1-vertex
$\vrt\in\grvd(\xD)$
we define
\qq
\qquad
\hiedg =
\begin{cases}
\himfh &\text{if $\edg$ is Cartan}
\\
\himfr &\text{if $\edg$ is root}
\\
\himfg & \text{if $\edg$ is total,}
\end{cases}
\qquad
\hvrt =
\begin{cases}
\hmfh &\text{if $\vrt$ is Cartan}
\\
\hmfr &\text{if $\vrt$ is root}
\\
\hmfg & \text{if $\vrt$ is total.}
\end{cases}
\label{2.13}
\qqq
Next, for a graph $\xD\in\sCGRL$ we define a special tensor $\tgCD\in\TgD$ by the formula
similar to\rx{2.6*}
\qq
\tgCD =
\fmfg^{\otimes\numv{\grvth(\xD)}}\otimes
\otmedD \hiedg
\otimes
\otmvrsD \hvrt.
\label{2.14}
\qqq
Finally, we define
the weight system, which is an algebra homomorphism
\qq
\xmap{\Tg}{\tcBCL}{\TxLCg},
\qquad\mbox{where $\TxLCg = \invT{\bigotimes_{\lA\in\xL} \aTA}$,}
\label{2.15}
\qqq
by the analog of \ex{2.7}
\qq
\Tg(\xD) =
\xCD\lrbc{\tgCD
}
\in\TxLCg
\label{2.16}
\qqq
and by $\Tg(\ccrl)=\dim\mfh$, $\Tg(\rcrl)=\dim\mfr$.
\subsubsection{\CCb, IHX and STU quotient}

\begin{figure}[hbt]
\begin{fmffile}{f1_3}
\begin{displaymath}
\begin{array}{cccc}
\begin{fmfgraph*}(25,5)
\fmfpen{thin}
\fmfleft{i1} \fmfright{o1}
\fmf{dots,tension=0.3}{o1,i1}
\end{fmfgraph*}
 &
\begin{fmfgraph*}(25,5)
\fmfpen{thin}
\fmfleft{i1} \fmfright{o1}
\fmf{dashes,tension=0.3}{o1,i1}
\end{fmfgraph*}
 &
\begin{fmfgraph*}(25,5)
\fmfleft{i1} \fmfright{o1}
\fmf{plain,tension=0.3}{o1,i1}
\end{fmfgraph*}
 &
\begin{fmfgraph*}(25,5)
\fmfpen{thin}
\fmfleft{i1} \fmfright{o1}
\fmf{dbl_dots,tension=0.3}{o1,i1}
\end{fmfgraph*}
\\
\mbox{Cartan} & \mbox{root} &\mbox{total} &\mbox{any}
\end{array}
\end{displaymath}
\end{fmffile}
\caption{Types of edges in graphs of $\sCGRL$}
\label{f1.3}
\end{figure}

\begin{figure}[hbt]
\begin{fmffile}{f1_4}
\begin{displaymath}
\begin{fmfgraph*}(20,20)
\fmfpen{thin}
\fmfsurround{v1,v2,v3}
\fmf{dots,tension=0.3}{v1,c}
\fmf{dots,tension=0.3}{v2,c}
\fmf{dots,tension=0.3}{v3,c}
\fmfdot{c}
\end{fmfgraph*}
 \qquad
\begin{fmfgraph*}(20,20)
\fmfpen{thin}
\fmfsurround{v1,v2,v3,v4,v5}
\fmf{dashes,tension=0.3}{v1,c}
\fmf{dots,tension=0.3}{v2,c}
\fmf{dots,tension=0.3}{v3,c}
\fmf{dots,tension=0.3}{v4,c}
\fmf{dots,tension=0.3}{v5,c}
\fmfblob{20}{c}
\end{fmfgraph*}
\end{displaymath}
\end{fmffile}
\caption{\CCo\ and \CCt\ subgraphs}
\label{f1.4}
\end{figure}

The kernel of the weight system homomorphism\rx{2.15} contains two ideals
$\tcBCLCCb$ and $\tcBCLIHX$, which we are about to describe.
A (1,3)-valent \xCLld\ graph is
called \CCo\ if it contains a 3-vertex all of whose incident edges are Cartan.
A graph is called \CCt\ if it contains a proper subgraph all of whose legs are
Cartan except for exactly one root leg.
The ideal $\tcBCLCCb$ is the spans of \CCo\ and \CCt\ graphs.

\begin{figure}[hbt]
\begin{fmffile}{f1_5}
\begin{displaymath}
\begin{array}{cccc}
\begin{fmfgraph*}(20,20)
\fmfleft{i1,i2} \fmfright{o1,o2}
\fmf{dbl_dots,tension=0.3}{i1,v,o2}
\fmf{dbl_dots,tension=0.3}{i2,v,o1}
\fmfdot{v}
\end{fmfgraph*}
, &
\begin{fmfgraph*}(20,20)
\fmfleft{i1,i2} \fmfright{o1,o2}
\fmf{dbl_dots,tension=0.3}{i1,v1,o1}
\fmf{dbl_dots,tension=0.3}{i2,v2,o2}
\fmf{plain,tension=0.3}{v1,v2}
\fmfdot{v1,v2}
\end{fmfgraph*}
, &
\begin{fmfgraph*}(20,20)
\fmfleft{i1,i2} \fmfright{o1,o2}
\fmf{dbl_dots,tension=0.3}{i1,v1,i2}
\fmf{dbl_dots,tension=0.3}{o1,v2,o2}
\fmf{plain,tension=0.3}{v1,v2}
\fmfdot{v1,v2}
\end{fmfgraph*}
, &
\begin{fmfgraph*}(20,20)
\fmfleft{i1,i2} \fmfright{o1,o2}
\fmf{dbl_dots,tension=1}{i1,v1}
\fmf{dbl_dots,tension=0.3}{v1,o2}
\fmf{dbl_dots,tension=1}{o1,v2}
\fmf{dbl_dots,tension=0.3}{v2,i2}
\fmf{plain,tension=0.3}{v1,v2}
\fmfdot{v1,v2}
\end{fmfgraph*}
\\
\xD & \xD_1 &\xD_2 &\xD_3
\end{array}
\end{displaymath}
\end{fmffile}
\caption{$\IHX$ graphs}
\label{f1.5}
\end{figure}


\begin{figure}[hbt]
\begin{fmffile}{f1_6}
\begin{displaymath}
\begin{array}{cccc}
\begin{fmfgraph*}(35,15)
\fmfleft{i1,i2} \fmfright{o1,o2}
\fmfpen{thin}
\fmf{dbl_dots,tension=0.3}{i2,v,o2}
\fmf{plain,tension=300,width=thick}{o1,v}
\fmf{plain_arrow,tension=300,width=thick}{v,i1}
\fmfdot{v}
\end{fmfgraph*}
, &
\begin{fmfgraph*}(35,15)
\fmfleft{i1,i2} \fmfright{o1,o2}
\fmf{plain,tension=300,width=thick}{o1,v1}
\fmf{plain_arrow,tension=300,width=thick}{v1,i1}
\fmf{dbl_dots,tension=0.3}{i2,v2,o2}
\fmf{plain,tension=0.3}{v1,v2}
\fmfdot{v1,v2}
\end{fmfgraph*}
, &
\begin{fmfgraph*}(35,15)
\fmfleft{i1,i2} \fmfright{o1,o2}
\fmf{plain,tension=300,width=thick}{o1,v1}
\fmf{plain,tension=300,width=thick}{v1,v2}
\fmf{plain_arrow,tension=300,width=thick}{v2,i1}
\fmf{dbl_dots,tension=0.3}{i2,v2}
\fmf{dbl_dots,tension=0.3}{o2,v1}
\fmfdot{v1,v2}
\end{fmfgraph*}
, &
\begin{fmfgraph*}(35,15)
\fmfleft{i1,i2} \fmfright{o1,o2}
\fmf{plain,tension=300,width=thick}{o1,v1}
\fmf{plain,tension=300,width=thick}{v1,v2}
\fmf{plain_arrow,tension=300,width=thick}{v2,i1}
\fmf{dbl_dots,tension=0.3}{i2,v1}
\fmf{dbl_dots,tension=0.3}{o2,v2}
\fmfdot{v1,v2}
\end{fmfgraph*}
\\
\xD & \xD_1 &\xD_2 &\xD_3
\end{array}
\end{displaymath}
\end{fmffile}
\caption{$\STU$ graphs}
\label{f1.6}
\end{figure}

The IHX ideal $\tcBCLIHX\subset\tcBCL$ is defined similarly to
$\tcBLIHX\subset\tcBL$. Let $\xD$ be a (1,3,4)-valent \xCLld\
graph with a single 4-valent vertex and incident vertices of any
type. Then
$\fCIHX(\xD) = \sIHXs{\xDv}\in\tcBL$,
where
$\xD_i$ ($1\leq i\leq 3$) are the graphs of $\sCGRL$ constructed
from $\xD$ by replacing its 4-valent vertex by two 3-valent
vertices connected by a total edge according to \fg{f1.5}.
Similarly, for $\xDp$ being a (1,2,3)-valent \xCLld\ graph with a
single 2-vertex labeled by an element of $\Urs$ and ordered
linearly together with other 1-vertices carrying the same label,
we define
$\fCSTU(\xDp) = \sIHXs{\xDv} \in\tcBL$,
where
$\xD_i$ ($1\leq i\leq 3$) are the graphs of $\sCGRL$ constructed
from $\xD$ by replacing its 2-valent vertex according to
\fg{f1.6}. The ideal $\tcBCLIHX$ is the span of $\fCIHX(\xD)$ and
$\fCSTU(\xDp)$ for all possible graphs $\xD,\xDp$. Now we define
\qq
\cBCL = \tcBCL/(\tcBCLCCb+\tcBCLIHX).
\label{2.17}
\qqq
\begin{remark}
\rm
Instead of taking a quotient over the \CCb\ ideal, we could exclude the \CCb\ graphs
from the set $\sCGRL$, whose elements span $\tcBCL$.
\end{remark}

The quotient $\cBCL$ again inherits
the properties of $\tcBCL$.
\begin{theorem}
\label{t2.2}
The algebra $\cBCL$ inherits the gradings $\dgrA$
($\lA\in\xLS\cup\xLW$) and $\dgrE$,
the gluing $\xmap{\glABm}{\cBL}{\cBL}$
and the weight systems
$\xmap{\Tg}{\cBL}{\TxLg}$ of $\tcBCL$.
Moreover, if $\xL\subset\xLp$, then the natural injection
$\tcBCL\hookrightarrow\tcBCLp$ descends to the injection
$\cBCL\hookrightarrow\cBCLp$.
\end{theorem}
\proof
The proof is essentially the same as that of Theorem\rw{t2.1}.
Note that \CCo\ graphs belong to $\ker\Tg$, because Cartan
subalgebra is commutative, and \CCt\ graphs belong to $\ker\Tg$,
because for any $n\geq 0$ there are no non-trivial $\mfT$-equivariant maps
$\mfh^{\otimes n}\longrightarrow\mfr$.\qed

\subsubsection{Injections $\cBLp\hookrightarrow\cBCL$}


Let $\xL=\xLt\cup\xLC\cup\xLr$ be a Cartan-split set of labels and
let $\xLp\subset\xLt$. Consider the following two injections
\qq
\ximap{\tftr,\tfC}{\sGRLp}{\sCGRL},
\label{2.19}
\qqq
where $\tftr$ maps a graph of $\sGRLp$ into the same graph of
$\sCGRL$, all of whose edges are declared total, while $\tfC$ does
the same, except that all legs of the graph in $\sCGRL$ are
declared Cartan. The injections\rx{2.19} generate the algebra
injections
\qq
\ximap{\tftr,\tfC}{\tcBLp}{\tcBCL}.
\label{2.20}
\qqq
Since both injections map $\tcBLpIHX$ into $\tcBCLIHX$, then they
generate the algebra homomorphisms
\qq
\xmap{\ftr,\fC}{\cBLp}{\cBCL}.
\label{2.21}
\qqq
\begin{theorem}
\label{t2.3}
The homomorphism $\ftr$ is an injection. If $\nxLpS = 1$,
while $\xLUp=\xLWp=\emptyset$, then $\fC$ is also an injection.
\end{theorem}

\begin{figure}[hbt]
\begin{fmffile}{f1_7}
\begin{displaymath}
\parbox{3cm}
{
\begin{fmfgraph*}(30,20)
\fmfleft{i} \fmfright{o}
\fmf{phantom,tension=0.3}{o,v}
\fmfblob{1.5cm}{v}
\fmf{plain,tension=0.3}{v,i}
\end{fmfgraph*}
}
=
\quad
\parbox{3cm}
{
\begin{fmfgraph*}(30,20)
\fmfleft{i} \fmfright{o}
\fmf{phantom,tension=1}{o,v1}
\fmf{plain,left=0.5,tension=0.45}{v2,v1}
\fmf{plain,right=0.5,tension=0.45,label=$\ast$}{v2,v1}
\fmf{plain,tension=1.5}{v2,i}
\fmfblob{1.5cm}{v1}
\fmfdot{v2}
\end{fmfgraph*}
}
=
\parbox{3cm}
{
\begin{fmfgraph*}(30,20)
\fmfleft{i1,i2,i3} \fmfright{o}
\fmf{phantom,tension=1}{o,v1}
\fmf{plain,tension=0.3}{v1,i3}
\fmf{plain,tension=0.3}{v1,v2}
\fmf{phantom,tension=1}{v2,i1}
\fmfblob{1.5cm}{v1}
\fmf{plain,right=90,label=$*$,label.side=left}{v2,v2}
\fmfdot{v2}
\end{fmfgraph*}
}
= 0
\end{displaymath}
\end{fmffile}
\caption{The IHX slide of the asterisk-marked edge onto itself}
\label{f1.7}
\end{figure}

\begin{figure}[hbt]
\begin{fmffile}{f1_8}
\begin{eqnarray}
&
\parbox{3.5cm}
{
\begin{fmfgraph*}(35,25)
\fmfstraight
\fmfleft{i1,i2,i3,i4,i5} \fmfright{o1,o2,o3,o4,o5} \fmfbottom{b1,b2,b3,b4,b5,b6}
\fmf{plain,tension=0.3}{i4,vi}
\fmf{plain,tension=0.3}{i5,vi}
\fmfdot{i4}
\fmfdot{i5}
\fmf{plain,tension=0.3}{o4,vo}
\fmf{plain,tension=0.3}{o5,vo}
\fmfdot{o4}
\fmfdot{o5}
\fmf{plain,tension=0.003}{b3,vb}
\fmf{plain,tension=0.003}{b4,vb}
\fmfdot{b3}
\fmfdot{b4}
\fmf{plain,tension=0.5}{vi,v}
\fmf{plain,tension=0.5}{vo,v}
\fmf{plain,tension=0.003}{vb,v}
\fmfblob{1cm}{vi}
\fmfblob{1cm}{vo}
\fmfblob{1cm}{vb}
\fmfdot{v}
\end{fmfgraph*}
}
\quad
=
\quad
\parbox{4cm}
{
\begin{fmfgraph*}(40,25)
\fmfstraight
\fmfleft{i1,i2,i3,i4,i5} \fmfright{o1,o2,o3,o4,o5} \fmfbottom{b1,b2,b3,b4,b5,b6}
\fmf{plain,tension=0.3}{i4,vi}
\fmf{plain,tension=0.3}{i5,vi}
\fmfdot{i4}
\fmfdot{i5}
\fmf{plain,tension=0.3}{o4,vo}
\fmf{plain,tension=0.3}{vo,v,o5}
\fmfdot{o4}
\fmfdot{o5}
\fmf{plain,tension=0.003}{b5,vb}
\fmf{plain,tension=0.003}{b6,vb}
\fmfdot{b5}
\fmfdot{b6}
\fmf{plain,tension=0.4}{vi,vo}
\fmf{plain,tension=0.003}{vb,v}
\fmfblob{1cm}{vi}
\fmfblob{1cm}{vo}
\fmfblob{1cm}{vb}
\fmfdot{v}
\end{fmfgraph*}
}
\quad
+
\quad
\parbox{4cm}
{
\begin{fmfgraph*}(40,25)
\fmfstraight
\fmfleft{i1,i2,i3,i4,i5} \fmfright{o1,o2,o3,o4,o5} \fmfbottom{b1,b2,b3,b4,b5,b6}
\fmf{plain,tension=0.3}{i4,vi}
\fmf{plain,tension=0.3}{i5,vi}
\fmfdot{i4}
\fmfdot{i5}
\fmf{plain,tension=0.3}{o5,vo}
\fmf{plain,tension=0.3}{vo,v,o4}
\fmfdot{o4}
\fmfdot{o5}
\fmf{plain,tension=0.003}{b5,vb}
\fmf{plain,tension=0.003}{b6,vb}
\fmfdot{b5}
\fmfdot{b6}
\fmf{plain,tension=0.4}{vi,vo}
\fmf{plain,tension=0.003}{vb,v}
\fmfblob{1cm}{vi}
\fmfblob{1cm}{vo}
\fmfblob{1cm}{vb}
\fmfdot{v}
\end{fmfgraph*}
}
\nonumber
\\
\nonumber
\\
\nonumber
\\
\vspace*{1cm}
&
\parbox{3.5cm}
{
\begin{fmfgraph*}(35,10)
\fmfstraight
\fmfleft{i1,i2,i3,i4,i5} \fmfright{o1,o2,o3,o4,o5} \fmfbottom{b1,b2,b3,b5,b6}
\fmf{plain,tension=0.3}{i3,vi}
\fmf{plain,tension=0.3}{i5,vi}
\fmfdot{i3}
\fmfdot{i5}
\fmf{plain,tension=0.3}{o3,vo}
\fmf{plain,tension=0.3}{o5,vo}
\fmfdot{o3}
\fmfdot{o5}
\fmf{plain,tension=0.5}{vi,v}
\fmf{plain,tension=0.5}{vo,v}
\fmf{plain,tension=0.003}{b3,v}
\fmfdot{b3}
\fmfblob{1cm}{vi}
\fmfblob{1cm}{vo}
\fmfdot{v}
\end{fmfgraph*}
}
\quad
=
\quad
\parbox{4cm}
{
\begin{fmfgraph*}(40,10)
\fmfstraight
\fmfleft{i1,i2,i3,i4,i5} \fmfright{o1,o2,o3,o4,o5} \fmfbottom{b1,b2,b3,b4,b5,b6}
\fmf{plain,tension=0.3}{i3,vi}
\fmf{plain,tension=0.3}{i5,vi}
\fmfdot{i3}
\fmfdot{i5}
\fmf{plain,tension=0.3}{o3,vo}
\fmf{plain,tension=0.3}{vo,v,o5}
\fmfdot{o3}
\fmfdot{o5}
\fmf{plain,tension=0.4}{vi,vo}
\fmf{plain,tension=0.003}{b5,v}
\fmfdot{b5}
\fmfblob{1cm}{vi}
\fmfblob{1cm}{vo}
\fmfdot{v}
\end{fmfgraph*}
}
\quad
+
\quad
\parbox{4cm}
{
\begin{fmfgraph*}(40,10)
\fmfstraight
\fmfleft{i1,i2,i3,i4,i5} \fmfright{o1,o2,o3,o4,o5} \fmfbottom{b1,b2,b3,b4,b5,b6}
\fmf{plain,tension=0.3}{i3,vi}
\fmf{plain,tension=0.3}{i5,vi}
\fmfdot{i3}
\fmfdot{i5}
\fmf{plain,tension=0.3}{o5,vo}
\fmf{plain,tension=0.3}{vo,v,o3}
\fmfdot{o3}
\fmfdot{o5}
\fmf{plain,tension=0.4}{vi,vo}
\fmf{plain,tension=0.003}{b5,v}
\fmfdot{b5}
\fmfblob{1cm}{vi}
\fmfblob{1cm}{vo}
\fmfdot{v}
\end{fmfgraph*}
}
\nonumber
\\
\nonumber
\\
\nonumber
\\
&
\parbox{3.5cm}
{
\begin{fmfgraph*}(30,15)
\fmfcurved
\fmfleft{i1,i2,i3}
\fmfright{o1,o2}
\fmf{plain,tension=0.3}{i1,vi}
\fmf{plain,tension=0.3}{i2,vi}
\fmf{plain,tension=0.3}{i3,vi}
\fmfdot{i1}
\fmfdot{i2}
\fmfdot{i3}
\fmf{plain,tension=0.5}{o1,v}
\fmf{plain,tension=0.5}{o2,v}
\fmfdot{o1}
\fmfdot{o2}
\fmf{plain,tension=0.5}{vi,v}
\fmfblob{1cm}{vi}
\fmfdot{v}
\end{fmfgraph*}
}
\quad =
\quad 0
\nonumber
\end{eqnarray}
\end{fmffile}
\caption{The IHX slide of a bridge and the IHX slide of a leg}
\label{f1.8}
\end{figure}
The proof of this theorem requires a couple of lemmas about the structure of $\cBLp$.
\begin{lemma}
\label{l2.1}
If a graph $\xD\in\sGRLp$ contains a proper subgraph
which has only one 1-vertex, then $\xD\in\tcBLpIHX$ and therefore
$\xD$ represents the 0-vector in $\cBLp=\tcBLp/\tcBLpIHX$.
\end{lemma}
\proof We follow the $\cBLp$ transformations of \fg{f1.7}. First,
we `pull out' a 3-vertex from $\xD$ (first equation). The IHX
relations in $\cBLp$ allow us to slide the asterisk-marked edges
all the way through the remaining subgraph onto itself (second
equation). Finally, any graph which has a proper subgraph
consisting of a loop with a single leg is equal to zero in
$\cBLp$ in view of the AS relation (third equation).\qed
\begin{lemma}
\label{l2.2}
If $\nxLpS = 1$, while $\xLUp=\xLWp=\emptyset$, then a graph $\xD\in\sGRLp$
which has a 3-vertex incident to 3 bridges, belongs to $\tcBLpIHX$ and therefore
represents the 0-vector in $\cBLp=\tcBLp/\tcBLpIHX$.
\end{lemma}
\proof Consider the IHX slides depicted in \fg{f1.8}. First, we
IHX-slide one bridge onto the legs. The summands in the \rhs of
the first equation of \fg{f1.8} have a 3-vertex which is incident
to a leg and two bridges. Next, we IHX-slide that leg onto the
legs as in the second equation of \fg{f1.8}. Each graph in the
\rhs of that equation has a 3-vertex which is incident to two
legs. Since both legs carry the same $\xLpS$ label, then such
graphs are zero due to the AS relation applied to that vertex.
\qed

\pr{Theorem}{t2.3}\footnote{We are very thankful to D.~Thurston who considerably
streamlined the proof of this theorem.}
Our strategy in proving the injectivity of the maps $\ftr,\fC$ is
to define the left-inverse maps of $\tftr$ and $\tfC$:
\qq
\xmap{\tftri,\tfCi}{\tcBCL}{\tcBLp},\qquad
\tftri\circ\tftr = \tfCi\circ\tfC = I.
\label{2.22}
\qqq
Then we will show that
\qq
\tftri(\tcBCLIHX),\tftri(\tcBCLCCb)\subset\tcBLpIHX,\qquad
\tfCi(\tcBCLIHX),\tfCi(\tcBCLCCb)\subset\tcBLpIHX.
\label{2.23}
\qqq
This implies that the maps $\tftri,\tfCi$ can be reduced to the
maps $\xmap{\ftri,\fCi}{\cBCL}{\cBLp}$, which in view of \ex{2.22}
are left-inverses of $\ftr$ and $\fC$:
\qq \ftri\circ\ftr = \fCi\circ\fC = I. \label{2.24} \qqq
The latter equation means that $\ftr$ and $\fC$ are injective, and
thus proves the theorem. Thus it remains to to define the maps
$\tftri,\tfCi$ which satisfy \ex{2.22}, and prove the
inclusions\rx{2.23}.

In order to define the maps $\tftri,\tfCi$ we pick a particular
basis in $\tcBCL$: it consists of graphs each of whose edges is
either root or Cartan (recall that a root edge is a difference
between a total edge and a Cartan edge). By definition, $\tftri$
maps a basis graph of $\tcBCL$ into the same graph of $\tcBLp$, if
all of its edges are total and all of its labels are from
$\xLp\subset\xL$, and $\tftri$ maps a basis graph to zero
otherwise. Then, obviously, $\tftri\circ\tftr=I$. The space
$\tcBCLIHX$ is equal to a span of the linear combinations of the
graphs of \fgs{f1.5} and\rw{f1.6} in which all those graphs are
the elements of the basis. Depending on the nature of their common
edges, $\tftri$ maps all graphs in each triplet either to zero or
to the corresponding graphs of $\tcBLp$, which form the
$\tcBLpIHX$ triplets. Therefore
$\tftri(\tcBCLIHX)\subset\tcBLpIHX$. Since $\tftri$ maps all
graphs, which have at least one Cartan edge, to zero, then
$\tftri(\tcBCLCCo)=0$. Almost all \CCt\ graphs have Cartan edges
and hence they are also mapped to zero. The only exception are
those \CCt\ graphs, whose \CCt\ subgraph has only one leg, which is
root. But $\tftri$ image of these graphs is proportional to a
graph of $\tcBLp$ which, according to Lemma\rw{l2.1}, belongs to
$\tcBLpIHX$. Thus we proved the inclusions\rx{2.23} for $\tftri$.

By definition, $\tfCi$ maps a basis graph of $\tcBCL$ to the same
graph of $\tcBLp$ if all of its non-bridge edges are total and all
of its labels are from $\xLp\subset\xL$. The image is $0$
otherwise. Obviously, $\tfCi\circ\tfC=I$. Also
$\tfCi(\tcBCLIHX)\subset\tcBLpIHX$ for the same reason as in the
case of $\tftri$. The $\tfCi$ image of a \CCo\ graph can be
non-zero only if all three Cartan edges of its \CCo\ subgraph are
bridges, but according to Lemma\rw{l2.2}, the image of such graphs
belongs to $\tcBLpIHX$, so $\tfCi(\tcBCLCCo)\subset\tcBLpIHX$.
Similarly, the $\tfCi$ image of a \CCt\ graph can be non-zero only
if all Cartan legs of its \CCt\ subgraph are bridges of the graph,
but then the root leg of that subgraph must also be a bridge.
However the $\tfCi$ image of a graph with a root bridge is zero.
Indeed, we can present it as a difference of two graphs in which
this bridge is either total or Cartan, but their $\tfCi$ images
are the same, since the action of $\tfCi$ on the basis elements
does not depend on the nature of their bridge edges. Thus we
proved the inclusions\rx{2.23} also for $\tfCi$.\qed

\subsection{Cohomology of graphs and Jacobi graph algebra $\cDCLc$}
\subsubsection{The space $\cHD$}

Let $\xD$ be a general graph.
We think of it as a $CW$-complex, its boundary $\boD$ being the set
of its
1-vertices. The space $\xCoD$ of 1-chains of
$\xD$ is a span of oriented edges of $\xD$ modulo the relation
that edges with opposite orientations represent opposite elements
of $\xCoD$. The rational relative cohomology $\cohbD$ can be
presented as a quotient
\qq
\cohbD = \xCoD/\xCmvD,
\label{2.25}
\qqq
where
\qq
\xCmvD =
\xspan\lrbc{
\seED \avedg\,\edg\; \Big|\;\vrt\in V_m(\xD), m\geq 2
}
\label{2.26}
\qqq
(here we fixed an orientation of each edge of $\xD$),
while $\avedg$ is an incidence number between a vertex $\vrt$ and
an oriented edge $\edg$:
\qq
\avedg =
\begin{cases}
0 & \text{if $\edg$ is not incident to $\vrt$ or if $\edg$ is a
loop}
\\
1 & \text{if $\edg$ goes into $\vrt$}
\\
-1 & \text{if $\edg$ goes out of $\vrt$}.
\end{cases}
\label{2.27}
\qqq
Univalent vertices were excluded from the sum of \ex{2.26}, because they
constitute the boundary of $\xD$.
Thus, in view of \ex{2.25}, the oriented edges, being the elements of
$\xCoD$, represent the integer elements of $\cohbD$: a pairing
between an oriented edge $\edg$ and a relative cycle of $\xD$ is
the coefficient at $\edg$ in the presentation of that cycle as a
linear combination of oriented edges.

Suppose that the edges of $\xD$ are labeled either as Cartan or root. Let
$\cohcD\subset\cohbD$ be the span of all Cartan edges of $\xD$. We denote
\qq
\cohrD = \cohbD/\cohcD.
\label{2.28}
\qqq
This quotient has an alternative presentation. Let $\xDr$ denote the graph
constructed from $\xD$ by removing all Cartan edges and 1-vertices
which are incident to them.
Then
\qq
\cohrD = \cohbDr.
\label{2.29}
\qqq
\begin{remark}
\label{r2.1}
\rm
A \xCLld\ graph is \CCt\ iff it has a root edge $\edg$ such that $\edg\in\cohrD$
(and hence $\edg=0$ as an element of $\cohrD$).
\end{remark}
In addition to $\xL$, we consider another label set $\xLH$,
whose elements we call \emph{\vrtl\ Cartan}
labels. All such labels are considered upper. The set $\xLH$ splits into the
subsets of \emph{\intgr} and \emph{\rtnl} labels: $\xLH = \xLZH\cup\xLQH$.

Let $\xD$ be a \xCLld\ graph.
For every label $\la\in\xLH$ we consider a separate copy $\cohrDa$ of the
space $\cohrD$ and the corresponding symmetric algebra $\tcHaD =
\Srst\cohrDa$. If $\xD$ is non-\CCt, then a root edge
%
%
$\edg\in\xEDr$ represents a non-zero element $\edg\in\cohrD$, and we will
denote by $\edga$ the corresponding elements in $\cohrDa\subset\tcHaD$. An element
$\zy\in\bpLQHaD$ is called \emph{\ebsd}, if it can be presented in the form
\qq
\zy = \peEDr p_{\edg}( (\edga)_{\la\in\xLQH}),
\label{2.68}
\qqq
where $p_{\edg}((\zx_\la)_{\la\in\xLQH})$ are non-zero formal power series of $\axLQH$ variables $\zx_\la$.
\begin{lemma}
\label{l2.8}
An \ebsd\ element\rx{2.68} can be zero only if $\xD$ is \CCt.
\end{lemma}
\proof
The product\rx{2.68} can be zero only if at least one factor $p_\edg$ is
zero, and a series $p_{\edg}( (\edga)_{\la\in\xLQH})$ can be zero only if
$\edg\in\SsthD$ is zero. However, according to Remark\rw{r2.1} the latter
happens iff $\xD$ is \CCt.\qed

The algebra $\SsthoLHD$ of a non-\CCt\ graph $\xD$ is
defined as a set of fractions
\qq
\SsthoLHD = \Big\{\;{\zx\over \zy}\;\Big|\;\zx\in\bpLHaD,\,\zy\in\bpLQHaD,\;
\text{$\zy$ is \ebsd}\Big\}.
\label{2.29*}
\qqq
Lemma\rw{l2.8} guarantees that the denominators $\zy$ are non-zero.

We define the symmetry group $\xGD$ of a \xCLld\ graph $\xD$ in the following
way: the elements of $\xGD$ map 3-vertices to 3-vertices, 1-vertices to
1-vertices with the same label and the edges to the edges of the same type,
while preserving the incidence between the edges and the vertices. The linear order of the
$\Urs$ 1-vertices has to be preserved, while the linear order of the $\Wrs$
vertices and the cyclic order of the edges at the multi-valent vertices may be
changed. For $\xg\in\xGD$, $\xsgng=\pm 1$ denotes the combined sign
associated with the change of order of $\Wrs$ 1-vertices and cyclic orders at
3-vertices produced by the action of $\xg$ on $\xD$. The group $\xGD$ acts
naturally on $\cohrD$, we denote this action as
$\xmap{\xpi}{\xGD}{\End(\cohrD)}$, $\xg\mapsto\xpg$. This action can
be extended as
algebra homomorphisms of $\SsthoLHD$.
We multiply this action by the factor $\xsgng$ and denote it as
$\xmap{\xpih}{\xGD}{\End(\SsthoLHD)}$.
We are
interested in the $\xGD$-invariant subalgebra
\qq
\cHLHD =
\begin{cases}
\invGD{\SsthoLHD}, & \text{if $\xD$ is non-\CCb,}
\\
\text{0-dimensional,} & \text{if $\xD$ is \CCb.}
\end{cases}
\label{2.30}
\qqq
There is a symmetrizing projector
\qq
\xmap{\prjGD}{\SsthoLHD}{\cHLHD},
\quad
\prjGD = {1\over \axGD} \sgGD \xphg.
\label{2.30*}
\qqq

If the label sets $\xL$ and $\xLH$ are fixed somehow, then we adopt the abbreviated
notation $\cHD$ for the space $\cHLHD$.

Let us consider three simple examples of the algebras $\cHLHD$.
\begin{description}
\item[Cartan strut] If $\xD=\cstab$, then $\xDr=\emptyset$, hence
$\cHLHv{\cstab}=\IR$ and we can symbolically present it as
\qq
\cHLHv{\cstab} = \stdf{ \ylab\stin{\cstab} }{\ylab\in\IR}
\label{2.30*x}
\qqq

\item[Root strut with different labels]
Since $\dim\cohrv{\rstxy}=1$ and for $\lx\neq\ly$ the graph
$\rstxy$ has no symmetry, then we can symbolically present
\qq
\cHLHv{\rstxy} = \stdf{ \ylaLHxy\stin{\rstxy} }{ \ylaLHxy\in\RLH },
\label{2.30*x1}
\qqq
where we used the multi-index notations for the lists of labels in
$\xLH$ and $\xLQH$
\qq
\mlaLH = \mltiv{\la}{\la\in\xLH}, \quad
\mlaLQH = \mltiv{\la}{\la\in\xLQH},\quad
\mlaLZH = \mltiv{\la}{\la\in\xLZH},
\label{2.30*x2}
\qqq
while
\qq
\RLH = \stdfl{{f(\mlaLH)\over g(\mlaLQH)}}
{f(\mlaLH)\in\RaLH,\;g(\mlaLQH)\in\RaLQH}.
\label{2.30*x3}
\qqq
\item[Root strut with the same labels]
The graph $\rstxx$ has a reflection symmetry, so
\qq
\cHLHv{\rstxx} = \stdf{ \ylaLHxx\stin{\rstxx} }{\ylaLHxx\in\RLHev},
\label{2.30x4}
\qqq
where $\RLHev\subset\RLH$ is the subalgebra of even functions,
that is $\yl(-\mlaLH) = \ylaLH$.

\end{description}

We define $\cHD$ also for four special `graphs': the empty unlabeled dot
$\emd$, the filled labeled dot $\fida$, $a\in\xLC$, the Cartan circle $\ccrl$
and the root circle $\rcrl$: by definition,
\qq
\cHav{\emd} = \Srs^2\, \IR^{\nxLH},\qquad
\cHav{\fida} = \IR^{\nxLH},\qquad
\cHav{\ccrl} = \IR.
\label{2.30*1}
\qqq
Equivalently we can present $\cHav{\emd}$ and $\cHav{\fida}$ as formal spans of
quadratic or linear monomials, the commutative variables being the labels of
$\xLH$:
\qq
\cHav{\emd} = \xspan(\lao\lat\,|\,\lao,\lat\in\xLH),\qquad
\cHav{\fida} = \xspan(\lb\,|\,\lb\in\xLH).
\qqq
The definition of $\cHav{\rcrl}$ is a bit more delicate. Following the
definitions for graphs we set $\cohrv{\rcrl}=\IR$ so that
$\xtcHav{\rcrl}=\Srst\IR$. Then we define
\qq
\SsthoLHv{\rcrl}
= \xspan\lrbcs{\log\zy,\,{\zx\over\zy}\;\Big|\;\zx\in\bopaLH\xtcHav{\rcrl}
,\,\zy\in\bopaLQH\xtcHav{\rcrl}}
\label{2.30*1x}
\qqq
modulo the usual properties of the logarithm:
\qq
\log(\zx\zy) = \log\zx + \log\zy,\quad\log(1+\zx) = -\snoi (-1)^n\zx^n /n.
\label{2.30*1x1}
\qqq
The symmetry group of $\rcrl$ is $\xGv{\rcrl}=\{1,-1\}$, so
\qq
\cHav{\rcrl} = \evbrs{\SsthoLHv{\rcrl}}.
\label{2.30*1x2}
\qqq
Equivalently,
\qq
\cHav{\rcrl} = \stdf{ \ylaLH\stin{\rcrl} }{ \ylaLH\in\RlogLHev },
\label{2.30*1x3}
\qqq
where $\RlogLHev\subset\RlogLH$ is the even part of the space
\qq
\RlogLH = \xspanv{ \log g(\mlaLQH),\,{f(\mlaLH)\over g(\mlaLQH)}}
{f(\mlaLH)\in\RaLH,\;g(\mlaLQH)\in\RaLQH}.
\label{2.30*1x4}
\qqq
%


Let $\xDo,\xDt$ be two \xCLld\ non-\CCb\ graphs. 
Their $\tcBCL$ product $\xDot$
is also non-\CCb.
As a graph, $\xDot$ is a disjoint union of
$\xDo$ and $\xDt$, hence $\cohrDot = \cohrDo\oplus\cohrDt$ and
$\xGDot\supset\xGDo\times\xGDt$. As a result,
%
\qq
\cHDot = \invv{\cHDo\otimes\cHDt}{\xGDot}.
\label{2.31}
\qqq
We use this equation in order to define $\cHD$ for $\xD$ being a disjoint
union of a regular graph
and some special `graphs'.


Suppose that for two non-\CCt\ graphs $\xD,\xDp$ there is a linear map
\qq
\xumap{\cohrD}{\xf}{\cohrDp}
\label{2.31*}
\qqq
(for example, it may be a cohomology pull-back
coming from a homeomorphism $\xDp\longrightarrow\xD$ which maps root edges
to root edges). We can extend $\xf$ to the algebra homomorphism
$$\bpLHaD\longrightarrow\bpLHaDp.$$
We call $\xf$ \emph{\ecnsv} if it maps any root edge $\edg\in\xD$
into another root edge $\edgp\in\xDp$. Then the corresponding algebra
homomorphism maps \ebsd\ elements into \ebsd\ elements. Hence $\xf$ can be
extended to the algebra homomorphism
$${\SsthoLHD}\longrightarrow{\SsthoLHDp},$$ and composing the latter with the
symmetrizing projector $\prjGDp$ we get the linear map
\qq
\xumap{\cHLHD}{\xfh}{\cHLHDp.}
\label{2.31*1}
\qqq
We call $\xfh$ a \emph{\htxt} of $\xf$.

\subsubsection{Pre-Jacobi graph algebra $\tcDCLc$}
Consider the regular graphs of $\sCGRL$. We call two such graphs
\emph{essentially equivalent} if they are the same as \xCLld\ graphs and hence
differ only by their orderings. Let us
pick a representative graph in each non-\CCb\ class of essential equivalency. The set
$\sCCGRL$
consists of these graphs as well as of their disjoint unions with
arbitrary numbers of special graphs $\emd, \fid, \ccrl, \rcrl$. The
space $\tcDCLc$ is
\qq
\tcDCLc = \bigoplus_{\xD\in\sCCGRL} \cHD.
\label{2.32}
\qqq
For any non-\CCb\ graph $\xD\in\sCGRL$ there is a natural injection
\qq
\cHD\hookrightarrow\tcDCLc,
\label{2.33}
\qqq
which is defined in the following way. Let $\xDp\in\sDCGRL$ be the graph which
is essentially equivalent to $\xD$. Since $\xD$ and $\xDp$ are isomorphic as
\xCLld\ graphs, there is the isomorphism $\cohrD\cong\cohrDp$ which is unique up
to the action of the symmetry group $\xGD$. Therefore there is the
unique isomorphism $\cHD\cong\cHDp$ coming from the former
isomorphism supplemented by the sign factor reflecting the
difference between the orderings of $\xD$ and $\xDp$. Since
$\cHLHDp\subset\tcDCLc$,
then the latter isomorphism implies the
injection\rx{2.33}.
%

Now we define the multiplication in $\tcDCLc$. Let
$\zx_i\in\cHav{\xD_i}$ ($i=1,2$). Consider the graph $\xDot$ which represents
the product of $\xDo$ and $\xDt$ in $\tcBCL$. In view of \ex{2.31} there is a
natural symmetrization projection map
\qq
\xumap{\cHDo\otimes\cHDt}{\prjGDot}{\cHDot.}
\label{2.34}
\qqq
We define the $\tcDCLc$ product $\zx_1\zx_2$ as
\qq
\zx_1\zx_2 = \prjGDot (\zx_1\otimes\zx_2).
\label{2.35}
\qqq
If the ordered graph $\xDot$ does not belong to $\sCCGRL$, then we
use the injection\rx{2.33} to place the product $\zx_1\zx_2$ into
$\tcDCLc$.

$\tcDCLc$ is a graded algebra. First, the whole space $\cHD$ inherits the gradings $\dgrA(\xD)$,
$\lA\in\xLS\cup\xLW$ and $\dgrE(\xD)$ from the graph $\xD$. Second, a space
$\cHD$ inherits the gradings $\dgra$, $\la\in\xLZH$ from the
individual symmetric algebras $\tcHaD$, since the action of $\xGD$
preserves these gradings.
All gradings are additive under the multiplication\rx{2.35}.

Next, we define the leg gluing, which is a linear map
$$\xmap{\glABm}{\tcDCLc}{\tcDCLc},$$
where $\lA,\lB\in\xLS\cup\xLW$ are labels of opposite levels.
Let $\zx\in\cHD$ and consider the graphs
$\xDi$ from the \rhs of \ex{2.1x}. Each of these graphs is a result of gluing
together of $m$ pairs of $\lA$ and $\lB$ legs, hence there is a natural map
\qq
\xumap{\hmbDri}{\fflti}{\hmbDr},
\label{2.36}
\qqq
which cuts the cycles of $\xDri$ at the gluing points.
The dual map $\xmap{\ffltis}{\cohrD}{\cohrDi}$ is \ecnsv\ and  we use its \htxt\
map
%
\qq
\xumap{\cHD}{\fhfltis}{\cHDi}
\label{2.37}
\qqq
in order to define
\qq
\glABm(\zx) = \siochAB \ysi\;\fhfltis(\zx)
\label{2.38}
\qqq
(\cf \ex{2.1x}).

As an example, suppose that $\lx,\lz\neq\ly$ and consider the following gluing:
\qq
\ylaLHxy\stin{\rstxy}\;\gl{\ly,\lz}{1}\;\ylaLHyz\stin{\rstyz} =
\ylaLHxz\stin{\rstxz},
\label{2.38*}
\qqq
where
\qq
\ylaLHxz =
\begin{cases}
\ylaLHxy\,\ylaLHyz & \text{if $\lx\neq \lz$}
\\
\hlfv\lrbcs{\ylaLHxy\,\ylaLHyz +
\ylv{\lx\ly}(-\mlaLH)\,\ylv{\ly\lz}(-\mlaLH)}
& \text{if $\lx = \lz$}.
\end{cases}
\label{2.38*1}
\qqq
%




\subsubsection{Weight system}
Recall the definition\rx{2.11} of the spaces $\aTA$ for the labels
$\lA\in\xL$. Since we consider the elements of $\xLH$ to be the upper $\Srs$
Cartan
labels, we define $\aTa=\Sh$ for $\la\in\xLH$.
Next we define the algebra
\qq
\TxLHHh = \Big\{ {\zx\over\zy}  \;\Big|\; \zx\in\bopaLH\aTa,\;\zy\in\bopaLQH\aTa \Big\}
\label{2.38y1}
\qqq
Then the weight system for $\tcDCLc$ is the
algebra homomorphism
\qq
\xmap{\TgcD}{\tcDCLc}{\TxLLHCg},\quad
\text{where $\TxLLHCg = \TxLCg\otimes\TxLHHh$.}
\label{2.38x1}
\qqq

In order to construct the
homomorphism\rx{2.38x1}, we have to introduce a finer splitting of
the Killing metric than\rx{2.12}. Let $\Dert$ be the set of roots
of $\mfg$. Then the complexified Lie algebra $\mfgC$ splits into
its complexified Cartan subalgebra and the root spaces $\mfrl$:
%
$\mfgC = \mfhC \oplus \boplD \mfrl$.
%
The Killing form (extended complex-linearly from $\mfg$ to
$\mfgC$) and its inverse now split into the following blocks:
\qq
\begin{alignedat}{2}
\hmfg &= \hmfh\oplus\boplD \hl, &\quad& \text{where $\hl\in\mfrmlls$,}
\label{2.38x3}
\\
\himfg &= \himfh\oplus\boplD \hil, && \text{where
$\hil\in\mfrlml$.}
\end{alignedat}
\qqq

Let $\xD$ be a regular $\sCCGRL$ graph and let us choose an orientation of
its root edges.
A map $\xmap{\yac}{\xEDr}{\Dert}$ is called \emph{a \rtas}.
For an edge $\edg$ of $\xD$ we define
a tensor $\hiedgc$, which depends on $\yac$:
\qq
\qquad
\hiedgc =
\begin{cases}
\himfh &\text{if $\edg$ is Cartan}
\\
\himfce &\text{if $\edg$ is root}
\\
\himfg & \text{if $\edg$ is total,}
\end{cases}
\label{2.38x4}
\qqq
(\cf \ex{2.13}) and consider a related tensor
\qq
\tgCDc =
\fmfg^{\otimes\numv{\grvth(\xD)}}\otimes
\otmedD \hiedgc
\otimes
\otmvrsD \hvrt.
\label{2.38x5}
\qqq
The following lemma about the application of the contraction map\rx{2.6}
to $\tgCDc$ is an easy consequence of the $\mfT$
invariance of the tensor $\fmfg$ which participates in the
expression\rx{2.38x4}.
\begin{lemma}
\label{l2.3}
$\xCD(\tgCDc)=0$ unless the \rtas\ $\yac$ satisfies the condition
\qq
\seEDr \avedg\,\yac(\edg) = 0 \quad\text{for all $\vrt\in
V_2(\xDr)\cup V_3(\xDr)$}.
\label{2.38x6}
\qqq
(we call such $\yac$ \emph{\cstas}).
\end{lemma}

The condition\rx{2.38x4} implies that a \csta\ $\yac$ defines an
element $\yachs\in\hmbDr\otimes\mfhs$.
By using the
Killing metric we construct $\yach\in\hmbDr\otimes\mfh$, which is
the conjugate element of $\yachs$, and build the
corresponding algebra homomorphism $\xmap{\hfyach}{\SsthD}{\Sh}$.
We extend it to the algebra homomorphism
\qq
\xumap{\SsthoLHD}{\hfLHc}{\TxLHHh}.
\label{2.38x7}
\qqq
Now we can write a formula for the weight system
homomorphism\rx{2.38x1}. Let $\yaCD$ be the set of all \cstas, then
for any $\zx\in\cHD\subset\SsthoLHD$ we define
\qq
\TgcD(\zx) = \syacD \xCD(\tgCDc)\otimes\hfLHc(\zx)\in\TxLLHCg = \TxLCg\otimes\TxLHCh.
\label{2.38x9}
\qqq

It remains to define the weight system map for special graphs:
\qq
&\TgcD(\lao\lat) = \himfh &\quad \text{for $\lao\lat\in\cHLHv{\emd}$,}
\nonumber
\\
&\TgcD(\lb) =
\begin{cases}
\himfh, & \text{if $\la\in\xLSu\cup\xLWu$}
\\
I, & \text{if $\la\in\xLSd\cup\xLWd$}
\end{cases}
&\quad\text{for $\lb\in\cHLHv{\fida}$}
\nonumber
\\
&\TgcD(1) = \dim \mfh&\quad\text{for $1\in\IR=\cHav{\ccrl}$.}
\label{2.38x3*}
\qqq
For a root circle $\rcrl$ we define the weight system map in accordance with
the general formula\rx{2.38x9} except that we have to include the logarithms
in it. Thus we define 
\qq
\TxLHrHh = \xspan\lrbcs{\log\zy,\,{\zx\over \zy}\;\Big|\;
\zx\in\bopaLH \aTa,  \; \zy\in\bopaLQH\aTa }
\label{2.80p}
\qqq
For a root
$\vl\in\Dert\subset\mfhs$ let $\vls\in\mfh$ denote the its dual element. The
space $\cohrv{\rcrl}=\IR$ has a canonical element $\edg$ corresponding to
$\rocrl$ itself. Let $\xmap{\xfl}{\cohrv{\rcrl}}{\mfh}$ be a linear map such
that $\xfl(\edg) = \vls$ and let
$$\xmap{\xfhl}{\bopaLH\xtcHav{\rcrl}}{\bopaLH\aTa}$$
be the corresponding algebra
homomorphism. Then we define the homomorphism
\qq
\xmap{\TgcD}{\cHLHv{\rcrl}}{\TxLHrHh}
\label{2.80p1}
\qqq
by the formulas
\qq
\TgcD(\log\zy) = \svlDert \log\xfhl(y),\quad
\TgcD\lrbc{\zx\over\zy} = \svlDert {\xfhl(\zx)\over\xfhl(\zy)}.
\label{2.79}
\qqq
for $\zx\in\bopaLH\xtcHav{\rcrl}$ and $\zy\in\bopaLQH\xtcHav{\rcrl}$.

The weight system maps are algebra
homomorphisms for all graph algebras. In other words, the multiplication of graphs is converted
into the multiplication of tensors. Leg gluing $\glABm$ is also converted
into a natural tensor operation which we are about to describe.

Let $V$ be a linear space. Its dual space $\Vast$ is the space of
linear operators on $V$. These operators can be identified with
derivations of the symmetric algebra $\Srst V$. Similarly, $V$ is
naturally the space of derivations of $\Srst \Vast$ and
$\Vast\otimes V$ is the space of the second order bilinear differential operators on
$\Srst V\otimes\Srst\Vast$. There is a special element
$I\in\End(\Vast) = \Vast\otimes V$, which is the identity operator, let $\hI$ denote the corresponding
operator acting on $\Srst V\otimes\Srst\Vast$. The operator $\hI$
acting on $\Wrst V\otimes\Wrst\Vast$ is constructed in the same
way, except that the first order derivation operators have degree 1
and hence they satisfy the super-Leibnitz rule.
In case when the
spaces $V$ and $\Vast$ are associated with labels $\lA,\lB$ we
will denote this operator as $\hOdAB$.

\begin{theorem}
\label{t2.3*}
For all graph algebras, the weight system homomorphism converts the gluing of
legs $\glABm$ into $(\hOdAB)^m/m!$.
\end{theorem}
\skproof
Since $\glABm = (\glABo)^m/m!$, it is sufficient to prove the
theorem for the case of $m$=1. Consider the algebra $\cBL$ and
suppose for simplicity that a graph $\xD$ has only one $\lA$ leg and only
one $\lB$ leg. Then the claim of the theorem
that $\hOdAB\Tg(\xD) = \Tg\,\glABo(\xD)$
follows from a
general fact that $\hI$ acts on $\Srs^1 V\otimes
\Srs^1\Vast\subset \Srst V\otimes\Srst\Vast$ as index contraction
operator $V\otimes\Vast\rightarrow \IR$. The same proof works for
$\cBCL$.

Now consider the algebra $\cDCL$. Again, let $\xD$ be a graph, which has only
one $\lA$ leg and one $\lB$ leg, and let $\xDp$ be the graph constructed by
gluing these legs together. There is a natural injection of consistent root
assignments: $\yaCDp\hookrightarrow\yaCD$, its image being the root
assignments of $\yaCD$ which which assign the same root to the $\lA$ and $\lB$
leg. These are only assignments which contribute non-zero terms in the sum of
\ex{2.38x9}. If $\yac$ is a root assignment of $\xD$ coming from the root assignment
$\yacp$ of $\xDp$, then
$\hOdAB\xCD(\tgCDc) = \xCDp(\tgCDpc)$ (for the same reason as for $\cBL$ and
$\cBCL$), while $\hfLHc(\zx) = \hfLHcp (\glABo(\zx))$. Hence, according to
the definition of the weight system homomorphism\rx{2.38x9},
$\hOdAB\TgcD(\zx) = \TgcD \glABo(\zx)$. \qed



\subsubsection{Total edge and IS quotient}
\label{ss2.3.4}

Let $\xD$ be a \xCLld\ \xrd\ non-\CCt\ (1,3,4)-valent graph with a single 4-valent
vertex. The graphs $\xDicr,\xDirt$ ($1\leq i\leq 3$) are constructed from it by
replacing the 4-valent vertex by two 3-valent vertices connected by
either a Cartan or a root edge (see \fg{f1.5}), which we will call
$\edg$. A contraction of $\edg$ in $\xDrirt$ creates a
homeomorphism $\xmap{\xfirt}{\xDrirt}{\xDr}$. The corresponding
pull-back map
\qq
\xmap{\xfirtast}{\cohrD}{\cohrDirt},
\label{2.39p}
\qqq
is \ecnsv\
and its \htxt\ is the
map
\qq
\xmap{\xfhirtast}{\cHD}{\cHDirt}.
\label{2.39}
\qqq
The construction of
$\xDricr$ from $\xDicr$
produces two 2-vertices which were 3-vertices of $\xDicr$ incident to the
edge $\edg$.
By gluing together
these 2-vertices into a single 4-vertex
we get a homeomorphism $\xmap{\xficr}{\xDricr}{\xDr}$,
its pull-back map
%
\qq
\xmap{\xficrast}{\cohrD}{\cohrDicr},
\label{2.40p}
\qqq
is \ecnsv\ and its \htxt\ is
\qq
\xmap{\xfhicrast}{\cHD}{\cHDicr}.
\label{2.40}
\qqq
Composing the maps\rx{2.39} and\rx{2.40} with the natural
injections $\cHDirt,\cHDicr\hookrightarrow\tcDCLc$ we get the maps
\qq
\xmap{\fCDIHXrti,\fCDIHXcri}{\cHD}{\tcDCLc}
\label{2.40x}
\qqq
and
\qq
\xmap{\fCDIHXi}{\cHD}{\tcDCLc},\qquad
\fCDIHXi = \fCDIHXcri + \fCDIHXrti.
\label{2.41}
\qqq
Finally,
\qq
\fCDIHXD =
\sIHXs{\fCDIHXv}.
\label{2.44}
\qqq
\begin{remark}
\rm
\label{r2.2}
Since the homeomorphisms $\xfirt$ are homotopic to identity, then
\\
$\dim\cohrDirt = \dim\cohrD$ and the maps $\xfirtast$ are
isomorphisms. If $\xDirt$ is \CCt, then $\dim\cohrDicr =
\dim\cohrD$ and $\xficrast$ is an isomorphism. If $\xDirt$ is
\CCt, then $\dim\cohrDicr = \dim\cohrD-1$ and $\xficrast$ is
surjective.
\end{remark}

We define the map $\fCDSTUD$ in a similar way by using \fg{f1.6}
instead of \fg{f1.5}. Note that if $\xD$ is a (1,2,3)-valent graph
with a single $\Urs$-type 2-vertex, as depicted in \fg{f1.6}, then
the definition of the space $\cohrD$ assumes that this 2-vertex
is a part of the boundary of $\xDr$.

Now the ideal $\tcDCLcIHX\subset\tcDCLc$ is
the span of the images of these two maps for all the appropriate
graphs $\xD$, and the Jacobi graph algebra $\cDCLc$ is the quotient
\qq
\cDCLc=\tcDCLc/\tcDCLcIHX.
\label{2.45}
\qqq
%

\begin{theorem}
\label{t2.7}
The Jacobi graph algebra $\cDCLc$ inherits the gradings $\dgrA$
($\lA\in\xLS\cup\xLW\cup\xLZH$) and $\dgrE$, the gluing map $\xmap{\glABm}{\cDCLc}{\cDCLc}$ and the
weight system homomorphism $\xmap{\Tg}{\cDCLc}{\TxLLHCg}$ from the pre-Jacobi graph algebra
$\tcDCLc$. Moreover, if $\xL\subset\xLp$ and $\xLZH\subset\xLZHp$, while
$\xLQH = \xLQHp$,
then the natural injection $\tcDCLc\hookrightarrow\tcDCLcp$
descends to the injection $\cDCLc\hookrightarrow\cDCLcp$.
\end{theorem}
\proof
The proof of almost all the claims is similar to that of Theorem\rw{t2.1}. Let
us prove that the weight system $\Tg$ descends to $\cDCLc$. We have to show
that $\im \fCDIHXD,\im\fCDSTUD\subset \ker \Tg$. We will prove this for
$\fCDIHXD$, the proof for $\fCDSTUD$ is similar.

In is easy to see that the homeomorphisms $\xfirt$ and $\xficr$ generate the
injections of \cstas\ $\ximap{\xfirtasu}{\yaCDirt}{\yaCD}$ and
$\ximap{\xficrasu}{\yaCDicr}{\yaCD}$. Naturally, for any $\zx\in\cHD$, these injections satisfy the
relations
\qq
\hfLHc(\fCDIHXrti(\zx)) = \hfLHv{\xfirtasu(\yac)}(\zx),\quad
\hfLHc(\fCDIHXcri(\zx)) = \hfLHv{\xficrasu(\yac)}(\zx)
\label{2.45x1}
\qqq
Moreover,
\qq
\xfirtasu(\yaCDirt)\cap\xficrasu(\yaCDicr) = \emptyset, \qquad
\xfirtasu(\yaCDirt)\cup\xficrasu(\yaCDicr) = \yaCD.
\label{2.45*1}
\qqq
Hence, we can define the functions $\xmap{\cTfi}{\yaCD}{\TxLHCh}$ ($1\leq i\leq 3$) as
\qq
\cTfi(\yac) =
\begin{cases}
\xCDicr(\tgCDicr),\quad\text{if $\yac\in\xficrasu(\yaCDicr),$}
\\
\xCDirt(\tgCDirt),\quad\text{if $\yac\in\xfirtasu(\yaCDirt).$}
\end{cases}
\label{2.45*2}
\qqq
Then the Jacobi identity for $\mfg$ implies that
\qq
\sIHXs{\cTfv} = 0.
\label{2.45*3}
\qqq
Also for $\zx\in\cHD$
\qq
\lefteqn{\Tg(\fCDIHXi(\zx))}
\nonumber
\\
& = &
\Tg(\fCDIHXcri(\zx)) + \Tg(\fCDIHXrti(\zx))
\nonumber
\\
& = &
\syacDicr \xCDicr(\tgCDicr)\otimes\hfLHc(\fCDIHXrti(\zx))
+
\syacDirt \xCDirt(\tgCDirt)\otimes\hfLHc(\fCDIHXcri(\zx))
\nonumber
\\
& = &
\syacDicr \xCDicr(\tgCDicr)\otimes\hfLHv{\xficrasu(\yac)}(\zx)
+
\syacDirt \xCDirt(\tgCDirt)\otimes\hfLHv{\xfirtasu(\yac)}(\zx)
\nonumber
\\
& = &
\syacD \cTfi(\yac) \otimes \hfLHc(\zx).
\label{2.45*4}
\qqq
As a result,
\qq
\Tg(\fCDIHXD(\zx)) & = & \sIHXs{\Tg(\fCDIHXv}(\zx))
\nonumber
\\
& = &
\syacD \lrbc{ \sIHXs{\cTfv}(\yac) } \otimes\hfLHc(\zx) = 0.
\label{2.45*5}
\qqq
\qed

Finally, in order to make a connection between the definition of $\tcDCLcIHX$
and that of the IHX ideal in $\tcBCL$, we will introduce a notion of a total
edge.

Suppose that a graph $\xD\in\sCCGRL$ has a root edge $\edg$. We define the
subalgebra $\SsthoLHeD\subset\SsthoLHD$ as the set of fractions, whose \ebsd\
denominators do not include the factor coming from $\edg$:
\qq
\SsthoLHeD = \Big\{ {\zx\over\zy}
\;\Big|\; \zx\in\bpLHaD,\;\zy=
\pepEDr p_{\edgp}( (\edgpa)_{\la\in\xLQH})
\Big\}.
\label{2.45*5x}
\qqq
Then we define $\cHLHeD =\invGD{\SsthoLHeD}$.

If we declare the root edge
$\edg$ of $\xD$ to be Cartan, then we get another \xCLld\ \xrd\ graph
$\xDe$. Obviously, there is an embedding of the \castr\ graphs
\qq
\ximap{\xfe}{\xDre}{\xDr},
\label{2.45x}
\qqq
which generates the pull-back of the cohomology
\qq
\xmap{\xfes}{\cohrD}{\cohrDe}.
\label{2.45xy}
\qqq
This pull-back map is not \ecnsv, because $\xfes(\edg)=0$. However, since the
denominators of $\cHLHeD$ do not contain $\edg$, then we can still define the
\htxt\ map
%
\qq
\xumap{\cHLHeD}{\xfhes}{\cHLHDe.}
\label{2.46}
\qqq

Let $\xD$ be a \xCLld\ and \xrd\ graph, all of whose edges are either Cartan
or root except an edge $\edg$, which is declared total.
We are going to define the space $\cHD$ and its injection\rx{2.33}.
Let $\xDcr$ and $\xDrt$ be the
graphs constructed from $\xD$ by declaring the edge $\edg$ Cartan or root.
If the graph $\xDrt$ is \CCb, then we define $\cHLHD=\cHLHDcr$ with the
corresponding injection $\cHLHDcr\hookrightarrow\cDCLc$. If the graph $\xDrt$
is not \CCb, then we define $\cHLHD=\cHLHDrt$, while
%
the injection\rx{2.33} is the composition of two maps
\qq
\begin{CD}
\cHLHD @>{\xfhes\oplus I}>> \cHLHDcr\oplus\cHLHDrt
\hookrightarrow \tcDCLc,
\end{CD}
\label{2.47}
\qqq
in which $\xmap{\xfhes}{\cHLHD}{\cHLHDcr}$ is the map\rx{2.39}, while the second
map of the composition is the sum of standard injections\rx{2.33}.
The weight map $\xmap{\Tg}{\cHD}{\TxLLHCg}$ is the composition of the
injection\rx{2.47} and the weight map\rx{2.38x1}.

Now let $\xD$ be a (1,3,4)-valent \xCLld\ and \xrd\ non-\CCt\ graph with a single
4-vertex and let $\xDi$ ($1\leq i\leq 3$) be the graphs of \fg{f1.5}
constructed from $\xD$ by resolving the 4-vertex through an insertion of a
total edge. The previous definition allows us to use the maps\rx{2.39} in
order to define the maps
\qq
\xmap{\xfhiast}{\cHD}{\cHDi}.
\label{2.48}
\qqq
\begin{proposition}
\label{p2.1}
A composition of the map\rx{2.48} with the injection
$\cHDi\hookrightarrow\tcDCLc$ is equal to the map\rx{2.41}.
\end{proposition}
\proof
This equality follows from the fact that the composition of the homeomorphisms
$\ximap{\xfei}{\xDricr}{\xDrirt}$ and $\xmap{\xfirt}{\xDrirt}{\xDr}$ is equal
to $\xmap{\xficr}{\xDricr}{\xDr}$, so that
\qq
\xfhirtast\,\xfhesi = \xfhicrast.
\label{2.48*}
\qqq
\qed


\subsection{Conversion of legs}

\subsubsection{Conversion of integer virtual legs into real legs}

Let $\xLp$ and $\xLHp$ be the sets of labels and let $\lap\in\xLpSu$. Then let
$\xL = \xLp\setminus\{\lap\}$ and
$\xLZH=\xLHp\cup\{\la\}$, $\xLQH=\xLQHp$.
\begin{theorem}
\label{t2.4}
If $\xLU=\emptyset$, then there is a canonical algebra isomorphism
\qq
\xmap{\xhfaap}{\cDCLc}{\cDCLcp},
\label{2.49}
\qqq
which preserves the weight system homomorphism: the following diagram is
commutative
\qq
\begin{diagram}
\node{\cDCLc}
  \arrow{e,t}{\xhfaap}
  \arrow{s,r}{\TgcD}
\node{\cDCLcp}
  \arrow{s,r}{\TgcD}
\\
\node{\TxLLHCg}
  \arrow{e,=}
\node{\TxLLHpCg}
\end{diagram}
\label{2.49*}
\qqq
\end{theorem}

%

First, we define the (inverse) isomorphism $\xhfiaap$ for the special graphs. The spaces
$\cHLHpv{\ccrl}$, $\cHLHpv{\rcrl}$, $\cHLHpv{\emd}$ and $\cHLHpv{\fidb}$,
$\lb\neq\lap$ map identically to the same spaces in $\cDCLc$, while
$1\in\IR=\cHLHpv{\lb\cstr \lap}$ maps to
$\la\in\cHLHv{\fidap}$ and
\\
$1\in\IR=\cHLHpv{\lap\cstr\lap}$ maps to
$\la^2\in\cHLHv{\emd}$.

Now let us consider regular graphs.
A 3-vertex in a non-\CCb\ graph can have at most one incident Cartan edge, so a
3-vertex in a graph
$\xDp\in\sCCGRLp$,
which is incident to an $\lap$ leg, is also
incident to two root edges. Thus if we remove a $\lap$ leg and dissolve the
remaining 2-vertex, then we obtain another $\sCCGRLp$ graph.
For $\xD\in\sCCGRLp$, let $\xsDap\subset\sCCGRLp$ be the set of all graphs
such that the removal of all their $\lap$ legs produces $\xD$. We introduce
the notations
\qq
\tcHapD = \bopDpXDap\!\!\SsthoLHDp,\qquad
\cHapD = \bopDpXDap\!\!\cHLHDp.
\label{2.52}
\qqq
Since $\sCCGRLp = \bigcup_{\xD\in\sCCGRL}\xsDap$, then
\qq
\tcDCLcp =
\bopDsCCGRL
\cHapD.
\label{2.52*}
\qqq

Let $\tcDCLcIHXip\subset\tcDCLcIHXp$ ($0\leq i\leq 4$) be the ideals
constructed from (1,3,4)-valent graphs, whose 4-valent vertex is incident to
exactly $i$ of $\lap$ legs.
If the 4-vertex has at least 3 incident legs, then all graphs in
the corresponding IHX triplet of \fg{f1.5} are \CCb.
%
\begin{figure}[hbt]
\begin{fmffile}{f1_9}
\begin{displaymath}
\begin{array}{cccc}
\begin{fmfgraph*}(35,15)
\fmfleft{i1,i2} \fmfright{o1,o2}
\fmfpen{thin}
\fmf{dots,tension=0.3}{i2,v,o2}
\fmfdot{i2} \fmfdot{o2}
\fmflabel{$\lap$}{i2} \fmflabel{$\lap$}{o2}
\fmf{dbl_dots,tension=300,width=thin}{o1,v}
\fmf{dbl_dots,tension=300,width=thin}{v,i1}
\fmfdot{v}
\end{fmfgraph*}
, &
\begin{fmfgraph*}(35,15)
\fmfleft{i1,i2} \fmfright{o1,o2}
\fmfdot{i2} \fmfdot{o2}
\fmflabel{$\lap$}{i2} \fmflabel{$\lap$}{o2}
\fmf{dbl_dots,tension=300}{o1,v1}
\fmf{dbl_dots,tension=300}{v1,i1}
\fmf{dots,tension=0.3}{i2,v2,o2}
\fmf{plain,tension=0.3}{v1,v2}
\fmfdot{v1,v2}
\end{fmfgraph*}
, &
\begin{fmfgraph*}(35,15)
\fmfleft{i1,i2} \fmfright{o1,o2}
\fmfdot{i2} \fmfdot{o2}
\fmflabel{$\lap$}{i2} \fmflabel{$\lap$}{o2}
\fmf{dbl_dots,tension=300}{o1,v1}
\fmf{plain,tension=300}{v1,v2}
\fmf{dbl_dots,tension=300}{v2,i1}
\fmf{dots,tension=0.3}{i2,v2}
\fmf{dots,tension=0.3}{o2,v1}
\fmfdot{v1,v2}
\end{fmfgraph*}
, &
\begin{fmfgraph*}(35,15)
\fmfleft{i1,i2} \fmfright{o1,o2}
\fmfdot{i2} \fmfdot{o2}
\fmflabel{$\lap$}{i2} \fmflabel{$\lap$}{o2}
\fmf{dbl_dots,tension=300}{o1,v1}
\fmf{plain,tension=300}{v1,v2}
\fmf{dbl_dots,tension=300}{v2,i1}
\fmf{dots,tension=0.3}{i2,v1}
\fmf{dots,tension=0.3}{o2,v2}
\fmfdot{v1,v2}
\end{fmfgraph*}
\\
\xD & \xD_1 &\xD_2 &\xD_3
\end{array}
\end{displaymath}
\end{fmffile}
\caption{IHX graphs with two $\lap$ legs}
\label{f1.9}
\end{figure}
%
Consider the IHX triplet
of \fg{f1.9}, which comes from
a (1,3,4)-valent graph, whose 4-valent vertex is incident to two $\lap$ legs.
The graph $\xDo$ is
$\CCb$, while the graphs $\xDt$ and $\xDth$ are isomorphic, hence
$\fCDIHXt-\fCDIHXth=0$ in \ex{2.44}. Thus the spaces
$\xtcDCLcIHXip$
are 0-dimensional for $i\geq 2$,
and as a result,
\qq
\cDCLcp = \tcDCLc/ (\tcDCLcIHXzp + \tcDCLcIHXop).
\label{2.59}
\qqq
%

\begin{figure}[hbt]
\begin{fmffile}{f1_11}
\begin{eqnarray}
&
\begin{array}{cccc}
\begin{fmfgraph*}(35,15)
\fmfleft{i1,i2} \fmfright{o1,o2}
\fmfpen{thin}
\fmf{dashes,tension=0.3}{i2,v}
\fmf{dots,tension=0.3}{v,o2}
\fmfdot{o2}
\fmflabel{$\lap$}{o2}
\fmf{dashes,tension=300,width=thin}{o1,v}
\fmf{dashes,tension=300,width=thin}{v,i1}
\fmfdot{v}
\end{fmfgraph*}
, &
\begin{fmfgraph*}(35,15)
\fmfleft{i1,i2} \fmfright{o1,o2}
\fmfdot{o2}
\fmflabel{$\lap$}{o2}
\fmf{dashes,tension=300}{o1,v1}
\fmf{dashes,tension=300}{v1,i1}
\fmf{dots,tension=0.3}{v2,o2}
\fmf{dashes,tension=10}{v2,i2}
\fmf{dashes,tension=10}{v1,v2}
\fmfdot{v1,v2}
\end{fmfgraph*}
, &
\begin{fmfgraph*}(35,15)
\fmfleft{i1,i2} \fmfright{o1,o2}
\fmfdot{o2}
\fmflabel{$\lap$}{o2}
\fmf{dashes,tension=300}{o1,v1}
\fmf{dashes,tension=300}{v1,v2}
\fmf{dashes,tension=300}{v2,i1}
\fmf{dashes,tension=0.3}{i2,v2}
\fmf{dots,tension=0.3}{o2,v1}
\fmfdot{v1,v2}
\end{fmfgraph*}
, &
\begin{fmfgraph*}(35,15)
\fmfleft{i1,i2} \fmfright{o1,o2}
\fmfdot{o2}
\fmflabel{$\lap$}{o2}
\fmf{dashes,tension=300}{o1,v1}
\fmf{dashes,tension=300}{v1,v2}
\fmf{dashes,tension=300}{v2,i1}
\fmf{dashes,tension=0.3}{i2,v1}
\fmf{dots,tension=0.3}{o2,v2}
\fmfdot{v1,v2}
\end{fmfgraph*}
\\
\xD & \xD_1 &\xD_2 &\xD_3
\vspace*{1cm}
\end{array}
\nonumber
\\
&
\begin{array}{ccc}
\begin{fmfgraph*}(35,15)
\fmfleft{i1,i2} \fmfright{o1,o2}
\fmfpen{thin}
\fmf{dots,tension=0.3}{i2,v}
\fmf{dots,tension=0.3}{v,o2}
\fmfdot{o2}
\fmflabel{$\lap$}{o2}
\fmf{dashes,tension=300,width=thin}{o1,v}
\fmf{dashes,tension=300,width=thin}{v,i1}
\fmfdot{v}
\end{fmfgraph*}
, &
\begin{fmfgraph*}(35,15)
\fmfleft{i1,i2} \fmfright{o1,o2}
\fmfdot{o2}
\fmflabel{$\lap$}{o2}
\fmf{dashes,tension=300}{o1,v1}
\fmf{dashes,tension=300}{v1,v2}
\fmf{dashes,tension=300}{v2,i1}
\fmf{dots,tension=0.3}{i2,v2}
\fmf{dots,tension=0.3}{o2,v1}
\fmfdot{v1,v2}
\end{fmfgraph*}
, &
\begin{fmfgraph*}(35,15)
\fmfleft{i1,i2} \fmfright{o1,o2}
\fmfdot{o2}
\fmflabel{$\lap$}{o2}
\fmf{dashes,tension=300}{o1,v1}
\fmf{dashes,tension=300}{v1,v2}
\fmf{dashes,tension=300}{v2,i1}
\fmf{dots,tension=0.3}{i2,v1}
\fmf{dots,tension=0.3}{o2,v2}
\fmfdot{v1,v2}
\end{fmfgraph*}
\\
\xD
&\xD_2 &\xD_3
\end{array}
\nonumber
\end{eqnarray}
\end{fmffile}
\caption{Non-\CCb\ IHX graphs with one $\lap$ leg}
\label{f1.11}
\end{figure}
%

Let us consider the structure of $\tcDCLcIHXop$.
If at least two edges, incident to the 4-valent vertex, besides the $\lap$
leg, are Cartan, then all graphs of the corresponding IHX triplet are \CCb\
and we discard them. Thus there should be either three or two root edges
incident to the 4-valent vertex (see \fg{f1.11}), and the graph $\xDo$ in the
latter case is discarded, because it is \CCb. In both cases, all graphs of the
same triplet belong to the same set $\xsDap$, hence
%
if we denote
\qq
\ccHIHXapD = \tcDCLcIHXop\cap\cHapD,
\label{2.60p1}
\qqq
then
\qq
\tcDCLcIHXop = \bopDxsDCGRL \ccHIHXapD
\label{2.60p}
\qqq
and
\qq
\cDCLcp = \lrbc{ \bopDxsDCGRL \ccHapD } \Big/ \tcDCLcIHXzp.
\label{2.60}
\qqq
%
%

%
\begin{figure}[hbt]
\label{f1.10}
\qq
\begin{diagram}
\dgsquash[4/2]
\dgmag{800}
\node[2]{\SsthoLpHD\otimes\SxCoDr}
  \arrow{sw,t}{\prjGD}
  \arrow{se,t,1}{\xhtfaap}
  \arrow[2]{e}
  \arrow[2]{s,r,3}{\xfo}
\node[2]{\tcHapD}
  \arrow[2]{w,t}{\xhtfaaps,\xhtfbiaaps}
  \arrow{sw,b}{\bopDpXDap\prjGDp}
  \arrow{swww,t,1}{\xhtfbiaap}
\\
\node{\invGD{\SsthoLpHD\otimes\SxCoDr}}
  \arrow[2]{e,t}{\xhtfaap}
  \arrow[2]{s,r}{\xfo}
\node[2]{\cHapD}
  \arrow[2]{s,r}{\xft}
\\
\node[2]{\SsthoLHD}
  \arrow{sw,t}{\prjGD}
  \arrow{se,t}{\xhfaap}
\\
\node{\cHLHD}
  \arrow[2]{e,t}{\xhfaap}
\node[2]{\ccHapD}
\end{diagram}
\nonumber
\qqq
\caption{Monster diagram}
\end{figure}
%

Now let us consider the commutative monster diagram
of \fg{f1.9}
piece by piece.
Let us choose the orientation on the root edges of a graph $\xD\in\sCCGRL$.
For every monomial
$$\zp=\prod_{\edg\in\xEDr}\edg^{\nedg}\in\SxCoDr$$
there exists a unique graph $\xDpzp\in\xsDap$
constructed by attaching $\nedg$ of $\lap$ legs to every root edge $\edg$ of $\xD$.
Obviously, the graphs $\xDpzpr$ and $\xDr$ are isomorphic (up to the
dissolution of 2-vertices left by $\lap$ legs of $\xDpzp$), but this
isomorphism is not unique because of the symmetry of the graphs. Let $\zs$ be
a choice of the isomorphism
$\xDpzpr\longrightarrow\xDr$ for every
monomial $\zp\in\SxCoDr$ and let $\xmap{\xfs}{\SsthoLpHD}{\SsthoLpHDzp}$ be
the corresponding algebra homomorphisms. Then for $\zx\in\SsthoLpHD$ we define
\qq
\xhtfaaps(\zx\otimes\zp) = (-1)^{\nlft(\xDpzp)}\,\xfs(\zx),
\label{2.55}
\qqq
where $\nlft(\xDpzp)$ is the number of $\lap$ legs, which had to be attached
on the left side of the oriented root edges of $\xD$ in order to reproduce
$\xDpzp$. Note that $\xhtfaaps(\zx\otimes\zp)$ does not depend on the choice
of orientation of the root edges of $\xD$.

The `inverse' maps $\xhtfbiaaps$ are constructed in the similar way.
Let $\zsp$ denote a choice of an isomorphism $\xDpr\longrightarrow\xDr$ for
any $\xDp\in\xsDap$, it implies the choice of algebra isomorphisms
$\xmap{\xfbisDp}{\SsthoLpHDp}{\SsthoLpHD}$. We define the monomials
\qq
\zpsDp =(-1)^{\nlft(\xDp)}\prod_{\edg\in\xEDr} \edg^{\nedg},
\label{2.53}
\qqq
where $\nedg$ is the
number of $\lap$ legs attached to the edge $\edg$,
and then for $\zx\in\tcHapD$
\qq
\xhtfbiaaps(\zx) = \xfbisDp(\zx)\otimes\zpsDp.
\label{2.54}
\qqq

The compositions of the maps\rx{2.54} and\rx{2.55} with the
symmetry projectors
\qq
\xhtfaap = \lrbc{\bopDpXDap\prjGDp}\xhtfaaps,\qquad
\xhtfbiaap = \prjGD\,\xhtfbiaaps
\label{2.56}
\qqq
do not depend on the choices of identifications $\zs$, $\zsp$ and our goal is
to show that they descend to the isomorphism\rx{2.49} and its inverse.

\begin{lemma}
\label{l2.4}
The maps $\xhtfaap,\xhtfbiaap$ between
$\invGD{\SsthoLpHD\otimes\SxCoDr}$ and
$\cHapD$ are
inverses of each other and thus establish the isomorphism between
these spaces.
\end{lemma}
\proof
The products of maps $\xhtfaap\xhtfbiaap$ and $\xhtfbiaap\xhtfaap$
are equal to identity up to the action of the symmetry groups,
which account for the indeterminacy in identification between the
graphs. Therefore, it is easy to see that these products are equal
to identity when restricted to the symmetric parts of their
domains.\qed

Now let us consider the lower part of the monster diagram of
\fg{f1.9}. The map $\xft$ is the natural
surjections onto a quotient.
The map $\xmap{\xfo}{\SsthoLpHD\otimes\SxCoDr}{\SsthoLHD}$
is of the same
nature: since $\la$ is an integer virtual label, then $\SsthoLHD =
\SsthoLpHD\otimes\Srs^\ast\cohrD$, while according to \eex{2.29}
and\rx{2.25}, $\cohrD = \xCoDr/\xCmvDr$, so
\qq
\SsthoLHD = \SsthoLpHD\otimes\SxCoDr
\Big/\lrbc{\SsthoLpHD\otimes\xImvDr},
\label{2.62}
\qqq
where $\xImvDr\subset\SxCoDr$ is the ideal generated by $\xCmvDr$.

The map $\xfo$ is equivariant with respect to the action of the graph
symmetry group $\xGD$, therefore it maps the $\xGD$-invariant subspace
$\invGD{\SsthoLpHD\otimes\SxCoDr}$ into $\cHLHD$. In fact, this restriction
of $\xfo$ is again a surjection onto a quotient, because
\qq
\cHLHD = \invGD{\SsthoLpHD\otimes\SxCoDr}\Big/
\invGD{\SsthoLpHD\otimes\xImvDr},
\label{2.63}
\qqq
which follows from the next lemma.
\begin{lemma}
\label{l2.5}
Suppose that a finite group $G$ acts on a linear space $V$ and a subspace
$W\subset V$ is invariant under this action. Then $\invGVW =
\invGV/(W\cap\invGV)$.
\end{lemma}
\proof
Let $\xmap{f}{V}{V/W}$ be the surjection onto the quotient and let
$\prjG$ be the symmetrizing projection for the action of $G$.
The composition of
maps
$$
\begin{CD}
V @>f>> V/W @>\prjG>> \invGVW\subset V/W
\end{CD}
$$
and the composition of maps
$$
\begin{CD}
V @>\prjG>> \invGV @>f>> \invGV/(W\cap\invGV)\subset V/W
\end{CD}
$$
are both surjective.
Since $W$ is $G$-invariant, then $f$ commutes with $\prjG$,
hence
$\invGVW =
\invGV/(W\cap\invGV)$.\qed

Thus all three vertical maps of the monster diagram are surjections onto quotients.

\begin{lemma}
\label{l2.6}
The isomorphisms $\xhtfaap,\xhtfiaap$ between $\invGD{\SsthoLpHD\otimes\SxCoDr}$ and $\cHapD$
descend to the isomorphisms
$\xhfaap,\xhfiaap$ between $\cHLHD$ and $\ccHapD$.
\end{lemma}
\proof
It is enough to show that the isomorphisms $\xhtfaap,\xhtfiaap$
map the spaces
\\
$\invGD{\SsthoLpHD\otimes\xImvDr}$ and
$\ccHIHXapD$ into each other. We will prove it for $\xhtfaap$, the proof for
$\xhtfiaap$ is similar. According to the definition\rx{2.26},
\qq
\xImvDr = \xspan\lrbc{\zpp
\seEDr \avedg\,\edg\; \Big|\;\vrt\in V_m(\xDr), m=2,3; \text{$\zpp$ is a monomial in
$\SxCoDr$}
}.
\label{2.65}
\qqq
Let us take a 3-vertex $\vrt\in V_3(\xDr)$ and consider three
monomials $\zppei$, where $\edgi$ are three root edges incident to
$\vrt$. The graphs $\xDpzppei$ constructed by converting the factors
of $\zppei$ into $\lap$ legs, form an IHX triplet of graphs in the first line
of \fg{f1.11} coming from the (1,3,4)-valent graph constructed by attaching
the $\lap$ leg to the vertex $\vrt$ of the graph $\xDpzpp$. Similarly, the graphs arising from
2-valent vertices $\vrt$ of the span\rx{2.65} form IHX doublets of the second line of \fg{f1.11}.
Hence $\xhtfaap$ maps $\SsthoLpHD\otimes\xImvDr$ into
$\ccHIHXapD$.
\qed

Thus we proved that the spaces $\cHLHD$ and $\ccHapD$ are isomorphic. In view
of \ex{2.60}, the following lemma is the last step
in establishing the isomorphism\rx{2.49}.
\begin{lemma}
\label{l2.7}
The isomorphisms $\xhfaap,\xhfiaap$ between $\cHLHD$ and
$\ccHapD$
map the
spaces
$\tcDCLcIHX$ and $\tcDCLcIHXzp/(\tcDCLcIHXop\cap\tcDCLcIHXzp)$ into each other.
\end{lemma}
\skproof
We will prove the claim of the lemma for $\xhfaap$, the proof for $\xhfiaap$
is similar.
Consider the definition of $\tcDCLcIHX$. Let $\xD$ be a (1,3,4)-valent graph
and $\xDic, \xDir$ ($1\leq i\leq 3$) be the corresponding IHX triplets.
The claim of the lemma follows from the commutativity of the
diagram
\qq
\begin{diagram}
\node{\cHLHD}
  \arrow[3]{e,t}{\fCDIHXD}
  \arrow{s,r}{\xhfaap}
\node[3]{\tcDCLc}
  \arrow{s,r}{\xhfaap}
\\
\node{\ccHapD}
  \arrow[3]{e,t}{\bopDpXDap\fCDIHXDp}
\node[3]{\tcDCLcp/\tcDCLcIHXop}
\end{diagram}
\label{2.66}
\qqq
and we leave the details for the reader.
\qed

Thus we established the isomorphism\rx{2.49}. Now it remains to prove the
commutativity of the diagram\rx{2.49*}. The simplest way to do it is to use
Theorem\rw{t2.3*}. It is a straightforward exercise to check the
commutativity of the `inverse' diagram
\qq
\begin{diagram}
\node{\cDCLc}
  \arrow{s,r}{\TgcD}
\node{\cDCLcp}
  \arrow{s,r}{\TgcD}
  \arrow{w,t}{\xhfiaap}
\\
\node{\TxLLHCg}
  \arrow{e,=}
\node{\TxLLHpCg}
\end{diagram}
\label{2.67}
\qqq
for the special graphs as well as for a strut and for a graph, which
has a single 3-valent vertex and three legs (a calculation is only required
if one of these legs has label $\lap$, so that $\xhfiaap$ converts it into a
strut). Any other graph can be constructed by gluing legs in the products
of these elementary graphs, and Theorem\rw{t2.3*} guarantees that the weight
system homomorphisms for both $\cDCLc$ and $\cDCLcp$ behave in the same way
under leg gluing.

This completes the proof of Theorem\rw{t2.4}.

\subsubsection{Elementary leg conversion, gluing and renaming of virtual labels}

The isomorphism\rx{2.49} explains the meaning of the term `\vrtl\ label': if
$\laz\in\xLZH$,
then the elements of the algebra $\tcHazD$ represent \vrtl\ $\laz$ legs, which
can be converted into real legs by the isomorphism $\xhfazazp$. Therefore
we can define the gluing between an integer virtual label $\laz\in\xLZH$ and
a lower $\Srs$ Cartan label $\lb\in\xLCSu$
\qq
\xmap{\glazbm}{\cDCLc}{\cDCLc}
\label{2.67x}
\qqq
by the formula
\qq
\glazbm = \xhfiazazp\;\glazpbm\;\xhfazazp.
\label{2.67x1}
\qqq
The disadvantage of this formula is that it requires a total conversion of
$\laz$ legs as an intermediate step, and this makes it inapplicable to the
gluing of rational virtual labels. However, there is an alternative gluing
formula, requiring a conversion of only those $m$ $\laz$ legs which are about
to be glued. This formula coincides with \ex{2.67x1} for integer virtual labels
and at the same time defines the gluing map\rx{2.67x} also for rational virtual
labels.

Let the label $\laz\in\xLH$ be either integer or rational and let
$\xLppZH=\xLZH\cup\{\lazpp\}$, $\xLppQH=\xLQH$. We define a linear
operator
$\azdazpp\in\End\lrbc{\bsaLHpp\cohrDa}$ by the conditions
$\azdazpp(\edgaz)=\edgazpp$ and $\azdazpp(\edga) = 0$, if
$\la\neq\laz$. This operator can be extended to the derivation
$\drLazdazpp\in\Der\lrbcs{\cHLHppD}$.

\begin{theorem}
\label{t2.5}
If the label $\laz$ is integer, then
the virtual leg gluing operator\rx{2.67x1} can be presented as
\qq
\glazbm = \glazpbm\,\xhfazppazp\, (\drLazdazpp)^m/m!,
\label{2.67x4}
\qqq
where
$\xhfazppazp$
is the isomorphism, which converts the virtual
$\lazpp$ legs into the real $\lazp$ legs.
\end{theorem}
\skproof
Equation\rx{2.67x4} follows from the fact that if we convert the virtual
$\laz$ and $\lazpp$ legs into the real legs with the same labels, then the
operator $\drLazdazpp$ acts in the following way: it converts a graph $\xD$ into
a sum of all possible graphs constructed from it by relabeling one $\laz$
1-vertex as $\lazpp$.\qed

If the virtual label $\laz$ is rational, then we use \ex{2.67x4}
as the definition of the gluing operator\rx{2.67x}.

There is another useful operator which manipulates virtual labels.
For a pair of labels $\laz,\lao\in\xLH$ suppose that either
$\laz\in\xLZH$ or $\laz,\lao\in\xLQH$. Consider a linear operator
$\rnazo\in\End\lrbc{\bsaLHpp\cohrDa}$ defined by the conditions
$\rnazo(\edgaz) = \edgao$ and $\rnazo(\edga) = \edga$ if
$\la\neq\laz$. We extend $\rnazo$ to the algebra homomorphism
$\rnhazo\in\Hom(\cHLHD)$, which `renames' $\laz$ into $\lao$. In
fact,
\qq
\xmap{\rnhazo}{\cHLHD}{\cHv{\xLH\setminus\{\laz\}}(\xD)\subset\cHLHD},
\label{2.67xx5}
\qqq
and if $\laz,\lao\in\xLZH$ and we convert them into real labels, then
$\rnhazo$ becomes the operator, which relabels all $\laz$ 1-vertices into $\lao$.
One can verify that if $\laz\in\xLZH$ and $\zx\in\cHLHD$, $\dgraz\zx=m$, then
\qq
{(\drLazdazpp)^m\over m!}\; \zx = \rnhv{\laz\lazpp}\;\zx.
\label{2.67x5}
\qqq
This relation implies that for any $\zx\in\cHLHD$
\qq
\atv{\lrbcs{\exp(\drLazdazpp)\; \zx }}{\laz=0} =
\rnhv{\laz\lazpp}\;\zx,
\label{2.67x6}
\qqq
where the symbol $\atv{}{\laz=0}$ denotes the projection of
$\cHLHD$ onto the subspace of $\dgraz=0$.

\subsubsection{Integer and rational virtual legs}

Suppose that $\laz\in\xLZH$ and let
$\xLZHp=\xLZH\setminus\{\laz\}$, $\xLQHp = \xLQH\cup\{\laz\}$.
Then for every graph $\xD\in\sCCGRL$ there is a natural injection
of algebras
$\cHLHD\hookrightarrow\cHLHpD$, hence there is an injection
\qq
\tcDCLc\hookrightarrow\tcDCLcxp.
\label{2.70}
\qqq
\begin{conjecture}
\label{c2.1}
The injection\rx{2.70} descends to the injection
\qq
\cDCLc\hookrightarrow\cDCLcxp.
\label{2.71}
\qqq
\end{conjecture}

This conjecture is not necessary to achieve our main objective,
that is, to prove the 1-loop exactness of the formal Kirillov
integral. However, it allows us to define the integration over the
virtual rational labels, which will be a part of a surgery formula
for the universal $U(1)$-reducible rational connection contribution
invariant of links. Also, Conjecture\rw{c2.1} has a useful
\begin{corollary}
\label{cr2.1}
Let $\xLHpp = \xLZH\cup\xLQHp$, then the natural map
\qq
\cDCLcxpp\hookrightarrow\cDCLcxp
\label{2.x72}
\qqq
is an injection.
\end{corollary}
\proof
Theorem\rw{t2.7} implies the injection
$\cDCLcxpp\hookrightarrow\tcDCLc$, then we combine it with the
injection\rw{2.71}.\qed

\subsubsection{Oriented root circle}
In order to define the formal gaussian integration over the
$\Wrs$ labels, we have to modify the definition of the Jacobi graph algebra
$\cDCLc$ by replacing the root circle $\rcrl$ with the oriented
root circle $\rocrl$. We define the corresponding algebra
\qq
\cHLHv{\rocrl} = \SsthoLHv{\rcrl}
\label{2.x73}
\qqq
the latter being defined by \ex{2.30*1x}. Now $\rcrl$ becomes a
`secondary' graph through the injection
\qq
\cHLHv{\rcrl}\hookrightarrow\cHLHv{\rocrl},\quad\zx\mapsto 2\zx.
\label{2.x74}
\qqq
The factor 2 in this formula suggests that an unoriented circle is
a sum of two oriented circle with opposite orientations.

\begin{remark}
\rm
Since we
are not going to develop a complete theory of a Jacobi graph algebra
with oriented root edges, then we have to be careful in applying
the leg gluing operation to the elements of $\cHLHv{\rocrl}$ so as
not to encounter more general graphs with oriented root edges. We
will make sure that whenever gluing is necessary, it will be
reduced to the manipulations with $\rcrl$ through the
injection\rx{2.x74}.
\end{remark}

Now it remains to define the weight system map for
$\cHLHv{\rocrl}$. We set $\TxLHroHh = \TxLHrHh$.
Let $\Dertp\subset\Dert$ be the subset of
positive roots of $\mfg$, then for $\log\zy,\,{\zx\over\zy}\in\cHLHv{\rocrl}$
we define
\qq
\TgcD(\log\zy) = \svlDertp \log\xfhl(y),\quad
\TgcD\lrbc{\zx\over\zy} = \svlDertp {\xfhl(\zx)\over\xfhl(\zy)},
\label{2.x75}
\qqq
where $\xfhl$ is the same as in \ex{2.79}. Obviously, this definition is compatible with the
injection\rx{2.x74} and the weight system map\rx{2.79}, that is,
the following diagram is commutative:
%
%
\qq
\begin{CD}
\cHLHv{\rcrl} @>>> \cHLHv{\rocrl}
\\
@VV{\TgcD}V   @VV{\TgcD}V
\\
\TxLHrHh @= \TxLHroHh
\end{CD}
\label{2.x77}
\qqq

\nsection{Graph-based differential geometry}
We are going to construct a Jacobi graph algebra model of a local
differential geometry
on a space, which is a product of several copies of the spaces
$\mfgs$, $\mfhs$ and $\mfrs$. The
weight system homomorphism $\Tg$ should map the graph objects and operations into
the corresponding objects and operations of differential geometry
on that space. Locality means that instead of the general smooth
functions we will be dealing with either polynomials or formal
power series (or a field of their quotients) of the Lie algebra
variables.

Our
construction works for all graph algebras $\cBL$, $\cBCL$ and
$\cDCLc$, so we will try to keep our discussion as general as possible,
without committing ourselves to a particular algebra, or using
$\cDCLc$ as a basic example, whereas the restrictions to $\cBL$ and
$\cBCL$ are obvious.

The objects
of our differential geometry, such as functions, tensor fields and
diffeomorphisms, will be defined through a particular labeling of
the corresponding graphs. Thus, if we say that an object is
\emph{\xbsd} on a certain type of graphs, then in the case of
algebras $\cBL$ and $\cBCL$ this means that these objects are
linear combinations of these graphs while in the case of $\cDCLc$
this means that these objects are linear combinations of the
elements $\zx\in\cHLHD$, where $\xD$ are the graphs of that type.
The operations of differential geometry, such as differentiation
and integration, will be defined in terms of leg gluing, hence
these definitions will apply equally to all graph algebras.

\subsection{Functions, vector fields and differentiation}
\subsubsection{Functions, differential forms and the differential operator}
First, we select the (finite) set of real labels $\xVst$, whose elements we
will call \emph{variables}. The labels in $\xVst$ must all be of the
upper $\Srs$ type. They can be either Cartan, root, or total.
In
case of $\cDCLc$, the set $\xVst$ must include all virtual labels:
$\xLH\subset\xVst$.

We split $\xVst$ into two non-intersecting subsets
$\xVst=\xCst\cup\xPst$, whose elements are called
\emph{coordinates} and \emph{parameters} respectively.

The \emph{functions} (of the variables of $\xVst$) are \xbsd\ upon the graphs,
whose 1-vertices are labeled exclusively by the labels of $\xVst$.
The space of functions
is denoted as $\gRVst$. The
meaning of the definition becomes transparent, if we recall that for a
linear space $V$ there is a canonical isomorphism
\qq
\Srst V \equiv \IRfv{x},\qquad x\in V^\ast.
\label{3.1}
\qqq
Therefore, if we define
\qq
\VbVst = \bopAbVst\aVA,\qquad \VVLQH = \bopALQH\aVA,
\label{3.2}
\qqq
where
\qq
\aVA=
\begin{cases}
\mfgs &\text{if $\lA$ is total}
\\
\mfhs &\text{if $\lA$ is Cartan}
\\
\mfrs &\text{if $\lA$ is root}
\end{cases}
\label{3.3}
\qqq
then the weight system homomorphism $\Tg$ maps $\bgRVst$ into the algebra of
fractions (`formal functions' on $\VbVst$)
\qq
\xRfbVst = \stdfl{ {f\over g}
}{f\in\IRfx,\;x\in\VbVst;\,g\in\IRfx,\;x\in\VVLQH}.
\label{3.4}
\qqq
%


Let $m$ be a positive integer number. \emph{Differential $m$-forms} are based upon
graphs, all of whose 1-vertices are labeled by the labels of $\xVst$ except
for $m$ 1-vertices, which are labeled by the special upper $\Wrs$-type labels
$\ydA$ ($\lA\in\xCst$).
We denote the space of $m$-forms as $\gOmcVstm$.
Obviously, $\gOmcVstz=\bgRVst$.
The weight system homomorphism $\Tg$ maps $\gOmcVstm$
into the space of formal $m$-forms on $\VbVst$:
\qq
\xOmcVstm = \bigwedge\nolimits^m\VcVsts\otimes\xRfbVst.
\label{3.5}
\qqq

The usual graph algebra multiplication turns the space of all
forms
\qq
\gOmcVstt = \bopmzi \gOmcVstm
\label{3.5*}
\qqq
into a graded algebra. $\Tg$ converts the multiplication into the
wedge product of $\xOmcVstm$.

For a label $\lA\in\cxVst$ we introduce an `auxiliary' lower $\Srs$-type label
$\xalA$ and then define a partial differential operator
$\xmap{\xdA}{\gOmcVstm}{\gOmcVstmo}$ by the formula
\qq
\xdA\,\omega =
(\ydA\tstr\xalA)\;\glov{\xalA,\lA}\;\omega.
\label{3.6}
\qqq
Since the labels $\lA$ and $\xalA$ are of $\Srs$-type, while $\ydA$
is of
$\Wrs$-type, then
\qq
\xdA\xdB = - \xdB\xdA,\quad
\xdA^2=0,
\label{3.7}
\qqq
and we define the differential operator
\qq
\xmap{d}{
\gOmcVstm}{
\gOmcVstmo},\qquad
d = \sAcVst \xdA,
\label{3.8}
\qqq
which satisfies the condition $d^2=0$.

\subsubsection{Vector fields and Lie derivative}

The \emph{vector fields} are \xbsd\ upon the graphs, all of whose 1-vertices are
labeled by the elements of $\bxVst$, except for one vertex, which is labeled by
a special lower $\Srs$-type label $\ydlA$ ($\lA\in\cxVst$). We
denote the space of vector fields as $\VctcV$. The weight system
map $\Tg$ maps $\VctcV$ into the space of formal vector fields
over $\VcVst$:
\qq
\VctfbV = \VcVst\otimes\xRfbVst.
\label{3.9}
\qqq

For a vector field $\xxi\in\VctcV$ we define the contraction operator
\qq
\xmap{\ixi}{\gOmcVstm}{\gOmcVstmmo}
\label{3.10}
\qqq
by the formula
\qq
\ixi \omega =
\sAcVst \xxi\;\glov{\ydlA,\ydA}\;\omega.
\label{3.11}
\qqq
We also define the Lie derivative
\qq
\xmap{\drLxi}{\VctcV}{\VctcV},\quad
\xmap{\drLxi}{\gOmcVstm}{\gOmcVstm}
\label{3.12}
\qqq
by the formulas
\qq
\drLxi\,\xeta & = & \sAcVst(\xxi\;\glov{\ydlA,\lA}\;\xeta -
\xeta\;\glov{\ydlA,\lA}\;\xeta),\quad\xeta\in\VctcV,
\label{3.13}
\\
\drLxi\,\omega & = & \sAcVst\lrbs{\xxi\;\glov{\ydlA,\lA}\;\omega +
\sBcVst\lrbcs{(\ydB\tstr\xalB)\;\glov{\xalB,\lB}\;\xxi}
\;\glov{\ydlA,\ydA}\;\omega}.
\label{3.14}
\qqq
The weight system map $\Tg$ converts $\ixi$ and $\drLxi$ into the
familiar operators of differential geometry.

It is a straightforward exercise in combinatorics to verify the
following
\begin{theorem}
\label{t3.1}
The space $\VctcV$ is a Lie algebra with respect to the bracket
\qq
[\xxi,\xeta] = \drLxi\,\xeta.
\label{3.14*}
\qqq
The action\rx{3.12} provides a
representation of this Lie algebra, and for any $\xxi\in\VctcV$,
$\omega\in\gOmcVstm$
the `homotopy formula' holds:
\qq
\drLxi\omega = \ixi\,d\omega + d\,\ixi\omega.
\label{3.15}
\qqq

\end{theorem}

\subsubsection{\Xmflds\ and smooth maps}
For the rest of the section we assume that
$\xLH\subset\xPst$,
that is,
the virtual
labels of $\xLH$ are parameters. This is not a restrictive
assumption for the integer virtual labels, because, according to
Theorem\rw{t2.4}, they can be converted into the real ones, and we
will keep this possibility in mind. We will use notation
$\xPstr=\xPst\setminus\xLH$ for the set of parameters, which are real
variables.

Let $\cxVsto$ and $\cxVstt$ be two more sets of coordinates. A
\xmfld\ $\xmVotM$ is \xbsd\ on graphs, all of whose 1-vertices are
labeled by the labels of $\xVst$ except for two 1-vertices, one of
which is labeled by $\ydlAo$ ($\lAo\in\cxVsto$) and the other one is
labeled by $\ydAt$ ($\lAt\in\cxVstt$). When the
indices $\cxVsto$ and $\cxVstt$ are clear from the context, we may
drop them from the notation leaving just $\mM$.
The weight system map $\Tg$
maps \xmflds\ into the space of matrix-fields
over $\VcVst$: $\MatfbV = \VcVst\otimes\VcVsts\otimes\xRfVst$.
The \xmflds\ can be matrix multiplied:
\qq
\xmfv{\xCstth}{\xCstt}{\mMt}\times
\xmfv{\xCstt}{\xCsto}{\mMo} = \sACstv{2}
\xmfv{\xCstth}{\xCstt}{\mMt}
\;\;\glov{\ydAt,\ydlAt}\;\;
\xmfv{\xCstt}{\xCsto}{\mMo}.
\label{3.16}
\qqq

Let $\xCstp$ be a set of coordinates. A \emph{smooth map}
\qq
\xmap{\xF}{\xCst}{\xCstp}
\label{3.17}
\qqq
is \xbsd\ on graphs, all of whose
1-vertices are labeled by the labels of $\xVst$ except for one
1-vertex, which is labeled by a lower $\Srs$-type label $\yalAp$
($\lap\in\xCstp$). We will also denote the map\rx{3.16} as $\xCstp
= \xF(\xCst)$.
The \emph{derivative} of the smooth
map\rx{3.16} is the \xmfld\
\qq
\xmfFpVst = \sum_{\lA\in\xCst\atop\lAp\in\xCstp}
\stin{\ydlAp\tstr\yalAp}\;\gl{\yalAp,\yalAp}{1}
\;\xF\; \gl{\lA,\yalA}{1} \stin{\yalA\tstr\ydA}
\label{3.18}
\qqq
In the future we will drop the indices from the notation of
derivative.

For a set of labels $\xSst$ let $\mlAS=(\lA)_{\lA\in\xSst}$ be the
list of its elements. If the set $\xSst$ is clear from the
context, then we use a simplified notation $\mlA$.

Let $\xmap{f}{\xSsto}{\xSstt}$ be a
bijection between two label sets. Then we denote
\qq
\gl{\mlAv{\xSsto},\mlAv{\xSstt}}{m} =
\prod_{\lA\in\xSsto} \gl{\lA,f(\lA)}{m}.
\label{3.18*}
\qqq
For any
object
$\yT$ (such as a function, a differential form,
a vector field, a smooth map, \etc) which depends on the variables
$\xVstp=\xCstp\cup\xPst$ we
define the coordinate substitution $\yT(\xF)$ generated by\rx{3.17}
as
\qq
\yT(\xF) = \yT
\;\glLRmAyAp\;
e^{\xF},
\label{3.19}
\qqq
where the exponential $e^{\xF}$ is defined though the power series in
terms of graph algebra multiplication. The formula\rx{3.19} as
a whole just means that all $\lAp$ 1-vertices ($\lAp\in\xCstp$)
of $\yT$ are glued to the corresponding $\yalAp$ 1-vertices of
the graphs of $\xF$. The substitution\rx{3.19} defines the
composition of smooth maps: for two smooth maps
$\begin{CD}\xCsto @>\xF_{21}>>\xCstt@>\xF_{32}>>\xCstth\end{CD}$ we
define
\qq
\xF_{32}\circ \xF_{21} = \xF_{32}(\xF_{21}).
\label{3.20}
\qqq
The identity smooth map $\xmap{\xId}{\xCst}{\xCst}$
corresponding to this composition rule is
\qq
\xId = \sACst \yalA\tstr\lA.
\label{3.20*}
\qqq
The derivative of the composition is given by the `chain rule'
\qq
(\xF_{32}\circ \xF_{21})\p =
\xFp_{32}(\xF_{21})\times \xFp_{21}.
\label{3.21}
\qqq

The smooth maps\rx{3.17} have a (contragradient) action on functions of
$\xVstp$ through the
substitution\rx{3.19}: $f\mapsto f(\xF)$, where $f$ is a function of
coordinates $\xVstp$. They also act on $m$-forms as
\qq
\omega\mapsto
\omega(\xF)
\;\glLRydmAp\;
e^{\xFp},
\label{3.22}
\qqq
where the exponential $e^{\xFp}$ comes from the graph algebra
multiplication, rather than from that of \xmflds.

\subsubsection{Diffeomorphisms}

For any pre-Jacobi graph algebra element $\yT$ we define its
\emph{\xstp} $\xstrv{\yT}$ as the part of $\yT$ which is
\xbsd\ on strut graphs (that is, graphs consisting of a single
edge). Since struts do not participate in the IHX triplets, then
the notion of the \xstp\ extends to the Jacobi graph algebras.

Consider the \xstp\ of the smooth map\rx{3.17}. It splits into
a `linear' and `constant-parametric' parts:
\qq
\xstrF =
\sum_{\lA\in\xCst\atop\lAp\in\xCstp}
\yluFAtAp\stin{\yalAp\tstr\lA}
+
\sum_{\lB\in\xPstr\atop\lAp\in\xCstp}
\yvuFBtAp\stin{\yalAp\tstr\lB}
+
\sum_{\lB\in\xLH\atop\lAp\in\xCstp}
\yvuFBtAp\stin{\lB\fidv{\yalAp}}.
\label{3.23}
\qqq
%
Thus the \xstp\rx{3.23} is determined by two matrices
\qq
\ylF = \mtr{\yluFAtAp},\quad\yvF = \mtr{\yvuFBtAp}.
\label{3.24}
\qqq
The nature of the coefficients $\yluFAtAp$ and $\yvuFBtAp$
depends on the choice of
the Jacobi graph algebra and the labels $\lA$ and $\lAp$. If the
algebra is $\cDCLc$ and the labels are root, then according
to\rx{2.30*x1}, $\yluFAtAp=\ylLHuFAtAp\in\RLH$
and $\yvuFBtAp=\ylLHuFBtAp\in\RLH$. In all other cases
$\yluFAtAp,\yvuFBtAp\in\IR$.


Suppose that $\absv{\xCst}=\absv{\xCstp}$.
The smooth map\rx{3.17} is called a \emph{diffeomorphism}, if there exists an
inverse smooth map: $\xFi\circ\xF =\xId$, where
$\xmap{\xFi}{\xCstp}{\xCst}$.
\begin{theorem}
\label{t3.2}
The smooth map\rx{3.17} is a diffeomorphism iff
$\atv{\det\ylF}{\mlaLZH=0}\neq 0$.
\end{theorem}
\begin{lemma}
\label{l3.1}
The theorem holds in the case when $\xCstp=\xCst$ and
$\xstrF=\xId$.
\end{lemma}
\proof
We will construct $\xFi$ `perturbatively'.
Consider a sequence of smooth maps
$\xFin$ ($n\geq 1$) defined recursively
\qq
\xFio = \xId, \quad\quad \xFino = \xFin - \xFin\circ\xF.
\label{3.26}
\qqq
%
Let us prove by induction in $n$
that
\qq
\lrbc{\xId + \smon \xFim}\circ\xF = \xId - \xFino.
\label{3.27}
\qqq
For $n=0$ it is obvious. Suppose that we proved it for $n=k$, then
it holds for $n=k+1$, because
\qq
\lrbc{\xId + \smono\xFim}\circ\xF
& = &
\lrbc{\xId + \smon\xFim + \xFino}\circ\xF
\nonumber
\\
& = &
\lrbc{\xId + \smon\xFim}\circ\xF + \xFino\circ\xF
\nonumber
\\
& = &
\lrbc{\xId + \smon\xFim}\circ\xF + \xFino - \xFint
\nonumber
\\
& = & (\xId - \xFino) + \xFino - \xFint = \xId - \xFint.
\label{3.28}
\qqq

Since $\xF = \xId + \xDF$, where $\xDF$ consists of the graphs
with at least two edges, then
it is easy to prove by induction that $\xFin$ defined by
the recursion relations\rx{3.26} consists of the
graphs with at least $n$ edges. Hence the sum
\qq
\xtF = \smoi \xFim
\label{3.29}
\qqq
is well-defined and \ex{3.27} implies that we can take
$\xFi=\xtF$.\qed

\pr{Theorem}{t3.2}
We will prove it only one way (the proof of the inverse claim is simple).
Suppose that $\atv{\det\ylF}{\mlaLZH=0}\neq 0$, then $\det\ylF$ is
invertible in $\RLH$ and therefore there exists an inverse matrix
$\ylFi$. Consider a smooth map $\xmap{\xFistr}{\xCstp}{\xCst}$
containing only the struts with the matrices
\qq
\ylv{\xFistr} = \ylFi,\quad
\yvv{\xFistr} = - \ylFi\;\yvF.
\label{3.30}
\qqq
Since the matrices\rx{3.24} transform simply under the
composition\rx{3.20}
\qq
\ylzFv{31} = \ylzFv{32}\;\ylzFv{21},\quad
\yvzFv{31} = \yvzFv{32} + \ylzFv{32}\;\yvzFv{21},
%
\label{3.31}
\qqq
then
\qq
\xstrv{\xFistr\circ\xF} = \xId
\label{3.32}
\qqq
and, according to Lemma\rw{l3.1}, the smooth map
$\xmap{\xFistr\circ\xF}{\xCst}{\xCst}$ has an inverse. Now we can
construct the inverse of $\xF$:
\qq
\xFi = (\xFistr\circ\xF)^{-1}\circ\xFistr.
\label{3.33}
\qqq
\qed

All diffeomorphisms $\xmap{\xF}{\xCst}{\xCst}$ form a Lie group
$\xDiffC$. This group is connected. Indeed, the subgroup
$\xDiffstrC\subset\xDiffC$, which consists of purely strut
diffeomorphisms, is connected, and the map
$[0,1]\longrightarrow\xDiffC$, $t\mapsto \xstrF + t\,(\xF-\xstrF)$
connects $\xF$ with $\xstrF\in\xDiffstrC$.

In addition to the action\rx{3.22} on $\gOmcVstt$,
the group $\xDiffC$ acts also on $\VctcV$ according to he formula
\qq
\xxi\xmapt{\xF} \xxi(\xFi)
\;\glLRymA\;
e^{\xFp(\xFi)}.
\label{3.34}
\qqq

We can identify the tangent space at identity $\TId\xDiffC$ with
$\VctcV$ in the following way: for a one-parametric family of
diffeomorphisms $\xF(t)$ such that $\xF(0)=Id$ we consider the
smooth map $\atv{\del_t\xF(t)}{t=0}$ and convert it into a vector
field by replacing the $\xalA$ labels with $\ydlA$. We leave it
for the reader to verify that $\VctcV$ with the
commutator\rx{3.14*} is the Lie algebra of $\xDiffC$, that\rx{3.4}
is the adjoint action and that the Lie derivatives\rx{3.14}
correspond to the action\rx{3.22}.

\subsection{Integration and determinants}
\subsubsection{Gaussian integration}
A graph algebra element $\yT$ is called \emph{\sbstC}, if there exists
a number $N$ such that $\yT$ can
be presented as a linear combination of graphs, each of which has
at most $N$ $\xCst$-labeled struts as connected components. An
element $\yT$ is called \emph{\gsnC}, if it can be presented as
\qq
\yT = \ephvQP,
\label{3.35}
\qqq
where $\yP$ is \sbstC, while
\qq
\xQ =
\sABCst
\yquQAB\stin{\lA\tstr\lB}.
\label{3.36}
\qqq
As usual, we denote $\yqQ = \mtr{\yquQAB}$. We define the
conjugate matrix $\yqQs$ as
\qq
\yquQsAB =
\begin{cases}
\yquQsBA &\text{if $\yquQsAB\in\IR$}
\\
\yquQsBA(-\mlaLH) &\text{if $\yquQsAB\in\RLH$}.
\end{cases}
\label{3.37}
\qqq
%
Since
\qq
\yquQsAB\stin{\lA\tstr\lB} = \yquQBA\stin{\lB\tstr\lA},
\label{3.38}
\qqq
then for a fixed $\xQ$ we can impose a condition
\qq
\yqQs=\yqQ,
\label{3.39}
\qqq
after which
the graph algebra element $\xQ$ determines the matrix
$\yqQ$ uniquely.

If $\det\yqQ\xneq 0$, then we call $\xQ$ and $\yT$ \emph{\ndgn}.
In this case we define the inverse
\qq
\xQi = \sum_{\lA,\lB\in\xCst}\yquQiAB\stin{\yalA\tstr\yalB}.
\label{3.40}
\qqq
where $\yqQi = \yqQ^{-1}$.

For the graph algebra $\cBL$ we define
\qq
\log\det\xQ =
\log\det\yqQ\stsi{\tcrl}.
\label{3.40*}
\qqq
In the case of the Cartan-split graph algebras $\cDCL$ and $\cDCLc$
the matrix
$\yqQ$ consists of two blocks corresponding to Cartan and root
labels of $\xCst$: $\yqQC=\mtr{\yquQAB}_{\lA,\lB\in\xLC}$,
$\yqQr=\mtr{\yquQAB}_{\lA,\lB\in\xLr}$,
so we define
\qq
\log\det\xQ =
\log\det\yqQC \stsi{\ccrl} + \log\det\yqQr\stsi{\rcrl}
\label{3.41}
\qqq
We use \eex{3.40*} and\rx{3.41} in order to define
\qq
\det\xQ = e^{\log\det\xQ}.
\label{3.42}
\qqq
More generally, for any $\alpha\in\IR$ we define
\qq
(\det\xQ)^\alpha = e^{\alpha\log\det\xQ}.
\label{3.43}
\qqq

Let $\yT$ be a \ndgn\ \gsnC\ (\ndgC) graph algebra element of the form\rx{3.35},
then we define its (gaussian) integral as
\qq
\xintCv{\ephvQP} = \dtQmh \lrbc{\yP\;\glLRmAyA\;\ephvimQ},
\label{3.44}
\qqq
where $\mlAC = (\lA)_{\lA\in\xCst}$. Note that since $\yP$ is \sbstC, then the gluing in the \rhs of
this equation is well-defined.

Let us consider a simple but important example of the
integral\rx{3.44}. Let $\xCstp$ be another set of coordinate
labels and let
\qq
\xR = \sACpp \yquRAAp\stin{\lA\tstr\lAp}.
\label{3.44*}
\qqq
Then $\xP=e^{\xR}$ is \sbstC, and a combination of \eex{3.44}
and\rx{2.1x2} indicates that
\qq
\xintCv{e^{\hvQ+\xR}} = \dtQmh\,e^{-\hlfv\xRs\xQi\xR},
\label{3.44*1}
\qqq
where $\xRs\xQi\xR$ is a quadratic function of $\xCstp$ defined by
the condition $\yqv{\xRs\xQi\xR} = \yqRs\yqQ\yqR$.

\subsubsection{Properties of the gaussian integral}
Graph algebra gaussian integral\rx{3.44} shares many properties of the ordinary
integrals. However these properties can not be deduced from the Riemann
sums, but they have to be established combinatorially in line with the
nature of the definition\rx{3.44}.

\begin{lemma}
\label{l3.2}
The integration formula\rx{3.44} is determined by the following two
properties of the integral:
\begin{itemize}
\item
Consider two labels $\lB,\lC\not\in\xCst$ and let $\glBCarb$ denote either
$\glBCm$ or $\glLRBC$. Then the gluing
$\glBCarb$ applied to $\yP$ commutes with the integration:
\qq
\xintCv{\ephvQ \,\glBCarb(\yP) } =
\dtQmh \lrbs{\lrbcs{\glBCarb(\yP)}\;\glLRmAyA\;\ephvimQ}.
\label{3.47}
\qqq
\item
Denote
\qq
\ImAyA = \sACst \lA\tstr\yalA,
\label{3.45}
\qqq
then
\qq
\xintCv{\ephvQIm}
= \dtQmh\,\ephvimQ.
\label{3.46}
\qqq

\end{itemize}
\end{lemma}
\proof
It is easy to derive the first property from \ex{3.44} and the
second property from \ex{3.44*1}.
Conversely,
observe that if $\yhtP$ is the graph algebra element constructed
from $\yP$ by replacing all labels $\lA\in\xCst$ in its graphs with the
corresponding labels $\ydlA$, then
\qq
\yP =
e^{\ImAyA}\;
\glLRv{\yamlA,\ydmlA}\;\yhtP.
\label{3.48}
\qqq
Now the properties\rx{3.47} and\rx{3.46} imply that
\qq
\xintCv{\ephvQP} & = & \xintCv{\ephvQIm\;\glLRv{\yamlA,\ydmlA}\;\yhtP}
= \lrbc{
\xintCv{\ephvQIm}
}
\glLRv{\yamlA,\ydmlA}\;\yhtP
\nonumber
\\
& = &\dtQmh\,\ephvimQ \glLRv{\yamlA,\ydmlA}\;\yhtP,
\label{3.49}
\qqq
which is equivalent to \ex{3.44}.\qed

This lemma allows us to reduce the proof of some properties of
the general integral\rx{3.44} to the proof of these properties for
the integral\rx{3.46}.

Let us split the coordinate set $\xCst$ into two subsets
$\xCst=\xCsto\cup\xCstt$. Then the matrix $\yqQ$ splits into the
blocks
\qq
\yqQ =
\left(
\begin{array}{c|c}
\yqQoo & \yqQot \\
\hline
\yqQots & \yqQtt
\end{array}
\right),
\label{3.50}
\qqq
and the whole expression\rx{3.36} splits into the corresponding
sum
\qq
\xQ = \xQoo + 2\xQot + \xQtt.
\label{3.51}
\qqq
\begin{theorem}[Fubini]
\label{t3.3}
Suppose that both $\xQoo$ and $\xQ$ are \ndgn. Then the
integral\rx{3.44} is equal to the iterated integral
\qq
\xintCv{\ephvQP} =
\xintv{
\lrbc{ \xintv{\ephvQP}{\xCsto} }
}{\xCstt}.
\label{3.52}
\qqq
\end{theorem}
\proof
In view of Lemma\rw{l3.2}, it is sufficient to prove the theorem
only for the integral\rx{3.46}. Then all integrals involved in
calculating both sides of \ex{3.52} are of the type\rx{3.44*1} and
hence its proof becomes a straightforward exercise in linear
algebra.\qed

\begin{theorem}[differentiation over a parameter]
Suppose that a
\ndgC\
algebra element $\yT$ depends on a real
parameter $\xt$. Then
$\dlt\yT$ is \ndgC\ and
\qq
\xintCv{\dlt\yT} =
\dlt\lrbc{\xintCv{\yT}}.
\label{3.53}
\qqq
\end{theorem}
\proof
Obviously,
\qq
\dlt(\ephvQP) = \hlfv\;\ephvQP\, \dlt\xQ + \ephvQ\,\dlt\yP.
\label{3.54}
\qqq
If $\yP$ is \sbstC, then $\yP\,\dlt\xQ$ and $\dlt\yP$ are also
\sbstC, hence $\dlt(\ephvQP)$ is \ndgC. Since
\qq
\dlt \det\yqQ = \det\yqQ \Tr\lrbcs{\yqQi\,\dlt\yqQ},
\label{3.55}
\qqq
then, according to the definition\rx{3.44}
\qq
\dlt\lrbc{\xintCv{\ephvQP}}  & = &
\dtQmh \Big[(\dlt\yP)\;\glLRmAyA\;\ephvimQ
\label{3.56}
\\
&&
\qquad
-\hlfv\;\Tr\lrbcs{\yqQi\,\dlt\yqQ}
\lrbc{\yP\;\glLRmAyA\;\ephvimQ}
\nonumber
\\
&&
\qquad
+ \hlfv\;\yP\;\glLRmAyA\;\lrbcs{\yqQi(\dlt\xQ)\yqQi}\,\ephvimQ
\Big]
\nonumber
\qqq

\end{document}